# Unconventional Superconductivity

G. R. Stewart
Department of Physics, University of Florida, Gainesville, FL USA 32611

**Gregory Randall Stewart**
**Department of Physics**
**2001 Museum Road**
**University of Florida**
**Gainesville, FL 32611**
**USA**
**\*Email: stewart@phys.ufl.edu**
**001 352 3929263**

**Acknowledgements:** This work was supported by the United States Department of Energy, Office of Basic Energy Sciences, under contract number DE-FG02-86ER45268.

Abstract:

"Conventional" superconductivity, as used in this review, refers to electron-phonon coupled superconducting electron pairs described by BCS theory. Unconventional superconductivity refers to superconductors where the Cooper pairs are not bound together by phonon-exchange but instead by exchange of some other kind, e. g. spin fluctuations in a superconductor with magnetic order either coexistent or nearby in the phase diagram. Such unconventional superconductivity has been known experimentally since heavy fermion $CeCu_2Si_2$, with its strongly correlated 4f electrons, was discovered to superconduct below 0.6 K in 1979. Since the discovery of unconventional superconductivity in the layered cuprates in 1986, the study of these materials saw $T_c$ jump to 164 K by 1994. Further progess in high temperature superconductivity would be aided by understanding the cause of such unconventional pairing. This review compares the fundamental properties of 9 unconventional superconducting classes of materials - from 4f-electron heavy fermions to organic superconductors to classes where only three known members exist to the cuprates with over 200 examples – with the hope that common features will emerge to help theory explain (and predict!) these phenomena. In addition, three new emerging classes of superconductors (topological, interfacial – e. g. FeSe on $SrTiO_3$, and $H_2S$ under high pressure) are briefly covered, even though their "conventionality" is not yet fully determined.

# 1. Introduction

Superconductivity, the phenomenon where the resistivity falls to zero below a certain critical temperature $T_c$ (discovered in Hg at 4.2 K in 1911) and the magnetic flux is expelled from the bulk of the superconductor (a phenomenon discovered in 1933), was explained as due to electron-phonon coupling by Bardeen, Cooper, and Schrieffer (BCS) in 1957. The BCS weak-coupled theory describes the condensation into the superconducting state as due to the exchange of phonons between electrons of opposite spins, an s=0 singlet ground state. Thus the average phonon frequency, $<\omega>$, or equivalently the characteristic Debye temperature, $\Theta_D$, (a phenomenological cutoff of the phonon frequencies) plays an important role in the BCS expression for $T_c$

$$T_c^{BCS} \propto <\omega>\exp(-1/N(0)V) \qquad \text{eq. 1}$$

where $N(0)$ is the electronic density of states at the Fermi energy and V is an average electron-phonon coupling strength parameter. The BCS weak coupling theory has $N(0)V<1$. Many superconductors (e. g. the over 25 superconducting elements [1] and various classes including A15 superconductors [2] - useful for high field magnets), are generally believed to be described by the BCS model, as extended by various improvements (called Eliashberg theory) that encompass stronger coupling.

Defining "Unconventional Superconductor" (UcS) as a material where the Cooper pairing deviates from the BCS description, where the attraction between pairs and driving mechanism for condensation into the superconducting state might come from, e. g., exchange of spin fluctuations, is not a definition new to this review. This is consistent, e. g., with the definition of UcS in the phenomenological theory review of UcS by Sigrist and Ueda [3] from 1991. There are several other ways (as discussed thoroughly in section 2) to define an UcS. Based on a discussion of symmetry [4] by Tsuei and Kirtley, *any* long range ordering transition is accompanied by a lowering of symmetry. In a BCS superconductor the only symmetry broken is the one dimensional global gauge symmetry, U(1), caused by the macroscopic phase coherence that occurs below $T_c$. In an UcS, one or more *additional* symmetries (e. g. time reversal symmetry or reflection symmetry) are broken. p-, d-, and f-wave as well as s± (theorized for iron based superconductors [IBS]) pairing symmetries all break reflection symmetry. Thus, another definition of UcS could be a superconductor in which at least two symmetries are broken at $T_c$. As we will see, although a majority of the 9 superconducting UcS classes discussed herein exhibit such additional symmetry breaking (e. g. time reversal and reflection symmetry breaking in the hole-doped cuprates), others (including recently discovered classes or classes with only a few members) do not. Thus, we will remain with the "non-BCS" definition of UcS as being the best representation of the fundamental theme of this review.

Here, in the Introduction, a short overview discussion will help the reader follow the discussion of the nine classes of UcS, and the three additional classes. As can be inferred from the short discussion above, it is not at present possible to state the cause of UcS, if indeed (as seems unlikely) there is only one such cause. The present review aims to summarize the properties of the various classes of UcS in a way that points to fundamental similarities.

One of the questions important for understanding UcS is: what characteristics of materials are causes of, or at least consistent with, UcS? 1.) As we will see in this review, most UcS have strong electron correlations through their d-electrons (cuprates, IBS, $Sr_2RuO_4$, cobalt oxide hydrates, layered nitrides) or f-electrons (heavy Fermions and coexistent superconductivity and ferromagnetism). Only the organic UcS rely on pairing between p-electrons.

2.) The main instability found in UcS is antiferromagnetism (as will be seen herein in phase diagrams and discussions for the heavy Fermion 115 structure, $CeCu_2Si_2$, $U(Pd,Ni)_2Al_3$, for both electron- and hole-doped cuprates, for the majority (but not all) of the IBS, the non-centrosymmetric superconductors like $CePt_3Si$ that are strongly correlated, organics, and cobalt oxide hydrates.) In addition there are coexistent superconducting and ferromagnetic compounds like $UGe_2$. Another instability found in some UcS is charge density waves (hole-doped cuprates, theorized in some heavy Fermion superconductors, and cobalt oxide hydrates.)

3.) Two dimensional behavior based on the crystalline structure plays an important role and is widespread in many of the UcS: cuprates (where the ratio of resistivities along the c-axis vs the ab-plane can be larger than 1000), IBS ($\rho_c/\rho_{ab} \sim 100$), $Sr_2RuO_4$, layered metal nitride halides, cobalt oxide hydrates, and the 115 structure heavy Fermion superconductors, while there is quasi-1d behavior in some of the organics. This d<3 behavior can play an important role in creating UcS since according to the Mermin Wagner theorem fluctuation effects for d<3 become more important. Such fluctuation effects are predicted (as discussed below) to be a possible superconducting pairing mechanism (superconducting 'glue') in these same aforementioned superconductors. 2d antiferromagnetic fluctuations are repulsive, so a Fermi surface with a sign change in the order parameter where the repulsive interaction is large can cause (see the s± model discussed below for the IBS) attraction and condensation into a superconducting state. Put another way, this sign change in the order parameter under translation by the magnetic ordering vector of the parent phase "tends to remove the detrimental effects of the on-site Coulomb repulsion between the electrons." (Norman [5]). Thus, details of the Fermi surface topology are important and 2d crystalline structures are indeed widespread in UcS. Of course, there are obvious counterexamples, for instance the indisputably 3d nature of the cubic unconventional heavy Fermion superconductor $UBe_{13}$.

With all of these named properties being 'consistent' with UcS, there still remains no ability to <u>a priori</u> predict unconventional superconductivity. Put another way, there is no 'microscopic' theory of UcS, like the BCS theory for conventional superconductivity, since - due to the strong correlations - solving the pairing in UcS is a non-perturbative problem. As Norman has put it [5] "developing a rigorous theory for any of these classes of materials [in the UcS] has proven to be a difficult challenge, and will continue to be one of the major problems in physics in the decades to come." It is the goal of this review to provide a comprehensive overview of UcS in one place to aid in this challenge.

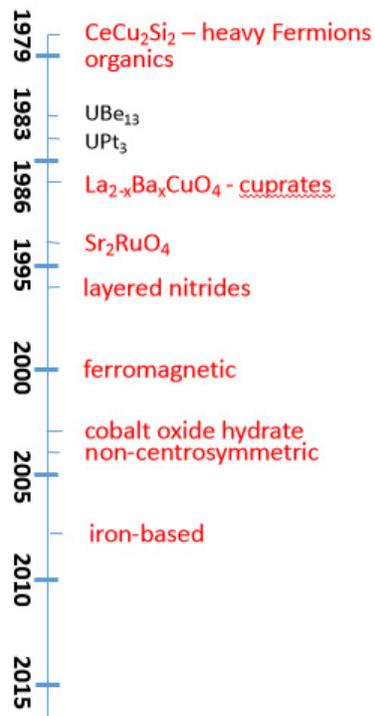

Fig. 1 (color online): Time line of the discovery of some unconventional superconductors

The first superconductor identified as unconventional was $CeCu_2Si_2$, $T_c$=0.6 K, reported [6] by Steglich et al. in 1979. Rather than electron-phonon coupling, the pairing in $CeCu_2Si_2$ has been described as due to some sort of fluctuations (either quantum critical and/or antiferromagnetic.) Steglich et al.'s discovery led to the discovery of other unconventional superconductors of the same class as shown in Fig. 1 ($UBe_{13}$ in 1983, $UPt_3$ in 1984, . . . ), where this class is known as heavy fermion superconductors (HFS) due to the large (sometimes > 100 rest mass of the electron) effective mass $m^*$ of the conduction electrons, where these large $m^*$ electrons are beyond a doubt involved in the superconductivity. $UPt_3$ is a very well studied material with numerous properties (including breaking of the time reversal and reflection symmetries at $T_c$) arguing for UcS. Organic superconductors were discovered [7] soon after $CeCu_2Si_2$ but were not so clearly identified as unconventional until after several years of investigation. Cuprate superconductors, starting [8] with the discovery by Bednorz and Mueller of $La_2CuO_4$ doped with Ba, were the next discovered class of UcS, followed by numerous others (see Fig. 1 for a partial timeline). Today, ~9 classes (perhaps 12 or more) of UcS have been identified, all with a pair binding interaction involving some other exchange than that of phonons in 'conventional' BCS superconductors.

This review attempts, for the first time, to bring together in one place a discussion and intercomparison of the properties of *all* of these UcS classes, spanning layered nitrides to heavy fermions, cuprates to organics, IBS to non-centrosymmetric, and other classes as well. There are of course numerous existing reviews devoted to *individual* UcS classes (e. g. there are over 20 reviews for HFS) as will be cited in the appropriate sections below. As well, there are compendia of reviews of individual superconducting classes (e. g. 24 chapters in *Superconductivity,* Bennemann. and Ketterson, ed. [9]; 32 articles on various superconducting classes, Hirsch, Maple and Marsiglio, ed. [10]; see also [11].) Although not their main purpose, these compendia sometimes provide useful intercomparisons between some of the classes, e. g Maple et al., Chap. 13 in [9] compare the organics, the cuprates, and HFS in their summary. In addition, coming closest to the intent of the current review, there are reviews with the specific intent to compare several UcS (typically including cuprates and heavy fermions) together, see, e. g., Refs. [12-14].

Discussions of the UcS experimental properties herein are intertwined with the associated theories with thorough references of the continuing effort to achieve theoretical understanding of UcS. UcS are after all the pathway (or one of the pathways, see section 13) to ever higher $T_c$ and perhaps eventually the 'holy grail' of room temperature superconductivity. As with any endeavor on the forefront of understanding, there remain unresolved controversies on whether particular classes exhibit UcS – indeed even the prototype UcS $CeCu_2Si_2$ has been recently reexamined in the light of data analysis indicating conventional behavior. We focus on the 9 classes of superconductors for which there is substantial evidence indicating UcS (with a discussion of three additional emerging, possible UcS classes in section 13), and omit the classes where the current consensus is "uncertain" (e. g. for the borocarbides, see ref. [15] for discussion).

As discussed below in section 13, beyond the 9 classes (counting electron-doped and hole-doped cuprates as one class) of superconductors reviewed herein as unconventional, there certainly exist other possibilities for UcS classes.

One class of current interest is the relatively young field of topological superconductors, which can be defined as superconductors with "non-trivial" topological states. One predicted possibility for finding topological superconductivity (which until now is mostly a theoretical study) is at the interface between a topological insulator (where the bulk is an insulator and there are conducting surface states that are symmetry protected) and a conventional superconductor. The first 3D topological *insulator* was discovered in 2008 [16]. The use of the term "topological" in describing either the insulating or theorized superconducting states is reserved for describing the special conducting surface states that are symmetry protected by particle number conservation and time reversal symmetry. The use of the word "topology" in describing simply the structure of, e. g., a superconducting Fermi surface, is unrelated to the concept of "topological" superconductor, even though indeed unusual Fermi surface topologies like in the s± states theorized for the IBS are certainly connected to unconventional superconducting behavior.

The two other classes of superconductor discussed in section 13 are the new high temperature superconductor monolayer FeSe on $SrTiO_3$ (interfacial superconductivity) as well as the 200 K $T_c$ superconductor $H_2S$ under pressure. Although the latter has been primarily described as a conventional superconductor, there are opposing viewpoints – which illustrates our point here that there continues to be discussion as to which classes of superconductors are unconventional. However, there exists ample evidence for a core understanding of UcS, which it is hoped will assist, among other things, in understanding the emerging new superconductors.

We discuss in the next section, section 2, 16 experimental measurements that either (relatively) definitively establish a given material to be an UcS or, again approximately, may be misleading in identifying UcS. When in sections 3-12 below the experimental evidences for UcS for the superconductors in the various classes are discussed, as each piece of evidence for UcS in a particular superconductor is discussed, it will be noted "✓#" in the text, where "#" is between 1 and 10 (i. e. the lower numbers), for 'strongly suggestive' characteristics and between 11 and 16 (the higher numbers) for indicative but possibly misleading characteristics to keep track of all the UcS features. At the end of each UcS class' section, a short summary of its properties and the current theoretical understanding and outstanding puzzles of interest will be appended. For some classes where the members are quite disparate, a table of the unconventional properties found for *each* superconductor discussed will be part of the summary.

It is this intercomparing of the classes, as well as identifying the important open questions ('puzzles') still present in many of the classes, that hopefully will help understanding of this disparate set of superconductors as an <u>interrelated</u> group. Although comparisons of cuprates, heavy fermions, and to a lesser extent the IBS have been rather common, the entire set (which we admit is likely still incomplete, see section 13) of UcS contains of the order of 9-12 members. Although the current review cites over 75 reviews of the individual classes of UcS and their experimental properties and the theories thereof, *no* review has so far attempted to discuss them all simultaneously. At the very least, bringing a description, primarily experimental but with attention paid to the theory as well, of these disparate UcS together here in one place should help further investigation into the important commonalities.

## 2. Experimental Evidence for UcS (non electron-phonon coupled)  (Note: many UcS show only a few of these properties)

### 2.1. *Strongly Suggestive Evidence*
2.1.1. *Characteristic energy $T_F$ (as obtained, e. g., from $\rho \sim (T/T_F)^2$) is much less than the Debye temperature, $\Theta_D$*

In the BCS theory, as already mentioned, the superconducting transition temperature involves two factors. One factor is an exponential containing the electronic density of states at the Fermi energy, N(0), and a parameter V expressing the strength of the electron phonon coupling. The other factor is some measure of the average phonon frequency which sets the energy scale for the strength of the electron-phonon coupling, as now discussed. Thus, all things being equal, a stiffer lattice leads to a higher $T_c$ in BCS theory.

One way to approximately quantify these quantities together in a more typical form (for a thorough discussion see Allen and Dynes, ref. [17]) is to identify N(0)*V as equal to λ, where λ can be calculated from the electron-phonon spectral function $\alpha^2(\omega)F(\omega)$ (F(ω) is just the phonon density of states as a function of energy ℏω determined, e. g., from neutron scattering):

$$\lambda = 2 \int \alpha(\omega)^2 \, F(\omega) \, d\omega/\omega \qquad \text{eq. 2}$$

The limits of integration are formally from 0 to infinity, however the upper bound can be set much lower. For example in the 22 K superconductor A15 $Nb_3Ge$ [18], where $\alpha(\omega)^2 F(\omega)$ was determined by electron tunneling, an effective cutoff for the integral is 33 meV or about 1.3 times $k_B \Theta_D$. Thus, using the Debye model cutoff temperature as the <u>energy scale</u> for electron phonon coupling is a good approximation.

The BCS model may be the correct description of a superconductor if the characteristic energy scales are arranged thusly:

$$T_c < \Theta_D < T_F \qquad \text{eq. 3}$$

where $T_F$ is the Fermi energy, $E_F$ (several to ~ 10 eV for most metals), divided by $k_B$. In the current context, we are of course interested in the *inverse* of this relationship, i. e. if a superconductor does *not* obey eq. 3, then it is *not* a BCS superconductor. (It still may be the case that a superconductor whose properties obey eq. 3 is not a BCS superconductor, but matches one or more of the other indicators for being an UcS discussed in the following sections.)

One useful rough approximation for $T_F$ is that it is inversely proportional to the coefficient of the normal state electronic specific heat, γ (=$C_{normal}/T$ as T→0). Another method for estimating $T_F$ is – in strongly correlated metals – to measure the low temperature resistivity, ρ, which in a Fermi liquid will follow a $T^2$ law:

$$\rho(T \to 0) = \rho_0 + AT^2 \qquad \text{eq. 4}$$

where $\rho_0$ is the 'residual' (or remnant) resistivity as T→0. The coefficient 'A' is proportional to $1/T_F^2$. (Other methods for estimating characteristic electronic energy scales in addition to $T_F$ are discussed below in the various UcS sections.)

As an example, Re, which is an elemental BCS superconductor with $T_c$=1.4 K, has [19] $\Theta_D$=416 K, $\gamma$=2.3 mJ/molK$^2$ and $E_F$ = 11.2 eV [20]. The quadratic temperature resistivity coefficient A is [21] 4 10$^{-6}$ μΩ-cm/K$^2$. The unconventional HFS CeCu$_2$Si$_2$ has [22] $T_c$=0.6 K, $\Theta_D$=310, $\gamma$=1000 mJ/molK$^2$, and the coefficient A is [23] 9.3 μΩ-cm/K$^2$, i. e. two million times that of Re. Finding $T_F$ for CeCu$_2$Si$_2$ by scaling $E_F$ of Re times $(A_{Re}/A_{CeCu2Si2})^{1/2}$ gives for CeCu$_2$Si$_2$ $T_F$ = 85 K and a characteristic energy of ~ 7 meV. Thus, Re – based on 1.4 K < 416 K < 1.3 10$^5$ K – would be consistent with the BCS model, whereas CeCu$_2$Si$_2$ – based on 0.6 K < 310 K > 85 K – clearly does not follow the BCS picture and would therefore fit one of the 'strongly suggestive' indicators (✓1) for UcS.

*2.1.2. Superconductivity forms out of a non-Fermi liquid normal state, implying that quantum critical fluctuations are involved in the Cooper pairing.*

When superconductivity in a phase diagram (where pressure, doping or magnetic field is the tuning parameter) occurs near where a second order phase transition (such as antiferromagnetism) is suppressed to T=0 with the tuning (see Fig. 2 for a schematic), then the possibility arises that the exchange of fluctuations that may arise near such a possible quantum critical point (QCP) could be responsible for the superconducting pairing. (Despite the common misperception, not every suppression of a second order phase transition to T=0 results in detectable quantum critical fluctuations, see ref. [24]. The thinking that a QCP and the associated fluctuations provide the superconducting pairing is ubiquitous in discussions of several UcS, particularly in the heavy Fermions (where the specific heat [25] of CeCoIn$_5$ actually obeys the quantum critical, non-Fermi liquid behavior C/T ∝ logT above $T_c$=2.3 K), in the cuprates, and in the IBS.

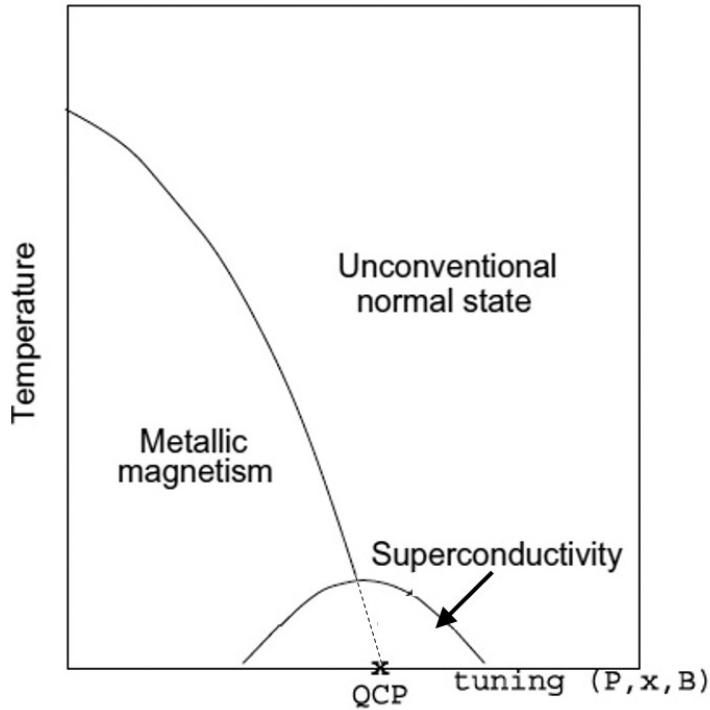

Fig. 2: The second order magnetic phase transition decreases in temperature with increasing the tuning parameter, e. g. by varying the composition, x, but is interrupted by the superconducting dome. The dashed line extrapolates $T_{mag}$ to T=0, where strong quantum critical fluctuations can occur at a QCP (a Quantum Critical Point).

Thus, the possibility that quantum critical fluctuations could be causing the UcS would be, as a first indication, raised by a phase diagram (such as that in Fig. 2) where superconductivity occurs at, or near a second order phase transition going to T=0. At such a quantum critical point, C/T may behave as logT but more often the resistivity ($\rho=\rho_0 + AT^2$ for Fermi liquid behavior, but $\rho=\rho_0 + AT^\alpha$, $\alpha<2$ for *non*-Fermi liquid behavior at a quantum critical point) is the deciding measurement. For example, see the case of $CePd_2Si_2$ [26]. The relevant UcS sections below contain thorough discussions. As already noted, not every suppression of a second order phase transition to T=0 results in fluctuations strong enough to cause such interesting effects.

2.1.3. *neutron spin resonance peak develops in the superconducting state*

A magnetic resonance in inelastic neutron scattering below $T_c$ is usually (but see counter-arguments for iron based UcS in Hosono and Kuroki [27]) considered to be definitive evidence for a sign change in the superconducting energy gap $\Delta$ on different parts of the Fermi surface, as would be consistent with d-wave or s± pairing. It is common to argue (see, e. g., P. C. Dai et al. [28]) that the properties of the resonance suggest that magnetism (e. g. exchange of spin fluctuations) plays an important role in the superconductivity, thus such a resonance serves as a strong indication (✓3) of UcS.

Inelastic neutron scattering experiments find a magnetic resonance below $T_c$ quite broadly in two of the classes of UcS discussed in this review, the cuprates (see the review [29] by Eschrig) and in the IBS (see the review [30] by Stewart). The usual understanding [31] of the magnetic resonance in the cuprates, with some continuing discussion in the IBS, is that it is an isotropic triplet excited state of the ground state Cooper pair singlet. As well, such a neutron spin resonance is found in two heavy Fermion UcS: $CeCoIn_5$ [32] and in $UPd_2Al_3$, $T_c$=1.8 K [33]. A broadened inelastic line that extends from the peak at 0.2 meV to at least 1 meV in the superconducting state has also been found [34] in $CeCu_2Si_2$. (The authors do not discuss the possibility of sample quality in the rather large (2g) single crystal of $CeCu_2Si_2$ playing a role in the peak width. As a reasonable indicator of the good quality of the sample, the transition width,

$\Delta T_c$, in the specific heat of the same 2g crystal used for the neutron scattering is [35] ~10% of $T_c$. A neutron spin resonance below $T_c$ has been searched for [36] in heavy Fermion $UPt_3$ and in the UcS $Sr_2RuO_4$ [37] without success.

When discussing the magnetic resonance in cuprates, it is common to point out that there is an approximately uniform scaling of the resonance energy with $T_c$, implying that the resonance is intimately connected to the superconductivity. In the cuprates, Hüfner et al. [38] state that $E_{resonance}$ is about $5k_BT_c$. Discussion of this scaling in the IBS [30] gives a similar ratio, while in $CeCu_2Si_2$ the ratio is 4 [34], in $CeCoIn_5$ the ratio is about 3.1 [32] and in $UPd_2Al_3$ the ratio is only 2.3 [33].

2.1.4. *Existence of a pseudogap (measured by, e. g., $\rho$, point contact, or tunneling spectroscopies above $T_c$)*

The pseudogap is well established as a property for underdoped ('*under*doped' implies doping - e. g. oxygen stoichiometry 'x' in $YBa_2Cu_3O_{6+x}$ - where x is *less than* the concentration needed for 'optimal' – or maximum – $T_c$) cuprates (see Fig. 3 for a phase diagram for hole doped), and is – as will be discussed - not well established in most of the other UcS classes (with perhaps the exception of the IBS, where work on the pseudogap continues.)

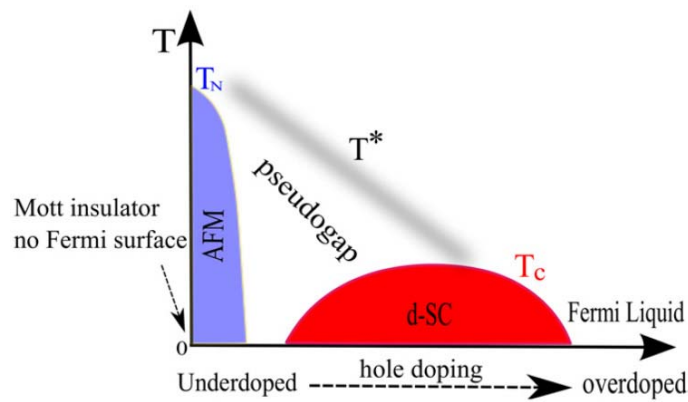

Fig. 3 (color online): In hole doped cuprates, below optimal (peak $T_c$) doping there appears below a temperature T* a pseudogap phase with properties as discussed in the text. The superconductivity under the red dome in hole-doped cuprates is understood to have d-wave pairing symmetry. The region between the blue antiferromagnetism (AFM) region in the phase diagram and the red superconductivity region is somewhat uncertain, with varying reports of some form (e. g. spin glass or spin density wave) of magnetic order (figure is reproduced from ref. [39]. © IOP Publishing. Reproduced with permission. All rights reserved.)

A nice overview of pseudogap physics in the hole-doped cuprates can be found in the recent theoretical review [39] by Rice, Yang, and Zhang. Fig. 3 shows a schematic of a hole-doped cuprate phase diagram. In the pseudogap region, anomalies at T* occur in a number of experimental measurements in the under hole-doped cuprates, including Knight shift in NMR, ARPES measurement of a shrinking (or partial gapping, whence the description 'pseudogap') of the full Fermi surface present in the overdoped regime down to four Fermi arcs, scanning tunneling microscopy, quantum oscillations, in-plane optical conductivity, specific heat, London penetration depth, Raman scattering, magnetic neutron scattering, and dc resistivity. (See also Timusk and Statt [40] for a somewhat older experimental review. For reviews of electron doped cuprates, with their different pseudogap behavior (e. g. the pseudogap in the optical spectrum in electron doped cuprates is more distinct [41]) in the underdoped regime, see Armitage, Fournier, and Greene [41] and, more recently, Fournier [42].

To give an example of a physical picture of the effect of a pseudogap, as the Fermi surface becomes partially gapped with increasing underdoping (see Fig. 3) in YBa$_2$Cu$_3$O$_{6+x}$, the normal state specific heat $\gamma_n$, proportional to the electronic density of states at the Fermi energy, N(0), decreases as does the superconducting condensation energy U (the pseudogap removes low energy spectral weight in the normal state spectrum) and the jump in the specific heat, $\Delta C$, at T$_c$ (since $\Delta C \propto U$). $\Delta C$(T$_c$) actually falls more than a factor of 5 below optimal doping [43]. The electrical resistivity in under-hole doped cuprates shows [44] a clear decrease at temperatures less than T* below the extrapolated linear-with-temperature behavior at high temperatures, ascribed to a decrease in the scattering by spin fluctuations caused by the opening of a gap in the spin excitation spectrum. To contrast, Zheng et al., using NMR measurements, found no evidence of such a spin pseudogap in an under-electron-doped cuprate [45].

The theoretical explanation for the pseudogap remains under discussion, and includes electron correlation effects, hidden order parameters due to various causes, and precursor ("preformed") pairing. See Timusk and Statt [40] for a discussion of the top competing theories as of 1999, and the review by Rice, Yang and Zhang [39] for an update.

Thus, although pseudogap behavior is common in both under- electron and hole doped cuprates, the behavior is not consistent between the two sides (electron- and hole-doped) of the phase diagram and may indeed have different origins. Further, although the pseudogap only occurs below optimal doping on both the electron- and hole-doped sides of the phase diagram, the superconductivity *throughout* the phase diagram is unconventional. Thus, the presence of a pseudogap in the phase diagram is an indication of UcS throughout the superconducting dome in a superconductor.

Indications of pseudogap behavior in point contract spectroscopy and resistivity exist in one of the HFSs (CeCoIn$_5$), which will be discussed below in section 4. Pseudogap behavior, as we will see, is under discussion in the IBS, section 6.

2.1.5. *Power law (instead of BCS exponential) temperature dependences in properties like penetration depth, $\lambda \propto T$, or NMR spin lattice relaxation time (1/T$_1$) but must be to very low temperatures to escape the possibility of deep minima in the gaps causing the observed power laws at temperatures > (energy of gap minimum)/k$_B$*

This is a very common type of experimental evidence used to determine UcS. A BCS superconductor has an energy gap open at T$_c$, thus giving many experimental properties (e. g. electronic specific heat, C$_{el}$) an exponential temperature dependence. (Note that since the size of the energy gap increases with decreasing temperature below T$_c$, the temperature dependence of experimental properties only fits a pure exponential form over limited temperature ranges, e. g. 0.4T$_c$ < T < 0.17T$_c$ [46].)

Before the 1957 BCS theory, other temperature dependences, motivated by early theories, were used. For example, the specific heat of Nb, T$_c$=9.2 K, fits the Gordon-Casimir model prediction of C$_{el} \propto T^3$ below T$_c$ down to almost 0.4 T$_c$ [47].

Since the advent of UcS in heavy Fermion $CeCu_2Si_2$ in 1979, power law temperature dependences have often been used to rule out BCS exponential-with-temperature behavior. In addition, *specific* power laws (e. g. $\propto T$ for the penetration depth) have been used to distinguish particular symmetries in the nodal behavior and thus in the pairing symmetry of an UcS. As clear from the above historical discussion, this involves various cautions. First, a pure power law must be seen over a large temperature region (not, e. g., just over half a decade in temperature), is only valid for $T \ll T_c$, and therefore extend down to a very low temperature, $T \ll T_c$. In addition to being measured in the low temperature limit where the simple power law is valid, there is another reason why low temperature measurements are critical. For example, distinguishing nodal UcS (where the energy gap, $\Delta$, goes to zero at points or lines on the Fermi surface) from deep minima behavior via, e. g., a finite thermal conductivity divided by temperature, $\kappa/T$, requires data as $T \to 0$. Thus, such data should be measured at dilution refrigerator (T<0.1 K) temperatures. Even then, the presence of impurities can mask the intrinsic behavior [48]. Fortunately, most of the materials discussed herein as candidates for UcS appear not to be dominated by impurity effects.

As an example, some typical power laws that have been used to argue for UcS are discussed briefly here (see also Table 1). When each class of superconductor is discussed more thoroughly in the following sections below, a more complete discussion will be offered.

Table 1 (From Heffner and Norman [48]) Theoretical temperature dependencies for several low temperature measurements in clean, defect-free samples, assuming a spherical Fermi surface and either line ('polar') or point ('axial') nodes in the superconducting gap structure

| **Measurement** | **Polar (lines)** | **Axial (points)** |
|---|---|---|
| Specific Heat (C) | $T^2$ | $T^3$ |
| NMR Relaxation ($1/T_1$) | $T^3$ | $T^5$ |
| Thermal Conductivity ($\kappa$) | $T^2$ | $T^3$ |
| Penetration Depth $(1/\lambda_\parallel)^2$ | $T^3$ | $T^2$ |
| Penetration Depth $(1/\lambda_\perp)^2$ | $T$ | $T^4$ |

The specific heat in the superconducting state of $CeCu_2Si_2$ follows [49] $C \sim T^2$ – consistent with line nodes in the superconducting gap - down to 0.1 $T_c$. As already mentioned, this old result has been superseded [50] by specific heat data below 0.15 $T_c$ that can be fit to a two band model, where both band gaps are fully gapped BCS-like.

As a second example, penetration depth, $\lambda$, measurements, for a fully gapped superconductor, $\Delta\lambda(T) \propto \exp(-\Delta/T)$, where $\Delta\lambda(T) = \lambda(T) - \lambda(T \to 0)$. $\Delta\lambda(T) \propto T$ is clear indication of nodes (e. g. lines nodes from d-wave pairing symmetry) and therefore UcS. For a review of penetration depth in UcS, see Prozorov and Gianetta [51]. In addition to the caveats mentioned above (measured over a large temperature range and including $T \ll T_c$), an additional caveat applies: it is theoretically possible [52] for $\Delta\lambda(T) \propto T$ to be caused by phase fluctuations. Experimentally this effect is too small, e. g. in the IBS [30] to compete with the large T-linear coefficients observed. $\Delta\lambda(T) \propto T^2$ at low temperatures for both d-wave parity in the presence of strong scattering, i. e. not the case covered in Table1 [53] as well as for a fully gapped s± state also with strong impurity scattering [54]. Thus, impurities/quality of sample can play an important role in being able to translate a 'simple' temperature dependence of $\Delta\lambda(T)$ into a firm conclusion as to the gap structure. As a further example of the difficulty in interpretation, $\Delta\lambda(T) \propto T^2$ has also been interpreted [55] – as will be discussed in the heavy Fermion section - as evidence for axial spin triplet, p-wave pairing in the HFS UBe$_{13}$ [30].

### 2.1.6. *Multiple superconducting phases with different order parameter symmetries*

Superfluid $^3$He is the prototypical material with two ordered phases, as discovered in 1972 (for a review, see D. M. Lee, ref. [56]). Upon cooling under ~33 bar pressure, first the anisotropic 'A' phase, described [57] by Anderson, Brinkman, and Morel (p-wave pairing), with nodal points, appears below 2.65 mK followed by the isotropic 'B' phase, described by Balian and Werthamer [58] (another type of p-wave pairing) below ~ 2 mK.

Amongst the UcS classes, heavy fermions have several examples of this metric for unconventional superconductivity with UPt$_3$, (T$_c$=0.54 K) the third discovered HFS, being the prototype. First discovered by ultrasonic attenuation measurements [59] and visible in good quality samples in the bulk specific heat (two distinct zero field transitions [60]), the *three* distinct superconducting phases (see Fig. 4) in UPt$_3$ have different nodal structures based on small angle neutron scattering off the flux line lattice [61] and based on theory [62].

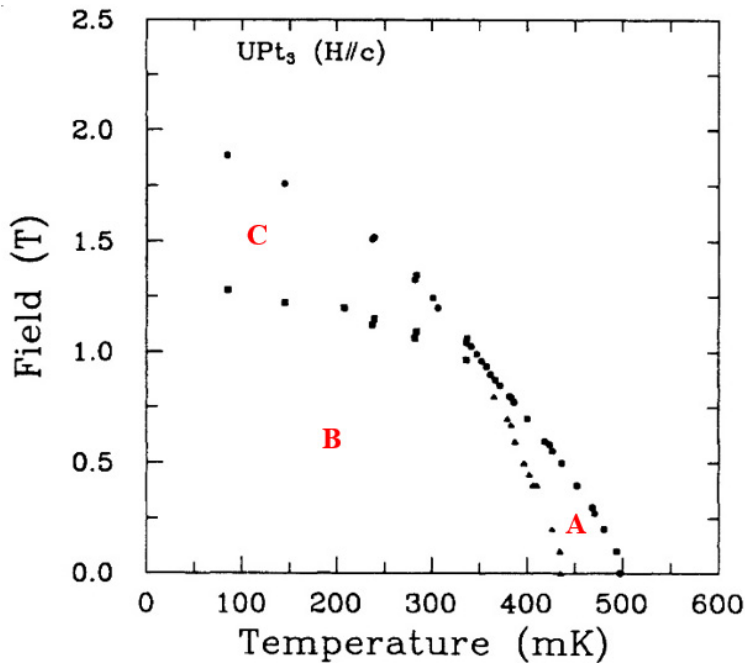

Fig. 4 (color online) Phase diagram of UPt$_3$ as determined by Adenwalla et al. [63] using ultrasonic attenuation. The three phases (A, B, and C) all have nodes in the superconducting gap and different gap structures as well. The zero field transitions into the A and B states can be seen in specific heat. Reprinted figure with permission from Adenwalla et al., Phys. Rev. Lett. 65 (1990), p. 2298. [63] Copyright (1990) by the American Physical Society.

Other UcS classes that show multiple phases, as will be discussed, might include Sr$_2$RuO$_4$ in magnetic field.

### 2.1.7. *Josephson tunneling/Phase sensitive measurement that shows non-BCS order parameter symmetry (e. g. triplet p-wave in Sr$_2$RuO$_4$)*

In UcS, one or more symmetries in addition to U(1), the one dimensional global gauge symmetry, are broken at the onset of superconductivity [4]. The amount of symmetry breaking involved in the Cooper pairing state is reflected in the symmetry properties of the order parameter. Phase insensitive measurements, such as the temperature dependence of low temperature penetration depth (see section 2.1.5. above), can prove the presence of line nodes but cannot distinguish between extended s-wave states (as found in some IBS) or d-wave pairing. d-wave pairing was first established in the cuprates by Tsuei, et al. [64] using the phase-sensitive measurement of Josephson tunneling. See section 5 on cuprates for a complete discussion of this technique. Josephson tunneling has also, as will be discussed in the respective sections below, helped establish the details of the order parameter symmetry in heavy fermions, $Sr_2RuO_4$, and IBS.

### 2.1.8. *Breaking of Time Reversal Symmetry: Polar Kerr Effect or µSR measured spontaneous appearance of an internal magnetic field below $T_c$*

In superconductors with time reversal symmetry breaking, the Cooper pairs have non-zero magnetic moments. These small moments align locally producing very small spontaneous magnetic fields which can be detected using µSR or magneto-optic measurements of the Faraday or Kerr effects [12]. Magneto-optic methods of demonstrating the breaking of time reversal symmetry (TRS) in UcS have been used in $Sr_2RuO_4$ and cuprates. More recently, the polar Kerr effect has also been observed in heavy fermion $UPt_3$ [65] and $URu_2Si_2$ [66]. Without time reversal symmetry, the Cooper pairs have to have some admixture of triplet pairing symmetry [67].

### 2.1.9. *Superconductivity is extremely sensitive to impurities, non-magnetic as well as magnetic*

It is accepted that magnetic impurities are harmful to BCS superconductors. For a review, see Maple [68] wherein the Abrikosov-Gorkov (AG) model for $T_c$ suppression by a local spin is discussed. The AG theory assumes randomly located magnetic impurity spins that do not correlate with one another. Plots of $\Delta C/\Delta C_0$ vs $T_c/T_{c0}$ (where the subscript '0' indicates the undoped superconductor) after the AG model lie below the straight line (law of corresponding states) derived from BCS, except where the two lines join at $T_c/T_{c0} = 1$ and 0. Also, conventional s-wave (s-wave ($\ell=0$)) superconductors are relatively insensitive to non-magnetic impurities [69]. Contrariwise, in an UcS the ability of the electrons to pair is a function of the momentum **k** over the Fermi surface, and impurity scattering mixes the **k** values – making UcS sensitive to either magnetic or non-magnetic impurities and impeding the superconducting pairing. Symptoms of such harmful effects on the superconductivity include lowering of $T_c$ (sometimes to T=0), incomplete opening of the superconducting gap (finite C/T as T→0 in the superconducting state) and smaller normalized jumps ($\Delta C$) in the specific heat at $T_c$.

In the early days of unconventional superconductivity in studies of HFSs, doping experiments were carried out to just qualitatively measure if a non-magnetic dopant had a slower depression rate of $T_c$ than a magnetic one, i. e. if the $T_c$ vs doping was consistent with non-BCS superconductivity. As with any doping experiment (in contrast to application of magnetic field or pressure), there were always concerns that the process of doping had a wider effect than simply placing a local-moment-carrying ion into the lattice. Since (see, e. g., Figs. 2 and 3) the superconductors under discussion here for their unconventional behavior very often lie close to (or coexistent with) magnetic order, any disturbance in the host lattice by the dopant (e. g. different ionic radius causes lattice expansion or contraction) can have secondary magnetic effects on the superconductivity independent of the presence or lack of a local moment on the dopant ion. Further, what might appear as a clearly non-magnetic dopant ion can, in a particular host lattice with the surrounding ligand atoms, be in actuality itself not as 'non-magnetic' as believed. (For a good general discussion of this point, see Pfleiderer [70]).

Good examples of checking for strong $T_c$ - sensitivity to non-magnetic dopants to determine where the superconductivity is unconventional exist, e. g., in the heavy fermions. For example, chemical substitution of Pt with Rh in $U_2PtC_2$ suppresses $T_c$, but not as rapidly as expected for a superconductor with spin-triplet pairing, which should be extremely sensitive to the presence of nonmagnetic impurities [71]. Triplet pairing in $UPt_3$ was previously inferred from the mean free path dependence after doping with selected impurities [72].

As another example, in organic superconductors there is a strong influence of non-magnetic impurities on $T_c$, see e. g. Joo et al. [73] and the review by Brown [74].

One interesting (reverse) example of magnetic vs non-magnetic dopants and their effect on $T_c$ in conventional and UcS is the case of $DyNi_2B_2C$, believed to be a conventional BCS superconductor [15]. This material is a magnetic superconductor, $T_c$=6 K, inside an antiferromagnetic phase, $T_N$=11 K where non-magnetic impurities (e. g. Y on the Dy site) suppress $T_c$ at almost the same rate as magnetic (Gd) substitution [15].

2.1.10. *Non-centrosymmetric crystal structure (usually) coupled with strong electronic correlations*

By the name, this indication of UcS is reserved for those superconductors that do not have a center of symmetry in their crystal structure. As discussed in Kneidinger et al. [67], only the non-centrosymmetric superconductors that also have strong electron-electron correlations exhibit UcS.

This observation can be understood as follows. As discussed by Mineev and Samokhin [75], mixed singlet and triplet pairing superconducting states can occur in the absence of crystalline inversion symmetry, which – due to the admixture of triplet – would mean UcS. However, Kneidinger et al. state that anomalous spin fluctuations caused by the lack of inversion symmetry stabilize the triplet pairing part of the superconducting condensate and that these spin fluctuations are enhanced in strongly correlated (high specific heat $\gamma$) systems. They then state that without strong correlations the triplet part of the condensate in non-centrosymmetric superconductors is not sufficiently stabilized and that weakly correlated superconductors without inversion symmetry have their physical properties dominated by the singlet part of the condensate and thus appear to be conventional superconductors. This is born out in numerous experimental systems.

We now discuss indicative, but possibly misleading, experimental evidence for UcS in sections 2.1.11. through 2.1.16.

2.1.11. *$C(H,\Theta)/\kappa(\Theta,H)$ shows evidence for nodal behavior, but can be due to deep gap minima (e. g. $YNi_2B_2C$)*

The measurement of the angular dependence of the electronic density of states at the Fermi energy in the superconducting state (determined from specific heat data) in zero and applied magnetic fields, coupled with theory, has been used to find nodal, unconventional superconductivity in several UcS. ($\kappa(\Theta,H)$ can be analyzed similarly.) Where this (technically challenging) method can give a false positive for UcS is when deep gap minima mimic at finite temperatures nodal behavior. One example of this behavior is in $YNi_2B_2C$, where $C(H,\Theta)$ data were measured at 2 K by Park et al. [76]. In a recent review [15] of the quaternary borocarbide superconductors, Mazumdar and Nagarajan argue based on a broad range of data, including isotope effect, that $YNi_2B_2C$ is a conventional, electron-phonon coupled superconductor. Another way to have the measurement of $C(H, \Theta)$ be misleading is to measure at too high a temperature, before the nodal and antinodal directions become inverted to their correct angular locations. The theory of this was developed by Vorontsov and Vekhter [77]. Thus, for example, measurements of $C(H, \Theta)$ in $CeCoIn_5$ originally concluded (see section 4 below) that the pairing symmetry was $d_{xy}$, rather than the currently accepted consensus view that it is $d_{x2-y2}$.

2.1.12. *See 2.1.5. above – power laws measured over a limited temperature range*

Power laws are not measured in the proper temperature range ($T<0.1\ T_c$) for the theory indicating unconventional behavior to apply. Certainly data only down to $T_c/2$ can be ignored.

2.1.13. *Isotope effect*

Elemental Ru (which is a conventional superconductor), when $T_c$ is measured as a function of isotope mass, does not show the BCS predicted $T_c \propto M^{-1/2}$, but instead shows essentially no mass dependence for $T_c$. [78] because of details of the screened Coulomb interaction $\mu^*$ between the electrons. Thus, the *lack* of an isotope effect on $T_c$ *can* be an indication of UcS, but is not definitive. Conversely, if a superconductor shows a (substantial) isotope effect then it is not unconventional. Here the size of the isotope effect measured must be considered. For example, in the IBS (certainly considered to be UcS) an isotope effect has been measured [30] on the Fe site ($T_c \propto M^{-0.35}$) but not on the O site. This implies that phonon modes that involve the Fe (a magnetoelastic effect) affect the spin fluctuations (theorized to provide the pairing mechanism) and thus the superconductivity.

2.1.14. *Large Finite γ in the Superconducting State Specific Heat*

Although nodes in the superconducting gap are indicative of UcS, it is a misapprehension to expect a γ measured in the superconducting state of much more than 2 mJ/molK$^2$ (the value in high quality nodal d-wave YBCO crystals discussed herein) to be due to nodes. For example, the value of the residual γ in $UPd_2Al_3$ of 24 mJ/molK$^2$ measured by Caspary [79] should not be considered as evidence for nodes in the gap function, but is rather due to some other cause. As will be discussed with the IBS, a finite $\gamma_{residual}$ can be a sample quality issue, where part of the sample remains normal due to defects, improper annealing, or similar issues. When the $\gamma_{residual}$ in a high quality (annealed, pure, well ordered) superconducting material is an abnormally large fraction of $\gamma_{normal}$ (e. g. in $UPt_3$ the fraction can be [80] as high as 45%), this is evidence (linked to the effect of non-magnetic impurities, Section 2.1.9 above) of UcS. However, the determination of the quality of the sample is clearly very important to this determination.

## 2.1.15. *Specific heat γ varies as $H^{1/2}$ in the Superconducting State*

Originally done by Moler et al. [81] for YBCO, $\gamma \propto H^{1/2}$ is, according to Volovik [82] indicative of lines of nodes in the gap. As was later pointed out [83], a behavior close to $\gamma \propto H^{1/2}$ can occur in superconductors with two fully gapped s-wave symmetry bands. Thus, $\gamma \propto H^{1/2}$ can only be taken as indicative of UcS.

## 2.1.16. *Plot of $T_c$ vs mean free path (proportional to sample quality through the Residual Resistivity Ratio, RRR) following Abrikosov-Gorkov model, e. g. in URhGe or UPt$_3$.*

For a conventional superconductor, $T_c$ is independent of the mean free path, $\ell$, while for UcS $T_c$ is suppressed as $\ell$ is suppressed down to of order the superconducting coherence length. (this argument is from Hardy and Huxley [84]). The superconductivity in various UcS, e. g. URhGe or UPt$_3$, is quite sensitive to the sample purity. With increasing residual resistivity, $\rho_0$, the $T_c$ decreases, and vanishes for low sample quality, consistent with unconventional superconductivity. Also: critical field data in URhGe strongly supports an odd-parity *p*-wave state with gap node parallel to the magnetic moments [70]. The dependence of $T_c$ on sample quality ($\ell \propto$ RRR $= \rho(300 \text{ K})/\rho_0$) in, e. g., URhGe provides evidence that the superconductivity is non-conventional [85]. The sensitivity of $T_c$ to the electronic mean free path (proportional to RRR or inversely proportional to the residual resistivity, $\rho_0$) is evidence for a nonconventional order parameter (having a vanishing angular average) [85].

## 3. Brief Summary of Status of Theories to Explain UcS

Theorists have been heavily involved in working to understand UcS since the first recognized discovery of unconventional behavior in heavy fermion CeCu$_2$Si$_2$ in 1979 (see Fig. 1). (Recognition of superconductivity - first discovered [86] by Bucher et al. in 1975 - as being a bulk property in heavy fermion UBe$_{13}$ had to wait until 1983.) However, no overarching common theory of all the various classes (9 in the current review, plus evolving discoveries at the time of writing) exists. Thus, the summary of theory given here (the interested reader may delve more deeply into the theory of the various classes by following the representative cited references below), in this primarily experimentally-oriented review, is perforce composed of parts not all that strongly interrelated. It is one of the goals of this review that, with the experimental situation of the many classes of UcS described herein, that more all-encompassing theoretical approaches may be inspired.

That said, there have been a few attempts to *begin* to form a common theory. Tohyama [87] discusses the similarities and differences of the IBS and the cuprates. Taillefer [88] pointed out a phenomenological correlation in the resistivity related to a nearby QCP in the cuprates, the organics, and the IBS. In the organic, IBS, and some heavy Fermion superconductors, Taillefer notes that "the central organizing principle is the presence of a QCP inside the superconducting dome, at which SDW order ends." Taillefer cites work by Bourbonnais and Sedeki [89] using a renormalization group approach that shows that low energy antiferromagnetic fluctuations are responsible for the superconducting pairing at least in the organic Bechgaard salts, T≤1 K.

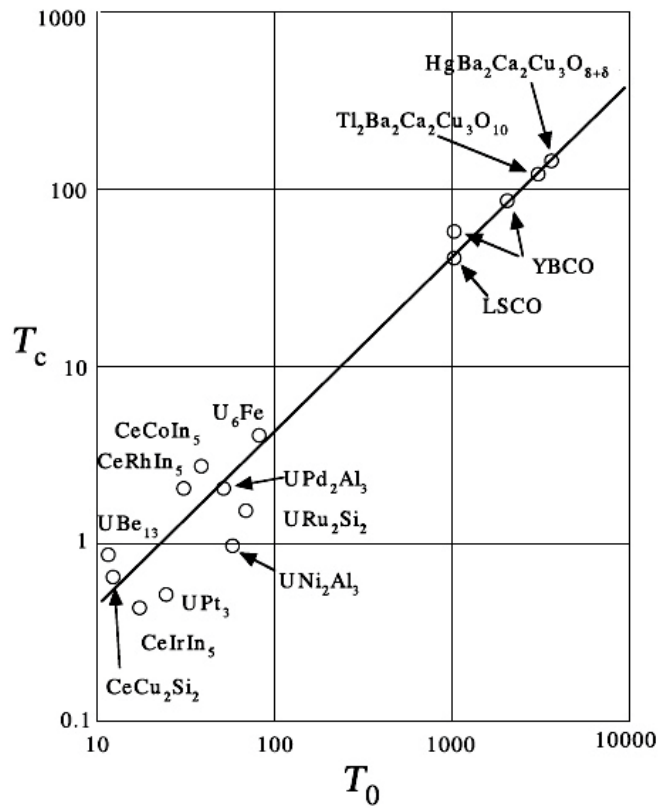

Fig. 5: From Moriya and Ueda [13], a plot of the superconducting $T_c$ vs a characteristic temperature/energy ($\propto$ energy spread) of the spin fluctuations, $T_0$, of various heavy Fermion and cuprate superconductors. According to Scalapino [14], Moriya and Ueda estimate $T_0 = 1.25 \cdot 10^4/\gamma$, with $\gamma$ in units of mJ/molK$^2$ for the heavy Fermions; for the high $T_c$ cuprates, Moriya and Ueda use the coefficient of the linear-in-temperature resistivity. (figure is reproduced from ref. [13]. © IOP Publishing. Reproduced with permission. All rights reserved.)

Scalapino [14], using "Hubbard-like" models, proposes that (antiferromagnetic) spin fluctuation mediated pairing can explain cuprates, IBS, and a number of heavy Fermion superconductors. As shown in Fig. 5, Moriya and Ueda in 2003 [13] showed a correlation between $T_c$ and the characteristic energy of spin fluctuations, $T_0$ (often referred to as $T_{SF}$) for cuprates and heavy Fermion superconductors.

The history of explaining unconventional superconductivity with spin fluctuations as mediating the pairing extends back to the discovery of UPt$_3$, the third heavy Fermion superconductor (see Fig. 1), where in the discovery paper [90] in 1984 (ferromagnetic) spin fluctuations (FMSF) were speculated to cause triplet superconductivity. This was based on a comparison with the specific heat of $^3$He (where FMSF lead to triplet pairing) and the theories of Doniach and Engelsberg [91] and Fay and Appel [92]. Later neutron scattering results [36] confirmed *anti*-ferromagnetic spin fluctuations in UPt$_3$.

A large number of early theories address this issue of spin fluctuation mediated superconductivity specifically in heavy Fermion superconductors: Anderson [93]; Varma [94]; Miyake et al. [95]; Scalapino et al. [96]; Norman [97].

Proposals that fluctuations near a spin density wave transition could cause UcS in the organic Bechgaard salts were also made [98-99].

After the cuprates were discovered, these ideas were extended to include this new superconducting class: [13], [100-101]. Anderson [102] expressed a contrary opinion about the utility of exchange-of-antiferromagnetic-spin-fluctuation theories for the cuprates. His point was that the Eliashberg extension of BCS theory discusses electrons coupled by exchange of *low frequency* bosons (phonons for BCS, antiferromagnetic spin fluctuations proposed for the cuprates). Exchange of *antiferro*magnetic spin fluctuations, in Anderson's view, is a high frequency interaction – only the exchange of *ferro*magnetic spin fluctuations (FMSF) offers the low frequency energy scale required by the Eliashberg formalism. Although UPt$_3$ is thought to be in a triplet (parallel spin) pairing state (as is $Sr_2RuO_4$ discussed below in section 7) as would result from FMSF exchange, the cuprates are known to have (discussed below in section 5) d-wave (antiparallel spin) pairing. Thus, according to Anderson, the cuprates, with their large (> 2 eV) Hubbard repulsion and large (~0.12 eV) antiferromagnetic exchange coupling, cannot be described by Eliashberg theory and low frequency "bosonic glue." In response, Scalapino [14] says that indeed "there is pairing glue in the Hubbard models." For an easily readable overview discussion on the theory of the cuprates, including the issue of the "glue", see Norman [103]. The issue of what provides the pairing interaction in cuprates remains under discussion, see e. g. Fanfarillo et al. [104] and an analysis of Raman spectra to determine the pairing interaction in Sr-doped 214 cuprate compounds.

For the IBS, Scalapino [14] notes that measurements support the importance of antiferromagnetic SDW-like fluctuations. Many of the IBS have SDW transitions in the undoped, non-superconducting parent compounds (e. g. in $BaFe_2As_2$, $T_{SDW}$=140 K) which are then suppressed with doping until the falling line describing $T_{SDW}$ vs doping concentration joins the superconducting dome ($T_c$ vs doping) in the phase diagram, see Fig. 2 above. (Stewart [30])

As discussed in the introduction to this section above, the theories of UcS in the other classes or even in individual compounds within a class, e. g. $UPt_3$ with its multiple superconducting phases, coexistent ferromagnet-superconductor $UGe_2$ (section 11) or triplet pairing $Sr_2RuO_4$ (section 7), do not exhibit much overlap with one another, and are best discussed below in their respective sections. As made clear here already for the heavy Fermion, cuprate, organic, and IBS, although there are theories that attempt to address a wider selection of superconducting classes, these theories are still very much under discussion and development. As Anderson said [102], "The Secret" of the pairing mechanism is still unknown.

## 4. Heavy Fermions ($CeCu_2Si_2$, $UBe_{13}$, $U_{1-x}Th_xBe_{13}$, $UPt_3$, $UNi_2Al_3$, $UPd_2Al_3$, $PrOs_4Sb_{12}$, $CeCoIn_5$, $CeIrIn_5$, $CeRhIn_5$, $PuCoGa_5$, $CePd_2Si_2$, $URu_2Si_2$, others)

Heavy fermions, and their unusual highly correlated f-electron-derived properties, were first discovered in non-ordered $CeAl_3$ in 1975 (Andres, Graebner, and Ott [105]). $CeAl_3$ has a very large C/T (T→0) ~ 1600 mJ/molK$^2$ ($\propto$ m*, the electron effective mass), along with the concomitant large magnetic susceptibility as T→0, $\chi(0)$, of 36 memu/mole and large $T^2$ coefficient in the low temperature resistivity ($\rho = \rho_0 + AT^2$), A=35 $\mu\Omega$-cm/K$^2$. What made the field of heavy Fermions much more exciting was the discovery of superconductivity in $CeCu_2Si_2$ (the first UcS) in 1979. Later discoveries included further superconducting heavy fermion systems (see Fig. 1 and below), another non-ordered example – $CeCu_6$, as well as several antiferromagnetic heavy fermion systems, e. g. $U_2Zn_{17}$ and $UCd_{11}$. For an early review, see Stewart [106].

Because of the long time frame of the ongoing studies, the field of UcS as it relates to heavy fermion materials is an extremely broad one, with disparate results because of the wide range of electronic states (for example the possibility of quadrupolar ordering discussed below in section 4.2.6. for $PrOs_4Sb_{12}$ oftentimes not easily classified in an all-encompassing fashion. (As will be discussed, this is also somewhat true for the extremely large body of work in the cuprates, although they at least have the commonality of the importance of Cu-O planes.) The present review thus not only attempts to meld the properties of heavy fermion UcS into the properties of the broad range of UcS now known, but also to provide some overview within the sub-field of unconventional superconductivity in heavy fermions alone.

Just because the low temperature value of C/T (T→0) (often just called γ) is large for a superconductor is not conclusive that the superconductivity is unconventional, although it is consistent with UcS. Thus $T_c < \Theta_D > T_F$ (section 2.1.1.) is satisfied (✓1) by *all* HFSs by virtue of their large γ values (since $T_F \propto 1/\gamma$) and will not be further mentioned in this section on heavy fermions. Also, the gray area of how to decide which γ values are 'large' is clearly somewhat arbitrary. A γ value (expressed as mJ/molCe/U/Pr-$K^2$, where the normalization scales the γ value per mole of contributing f-atoms) that is several times the largest d-electron metal value (e. g. γ for β-Mn is [107] ~70 mJ/mol$K^2$ and for [2] $V_3Ga$ ~35 mJ/molV-$K^2$) is a good candidate for the classification 'heavy fermion.' Although it would be defensible to restrict ourselves to γ≥100 mJ/molCe/U/Pr-$K^2$ in this heavy fermion section, it is common (see, e. g., Pfleiderer [70]) to include $PuCoGa_5$, $CePd_2Si_2$, and $URu_2Si_2$ (all with γ~60-70 mJ/molPu/Ce/U-$K^2$) as heavy fermions. Since they do not belong to any of the other UcS classes and are interesting examples of unusual superconducting behavior, we include them below, with the reminder that for these γ≤100 mJ/molCe/U/Pr-$K^2$ compounds $T_c < \Theta_D > T_F$ does not hold.

In terms of the 'strongly suggestive evidence' for determining UcS, the HFSs – being the first discovered – have a very rich set of such characteristics. Tunneling measurements (Section 2.1.7.), which were the first proof of d-wave pairing in the cuprates [64], are difficult to perform on heavy-fermion superconductors due to their short superconducting coherence lengths. However, the requisite clean sample surfaces have been achieved in several heavy fermion systems, including $UPt_3$, $CeCoIn_5$, and $UPd_2Al_3$. The only 'strongly suggestive of' UcS characteristic from section 2.1. missing is non-centrosymmetric crystal structure, which of course is its own unique class-determiner.

We will start our discussion chronologically – discussing in detail the first three discovered HFSs ($CeCu_2Si_2$, $UBe_{13}$ and $UPt_3$). Following a similarly thorough discussion of $PrOs_4Sb_{12}$ and $CeMIn_5$, with emphasis on M=Co, we will then cover the high points of several other HFS with γ < 100 mJ/mol$K^2$ and then attempt to provide a concise summary/overview of UcS in HFS. This review, in order to remain of quasi-readable length, will not discuss in depth *every* possible UcS. For example, nothing new would be gained in discussing $PuRhGa_5$, $T_c$=8.5 K, after $PuCoGa_5$, $T_c$=18.5 K, since they are essentially identical in their important properties. Similarly, $UNi_2Al_3$ is arguably very similar to $UPd_2Al_3$, which is more extensively studied.

### 4.1. *A Short Summary of the Theory of Heavy Fermion Superconductivity:*

Theory that is unique to a particular system will be discussed below with that compound, with this short introduction meant to provide an overview. Considering first the normal state out of which the superconductivity forms, heavy fermion compounds are highly correlated electron systems with both large low temperature normal state (T>$T_c$) specific heat γ values and large magnetic susceptibilities which reflect their nearness to magnetic behavior. As one example, textbook Curie Weiss behavior in the susceptibility (where $1/\chi \approx (8/\mu_{eff}^2)*(T+\Theta)$) – considered evidence for a local moment - in $PuCoGa_5$ extends [108] unchanged from 300 K <u>all the way down</u> to $T_c$=18.5 K, with an effective moment of 0.68 $\mu_B$ and $\Theta$=2 K. (See Flint, et al. [109] for a theoretical discussion of the implications and section 4.3.1. for discussion of sample dependence.) The theoretical understanding of a localized spin with an antiferromagnetic interaction (1/χ extrapolated from higher temperature → 0 for T<0) with the conduction electrons in the normal state is based on Kondo physics. However, the problem of going from the (understood theoretically) single ion Kondo model to the case of the heavy fermion materials, where each unit cell contains a localized spin on the f-ion site, is still the subject of approximations rather than an established theory. Thus, despite a few scattered results to the contrary, heavy fermion normal state physical properties such as $\chi(T\rightarrow 0)$ and $\gamma$ do *not* scale with f-atom concentration (as they would have to in any so-called "Kondo lattice" picture) but instead demonstrate a wealth of correlation effects, see e. g. Satoh et al.[110] for studies of $Ce_{1-x}La_xCu_6$ and Kim et al. [111] for studies of $U_{1-x}M_xBe_{13}$.

In the superconducting state, although exchange of spin fluctuations (ferromagnetic spin fluctuations, Berk-Schrieffer [112]; antiferromagnetic spin fluctuations, Scalapino [14]) as a pairing mechanism in the nearly magnetic heavy Fermions is an attractive theory, there are others in competition as will be discussed below in the individual cases. The search for a unified theory of superconductivity in heavy fermion systems (Grewe and Steglich, 1991 [113]; Sigrist and Ueda, 1991 [3]; Mineev and Samokhin, 1999 [75]; Thalmeier and Zwicknagl, 2005 [114]; Scalapino, 2012 [14]) or for a theory for a particular compound ($UPt_3$ Sauls, 1994 [62] and Joynt and Taillefer, 2002 [115]; $U_{1-x}Th_xBe_{13}$ Kumar and Wölfle, 1987 [116] and Sigrist and Rice, 1989 [117]; skutterudites Maple et al., 2008 [118]) is made more difficult by the very different microscopic natures of the various discovered compounds (although see a recent paper (Kim, Tam, and Stewart, 2015 [119]) for a scaling law for the superconducting condensation energy, U, that shows universal behavior for *all* superconductors vs $T_c$). For example, the nature of the multiple (three in number) superconducting phases (✓7) in $UPt_3$ seems, after a long struggle, to be solved while in $U_{1-x}Th_xBe_{13}$ (with two phases) there is still debate. The pairing symmetry of the various UcS heavy fermions (discussed below), thanks to the powerful experimental tools that have been developed, is mostly agreed upon. Sigrist and Ueda [3] point out that in heavy-fermion systems that have ions with heavy mass like Ce or U, spin-orbit coupling should also be important for the Cooper pairs (Anderson, 1984 [93], [120]). An important consequence of group theory is that, with this non-negligible spin orbit coupling, line zeros are not allowed for odd-parity (l=1 or 3, p- or f- wave ) superconductors (Volovik and Gor'kov, 1984 [121] and 1985 [82]; Blount, 1985 [122]; Ueda and Rice, 1985a and 1985b [123-124]). Blount, in particular, gave a general proof of this. Therefore at very low temperatures pure samples should obey power laws corresponding to the point zeros when they are odd parity superconductors., e. g. $1/T_1 \sim T^5$ point zeros, not $T^3$ (line zeros) and $C \sim T^3$ (point zeros), not $T^2$ (line zeros).

As will be seen below, the use of group theoretical arguments to restrict the possible gap structures (Anderson [120] and Volovik and Gorkov [82]) in the various crystal symmetries (tetragonal, e. g. $CeCu_2Si_2$; cubic, e. g. $UBe_{13}$; and hexagonal, e. g. $UPt_3$) provided an important set of guidelines in distinguishing which theories were appropriate.

### 4.2. *Heavy Fermion Superconductors with γ > 100 mJ/molK$^2$*

4.2.1. *$CeCu_2Si_2$, $T_c \approx 0.5$-0.6 K:*

The distinguishing feature of HFSs is that at the second order superconducting phase transition where the discontinuity/broadened transition in the specific heat occurs, the size of this anomaly, $\Delta C$, is roughly equal to the electronic normal state specific heat, $\gamma T_c$. (In weak coupling BCS theory, $\Delta C/\gamma T_c = 1.43$.) Since the definition of a heavy fermion system is that the electronic specific heat coefficient $\gamma$ ($\propto m^*$, the electron effective mass) is large (as just discussed primarily

$\geq 100$ mJ/molK$^2$), this therefore implies that $\Delta C/T_c$ is also large. This of course was the initial exciting development in CeCu$_2$Si$_2$ [6] that the heavy Fermion superconducting electrons have large m* ($\gamma \approx 1000$ mJ/molK$^2$ which is *still increasing* with decreasing temperature down to $T_c$) and strong correlations ("probably magnetic in origin") (✓2), contrary to the existing weak electron-phonon coupled theories. Of course, another characteristic property of UcS, $T_c < \Theta_D > T_F$ (section 2.1.1.) is satisfied (✓1) by all HFSs by virtue of their large $\gamma$ values since $T_F \propto 1/\gamma$.

When Steglich et al. first published [6] that heavy Fermion superconductivity occurred in CeCu$_2$Si$_2$, sample quality problems had required [125] approximately a year of effort to get samples with $\Delta C/\gamma T_c$ values anywhere near 1 (see Fig. 6 for a comparison between the specific heat of their superconducting polycrystalline sample and a non-superconducting single crystal). Samples with resistive and inductive indications of superconductivity were produced early in their studies, but refinement of the sample preparation was required to improve $\Delta C/T_c$. Now these many years hence it is known that the detailed Cu stoichiometry is important, with Cu above the 1:2:2 'stoichiometric' ratio required for a bulk $\Delta C/T_c$ anomaly in the specific heat. Although this extreme sample dependence (where the ground state depends "very delicately on the actual stoichiometry" –Stockert et al. [126]) (✓9) in CeCu$_2$Si$_2$ has not been quantified, it is fundamentally related to the discussion in section 2.1.9. above about the extreme sensitivity of superconductivity in UcS to non-magnetic defects. A more quantifiable measure of the sensitivity of CeCu$_2$Si$_2$ to non-magnetic defects is the measured rapid suppression of $T_c$ in Ce$_{1-x}$La$_x$Cu$_2$Si$_2$ ($T_c$ decreases by ~ 50% by x=0.06.) [127].

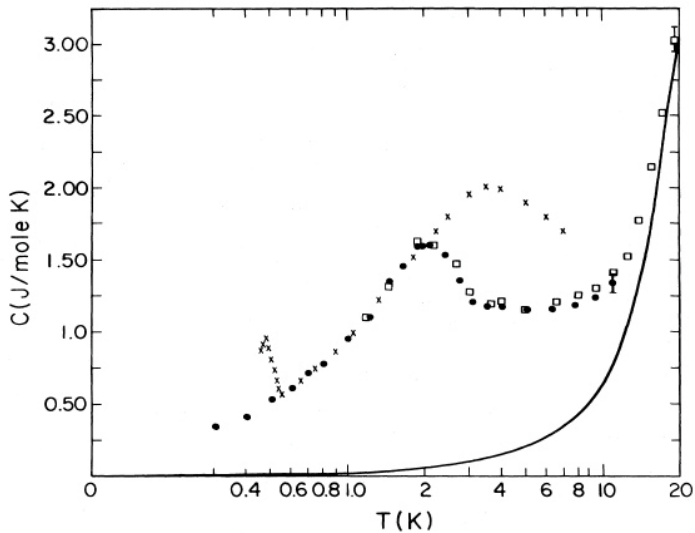

Fig. 6: Specific heat (x's, superconducting polyxtal from Steglich et al. [6], circles and squares – two different non-superconducting single crystal samples from Stewart, Fisk, and Willis [22] vs logT showing the same normal state behavior just above $T_c$, but with the superconducting sample having a Kondo peak higher in temperature – 4 K vs 2 K. Both samples have an extrapolated normal state $\gamma$ of about 1.05 J/molK$^2$, thus fulfilling (✓1) $T_F$ ($\propto 1/\gamma$)$< \Theta_D$ (section 2.1.1.) for UcS. Figure from ref. [22].

Work later established the presence of an antiferromagnetic ('A') phase as a function of stoichiometry (slight Cu deficit) in CeCu$_2$Si$_2$, whose proximity to superconductivity in the phase diagram is thought (see section 2.1.2. above) [126], [128-129] to cause a quantum critical point (✓2) important for the unconventional superconductivity (section 2.1.2.) in CeCu$_2$Si$_2$ at a nearby stoichiometry. Inelastic neutron scattering (INS) measurements [35], which supersede an earlier work [34], on a large, 2g superconducting crystal revealed spin excitations (broad in energy and not a narrow spin resonance as found in the cuprates and the IBS as sketched in section 2.1.3.) at the same Q vector associated with the antiferromagnetic quantum critical point found in the A phase. These INS measurements identify the antiferromagnetic fluctuations "as a main driving force" for the superconductivity. A recent analysis of the dynamical spin response and thermodynamic properties of CeCu2Si2 near its quantum-critical point is consistent with antiferromagnetic excitations being the primary pairing mechanism. [35]

Despite all these evidences for UcS in CeCu2Si2, one of the standard methods for determining UcS (that of non-BCS power laws in, e. g., the specific heat, section 2.1.5.) has recently revealed conventional behavior. In earlier studies on the specific heat of CeCu2Si2, $T_c$=0.63 K, Lang et al. [49] found C ~ $T^2$ down to approximately 0.1 K. Now, in 2014 Kittaka et al. [50], in a slightly less ideal crystal ($T_c$=0.6 K, $\Delta C/T_c$ smaller by 20%) analyze specific heat down to 0.04 K. In this altered analysis using more modern theoretical insight, they fit their data over the whole temperature range of 0.04 to 0.6 K to a two band, fully gapped BCS model (with energy gaps $\Delta_1/k_BT_c$=1.76 and $\Delta_2/k_BT_c$=0.7) as has been done routinely to both IBS and MgB2. Thus, partly due to improved analysis, and partly due to data slightly lower in temperature than previously, the conclusion of Kittaka et al. is that superconductivity in CeCu2Si2 is fully gapped, non-nodal. In addition to the temperature dependence, they also confirm the lack of nodes by finding $\gamma \propto H$ (see section 2.1.15.) and no angular dependence in C($\Theta$) (section 2.1.11.). Another recent report [130] involving specific heat, thermal conductivity, and penetration depth measurements on CeCu2Si2 down to 0.06 K also found behavior consistent with non-nodal behavior: both C and $\lambda$ varied exponentially with temperature at low T and thermal conductivity $\kappa/T \approx 0$ (no linear term). Contrasting with the modelling of Kittaka et al. [50] and Yamashita et al. [130], G. M. Pang et al. [131] - with measurements of the penetration depth of CeCu2Si2 down to 0.04 K and fitting of the specific heat data of Kittaka et al. – propose another model. Pang et al. propose that indeed the new low temperature measurements in CeCu2Si2 are consistent with nodeless behavior, but offer the novel explanation that this is the result of the summing of $d_{x2-y2}$ and $d_{xy}$ pairing, which add in quadrature, producing an s-wave (nodeless) like result. (For a deeper discussion of the theory behind this model, see ref. [132].) Qualitatively, the d+d model of Pang et al. for the Kittaka et al. CeCu2Si2 specific heat data appears to better fit the data than the two gap BCS model of Kittaka et al.

The final resolution of these new low temperature results, which imply nodeless behavior, requires some further time to digest and analyze the data. It would certainly be a surprise if the first UcS in the end was determined to have s-wave symmetry. Such a large specific heat $\gamma$, the strong sensitivity to stoichiometry, a nearby magnetic phase in the phase diagram and evidence for quantum criticality, still argue for a pairing mechanism other than via electron-phonon coupling, but we now understand the correct description as no longer the old picture of CeCu2Si2 as d-wave nodal.

4.2.2. *UBe13*:

Fortunately for the sample growers, although superconductivity in UBe13 is sensitive to defects, both magnetic and non-magnetic, (however superconductivity is not destroyed by grinding [86] – unlike UPt3, see below) this binary compound is alone in its phase diagram with only the pure U and the pure Be endpoints in addition to the 1:13 line compound. Thus – contrary to the experience with CeCu$_2$Si$_2$ - the discovery of bulk superconductivity in UBe$_{13}$, with a large $\Delta C/T_c$, was made in the first sample whose specific heat was measured to a low enough temperature. [133] In order for the superconducting and normal state entropies at T$_c$ (S$_s$(T$_c$) and S$_n$(T$_c$) to match in UBe$_{13}$, the extrapolation of the normal state specific heat divided by temperature, C$_n$/T, to T=0 ($\equiv\gamma$) has to *increase* with decreasing temperature below T$_c$, to about 1000 mJ/molK$^2$. A rising C$_n$/T as T$\rightarrow$0 is also observed with an applied magnetic field to suppress T$_c$. This is a definition of non-Fermi liquid (nFl) behavior (section 2.1.2. ✓2), even though there is no magnetic phase in the zero field phase diagram to create a QCP to produce such nFl behavior. Gegenwart et al. [134] however find evidence ($\rho \sim T^{3/2}$ between 0.4 and 1.5 K in 8 T; C/T ~ -logT between 0.15 and 3 K in 12 T) in magnetic field which they speculate implies a *field-induced* QCP at 4.5 T.

The critical field of UBe$_{13}$ displays unusual behavior, with H$_{c2}$(0)=14 T [135] – a record high for HFSs. Another record high for HFSs, and superconductors in general, found in UBe$_{13}$ is the slope of H$_{c2}$ at T$_c$, H$_{c2}$', which is -45 T/K [136-137]. Finally, there is an inflection point in H$_{c2}$ vs T/T$_c$ at about T/T$_c$=0.5. Y. Shimizu, et al. [138] have proposed that the upward curvature of H$_{c2}$ around T/T$_c$ = 0.5 is simply due to anisotropy rather than very strong coupling and an FFLO state.

In penetration depth measurements from T$_c$=0.86 K down to ~ 0.060 K [55] $\Delta\lambda$ was found to vary approximately at T$^2$. These data allowed for "a fairly good fit" to an axial (point nodes) p-wave gap function. This is consistent with the power law found in the specific heat, C~T$^3$ (but only between 0.2 and 0.9 K) by Jin et al. [139] which also implies point nodes and an axial state. In *apparent* contradiction, NMR measurements [140] find that 1/T$_1$, the spin-lattice relaxation rate, is proportional to T$^3$ between T$_c$ and 0.2 K, with a deviation at lower temperatures. These data were interpreted as being consistent with a polar (line node) p-wave state. However, it is worth noting here the argument of Heffner and Norman [48], where they point out the ubiquity of 1/T$_1$ $\propto$ T$^3$ in HFSs may be masking point node behavior because of a relative smaller phase space occupied by the latter behavior. In summary, these power law results can be used to infer UcS in UBe$_{13}$ (✓5) but, because of their internal inconsistencies (and, in the case of the NMR data, because of their limited temperature range) they are not conclusive.

As discussed above in section 2.1.7., Josephson tunneling can help establish the symmetry of the order parameter in a superconductor *if* high quality surfaces are available. Han et al. [141], in a rather clever tunneling experiment, established that the induced s-wave order parameter in UBe$_{13}$ around the contact region with the BCS superconductor Ta is suppressed when UBe$_{13}$ becomes superconducting. They then inferred that this "negative s-wave proximity effect" implies that UBe$_{13}$ might have an odd (triplet) parity order parameter.

Multiple phases in the superconducting state as an indication of UcS (section 2.1.6.) in UBe$_{13}$ are in general identified with Th-doped samples (discussed below.) However, after the discovery of multiple superconducting phase in doped U$_{1-x}$Th$_x$Be$_{13}$, Rauchschwalbe et al. [142-143] found an indication of a second transition in the specific heat of *pure* UBe$_{13}$, T$_c$=0.87 K, at around 0.5 K. This has been confirmed in various other measurements, including field sweeps of the specific heat at fixed temperatures below T$_c$ [144] and the thermal expansion [145].

Although, as will be discussed, there is a good amount of theory to describe the two UcS transitions in U$_{1-x}$Th$_x$Be$_{13}$, the anomaly in pure UBe$_{13}$ at a temperature not unlike that of the second, lower superconducting transition in $U_{1-x}Th_xBe_{13}$ remains a puzzle. It is an important function of this review to point out such interesting, but either unaddressed or poorly understood features of UcS. Thus, in addition to the contradiction discussed above for $CeCu_2Si_2$ between the low temperature temperature dependence of C/T (fully gapped) and the other measurements (e. g. $1/T_1$ and thermal expansion) which indicated nodal behavior, the ~0.5 K anomaly in C/T in pure $UBe_{13}$ and its possible connection to the second superconducting transition in $U_{1-x}Th_xBe_{13}$ joins the 'to-be-understood' list of this review.

### 4.2.3. $U_{1-x}Th_xBe_{13}$:

Although this subject is properly a part of the discussion of $UBe_{13}$, since it is one of the more fascinating cases of UcS in the heavy Fermions, we will devote a separate section to it.

Smith et al. [146], in doping $UBe_{13}$ with eight different elements to try to better understand the superconductivity in the parent compound, discovered an unusual non-monotonic decrease in $T_c$ (measured with ac susceptibility and resistivity) with increasing Th content. Ott et al. [147] then discovered that a second transition ΔC in the bulk specific heat of $U_{1-x}Th_xBe_{13}$ appears with small amplitude with increasing Th content at x=0.0216 (nominal composition), while this second ΔC becomes quite large and distinct around x≈0.03, and disappears around x=0.04. An example of the specific heat, C, for two samples (annealed and unannealed) of

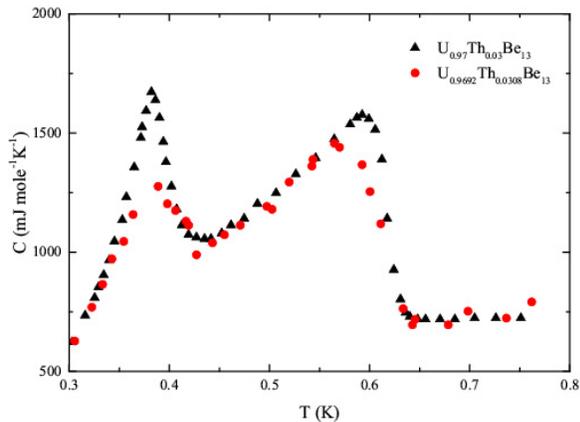

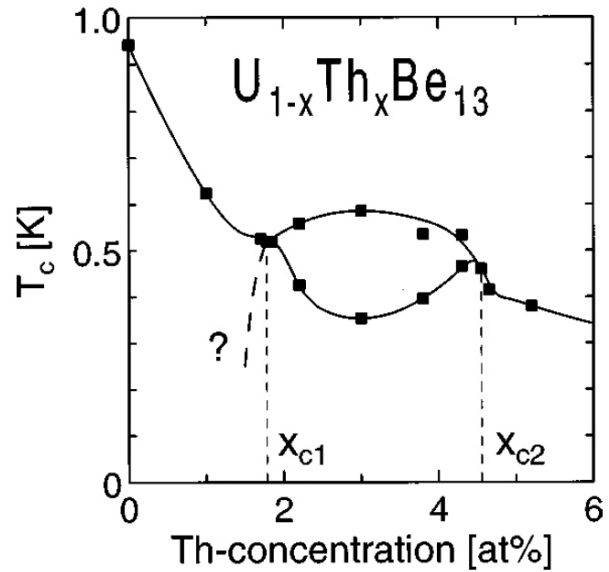

Fig. 7 (color online) Kim et al. [148] data are black triangles (high purity sample annealed at 1400 C for 1220 hrs) and Ott et al. [147] unannealed data (red circles). The lower transition appears to be more sensitive to defects removed by annealing than the upper transition. Figure from [148].

Fig. 8 Figure from Scheidt et al. [149]. The non-monotonic-with-x resistive $T_c$ transition at $T_{c1}$ in $U_{1-x}Th_xBe_{13}$ [146] was the first clue that something unusual was happening.

$U_{0.97}Th_{0.03}Be_{13}$ is shown in Fig. 7, with a phase diagram of $T_{c1}$ (upper transition) and $T_{c2}$ as a function of Th composition shown in Fig. 8.

The origin of the lower transition has been the subject of numerous experiments and theories. Since the upper superconducting transition blocks many of the probes (e. g. resistivity, $\chi_{ac}$) that might have been applied to characterizing the lower transition at $T_{c2}$, progress was at first slow. Batlogg et al. [150] measured a very large peak in the ultrasonic attenuation at $T_{c2}$ and inferred that the lower transition was antiferromagnetic, rather than an unconventional superconducting phase as speculated in the discovery paper [147] by Ott et al. Thinking that the behavior of the lower transition vs that of the upper transition with field might provide a useful comparison, Mayer et al. [151] measured the specific heat of the two peaks and remarked that the two transitions were suppressed "at a qualitatively similar rate" with applied magnetic field. However, Ott et al. [152] found that field suppressed the specific heat transition at $T_{c2}$ at a "somewhat smaller" rate. (This controversy was later resolved using high quality annealed samples with sharper transitions as shown in Fig. 7 by Kim et al. [148], who found that the $H_{c2}(T)$ curves for the two transition are indeed parallel for H≤1.25 T, which a priori argues against the two transitions being of radically different (e. g. antiferromagnetic vs superconducting) nature. Susceptibility measurements [153] of $T_c$ as a function of pressure in various compositions of $U_{1-x}Th_xBe_{13}$ are analyzed as also consistent with two superconducting states. Rauchschwalbe et al. [154] discovered that the slope of the lower critical field, $H_{c1}$, markedly increased as temperature was lowered through $T_{c2}$, indicating a strong increase in the superfluid density. They proposed that a second portion of the Fermi surface becomes superconducting below $T_{c2}$. Heffner et al., [155] and references therein, using µSR measurements, found that quasistatic magnetism, with a moment in the range of $10^{-3}$ to $10^{-2}$ $\mu_B$/U atom, appears below $T_{c2}$. Whether this result implies time reversal symmetry breaking (section 2.1.8.) depends on determining the nature of the second transition at $T_{c2}$. The entropy associated with the lower transition (in unannealed material like that measured by Heffner et al. [155], see Fig. 7) would be consistent with a local magnetic moment of ~ 0.01 $\mu_B$/U atom.

Various theories have been proposed to explain these seemingly contradictory results. Kumar and Woelfle [116] proposed two different superconducting symmetries, d-wave at $T_{c1}$ and s-wave below $T_{c2}$, with a mixture of the two for $T_{c2} < T < T_{c1}$. In their theory, the muon result of magnetism below $T_{c2}$, discovered later, is not addressed. Sigrist and Rice [117] also predict two superconducting symmetries being present, with non-unitary (unconventional) superconductivity below $T_{c2}$ (which would be consistent with the greater sensitivity to annealing/removal of defects of the lower transition shown in Fig. 7), where such non-unitary pairing creates [82] a finite local spin polarization – consistent with the result [155] of Heffner et al. Concerning the ultrasonic attenuation peak at $T_{c2}$, this is ascribed to dissipative domain wall motion. A non-unitary order parameter (Ohmi and Machida [156]) below $T_{c2}$ would also be consistent with the large residual γ in the low temperature superconducting state data of Jin et al. [139], where for a high purity, long term annealed sample of $U_{0.97}Th_{0.03}Be_{13}$ they found $\gamma_{residual}$=750 mJ/molK$^2$. This $\gamma_{residual}$ is 30% of $\gamma_{normal}$ extrapolated [157] from above $T_{c1}$ to match the superconducting and normal state entropies at $T_c$: $S_{sc}(T_c)=S_n(T_c)$. Such a large residual γ (✓14) in the superconducting state is also found in UPt$_3$, see below, while $\gamma_{residual}$ is only [139] 7% of $\gamma_{normal}$ in undoped UBe$_{13}$.

Due to the limited temperature range available below $T_{c2}$ in $U_{1-x}Th_xBe_{13}$, there are not a great deal of power law determinations. Jin et al. [139], whose lowest temperature of measurement (0.09 K) was determined by the radioactive self heat from the depleted U, observed that - for all three of their measured concentrations of Th with x>$x_{c1}$ (see Fig. 8) – the low temperature specific heat can be fit from 0.09 to 0.32 K by a fully gapped BCS temperature dependence with an adjustable size of the energy gap, Δ - rather than $C \sim T^3$ as they found for pure UBe$_{13}$. The value of the gap that gave the best fit to the data for all three compositions was Δ/k$_B$T$_c$=2.7, i. e. much larger than the weak-coupled BCS value of 1.76. Recall the result of Kittaka et al. [50] discussed above for CeCu$_2$Si$_2$, where they measured C down to 0.04 K (no self heating problem in Ce) and also found a BCS model, fully gapped fit to be appropriate, with however two gaps, both of which were weak coupled.

More recently, in a theory review by Thalmeier and Zwicknagl [114], the extent of theoretical understanding of U$_{1-x}$Th$_x$Be$_{13}$ was summarized "there is no developed microscopic theory for this complex behavior." Thus, the precise nature of the UcS lower transition and its pairing symmetry, as well as its cause, in U$_{1-x}$Th$_x$Be$_{13}$ joins our unsolved puzzle list.

4.2.4. *UPt$_3$:*

In its discovery as the third known HFS (Stewart et al. [90]), the specific heat transition ΔC of a collage of annealed needle single crystals of UPt$_3$ was broad (ΔT$_c$ ≈ 0.1 K with T$_c$$^{onset}$ = 0.54 K) and somewhat smaller than the BCS predicted size. ΔC/γT$_c$, γ≈450 mJ/molK$^2$, extrapolated to an idealized sharp transition was only 1.0 vs 1.43 for BCS (one of the signs as discussed in section 2.1.9. of sensitivity to defects). Since the normal state specific heat could be fit [90] to the spin fluctuation form (C=γT + βT$^3$ + δT$^3$logT), and since the superconductivity was destroyed by grinding (sensitivity to defects, ✓9), UPt$_3$ was from its discovery as a superconductor believed to be unconventional in its pairing, with exchange of spin fluctuations as a possible pairing mechanism (see, e. g., M. R. Norman [158]).

Besides the normal state temperature dependence of the specific heat (implying the presence of Fermi liquid spin fluctuations) and the T$^2$ temperature dependence (Willis et al. [159]) (Fermi liquid behavior) of resistivity (ρ=ρ$_0$ + 0.5 μΩ-cm/K$^2$ T$^2$), the temperature dependences of various quantities in the superconducting state in UPt$_3$ are of course of interest for confirming UcS and drawing inferences about nodal structure. In the superconducting state specific heat, this goal is somewhat thwarted since there is (Schuberth et al. [160]) an enormous peak in the specific heat below 0.1 K of unknown origin (although see Brison et al. [161]). As discussed already (Heffner and Norman [48]), the NMR 1/T$_1$ temperature dependence in UPt$_3$ is T$^3$ (measured down to 0.1 K, Kohori et al. [162]), typical of heavy fermion systems and lines of nodes. Since the zero field, low temperature B phase (see Fig. 4) is thought (Sauls [62]) to have both line and point nodes (anisotropic gap structure), knowledge of the temperature dependence of the ultrasonic attenuation and the penetration depth in the B phase is of less utility than a measure of the anisotropy (see Joynt and Taillefer [115] for a detailed discussion.) A very thorough study of the low temperature thermal conductivity (down to 0.016 K) by Suderow et al. [163] points up the difficulties of obtaining clean, definitive power laws in UPt$_3$ with its multiple phases and its anisotropy.

Mueller et al. [59], in 1987, using a single crystal of UPt$_3$ and ultrasonic attenuation measurements as a function of field, found three (later named A, B, and C) distinct phases (✓6) in the superconducting state. These data were later refined by Adenwalla et al. [63], also using ultrasonic attenuation, resulting in the phase diagram in Fig. 4 above. Aeppli et al. [36] found a small (0.02 μ$_B$) fluctuating antiferromagnetic moment in the basal plane of UPt$_3$ starting below 5 K. The magnetic order parameter stops evolving at T$_c$, establishing a link between the

superconducting and magnetic orders. Fisher et al. [60], in polycrystalline samples, and later Hasselbach et al. [80] in a 270 mg single crystal could follow the field evolution of the A and B phases (present in zero field as shown in Fig. 4) in a bulk measurement, see Figure 9. Thus, UPt$_3$ joined U$_{1-x}$Th$_x$Be$_{13}$ as an unconventional superconductor with multiple phases (see CeCoIn$_5$ below for the third example of multiple superconducting phases in the heavy fermions). While some of the indications of UcS listed in section 2 above are only suggestive (e. g. temperature dependences not measured over a decade of temperature and down to at least $T_c/10$), the presence of multiple superconducting phases is certain proof of UcS.

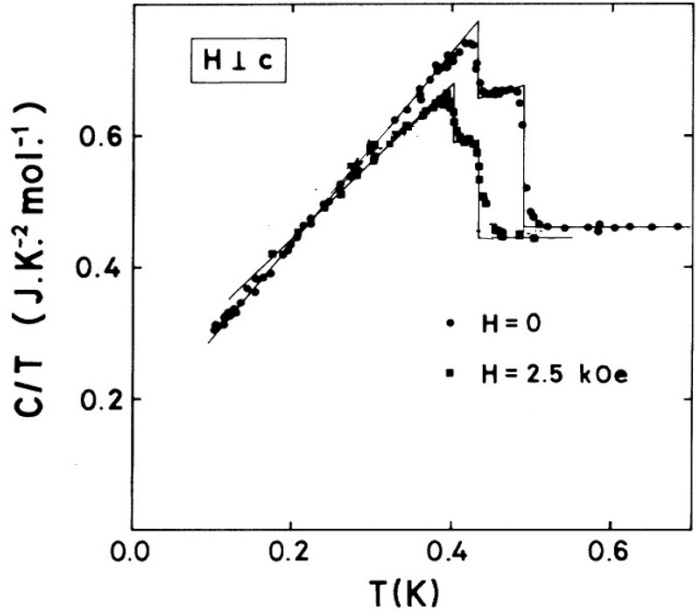

Fig. 9 Two representative fields, 0 and 0.25 T, are shown (in [80] are five fields), which make it clear that the decrease of $T_{c2}$ with field is slower than for the upper, $T_{c1}$ transition, unlike the behavior found for U$_{1-x}$Th$_x$Be$_{13}$. The upper transition is 0.015 K wide; the lower transition is 0.010 K wide. The large residual $\gamma$ value (✓14) in the superconducting state ($\approx$190 mJ/molK$^2$), almost ½ of $\gamma_{normal}$, in this high quality sample is discussed in the 1D, spin triplet model by Ohmi and Machida [156]. Reprinted figure with permission from Hasselbach et al., Phys. Rev. Lett. 63 (1989), p. 93. [80] Copyright (1989) by the American Physical Society.

The discovery of multiple superconducting phases in UPt$_3$ by Mueller et al. and the subsequent further experimental investigations led, just as in U$_{1-x}$Th$_x$Be$_{13}$, to intense theoretical efforts to explain the UcS. Starting from the group theoretical discussions of possible gap symmetries by Anderson [120] and Volovik and Gorkov [82], numerous theories were proposed to explain the complex UcS in UPt$_3$ (for a review, see Joynt and Taillefer [115]).

Although there is no complete theoretical understanding of this complex behavior, Joynt and Taillefer [115] argue for a two component order parameter belonging to either the $E_{1g}$ or the $E_{2u}$ 2D group theoretical representations. Sauls [62] gave a thorough discussion of the so called $E_{2u}$ odd parity, triplet (S=1 or 3, p- or f-wave) 2D representation order parameter for hexagonal UPt$_3$ (see Fig. 10), where the order parameter determines the gap symmetry and any nodes (see Table V in Volovik and Gorkov [82]). The $E_{1g}$ representation (see, e. g., Putikka and Joynt [164] and Park and Joynt [165]) is a singlet, even parity representation, s- or d-wave. In the $E_{2u}$ model, the coupling (Aeppli et al. [36]) of the small antiferromagnetic moment to the superconducting order parameter provides a symmetry breaking field which is responsible (Sauls [62]) for the apparent tetracritical point and the two transitions in zero field, Figs. 4, 9, and 10. For competing proposed models, see Chen and Garg [166] – with two order parameters that are degenerate), as

well as Machida and Ozaki [167] and Ohmi and Machida [156] – both 1D representations.) The 1D model of Ohmi and Machida has been used to explain the large residual specific heat γ (up to ≈45% of $\gamma_n$) observed in the superconducting state (see Fig. 9). Note that this large $\gamma_{residual}$ is strongly sample dependent, with, e. g., Ott et al. [168] reporting $\gamma_{residual}$ ≈ 0 (measured down to 0.07 K), Sulpice et al. [169] reporting $\gamma_{residual}$ ≈ 110 mJ/molK$^2$, and Fisher et al. [60] reporting

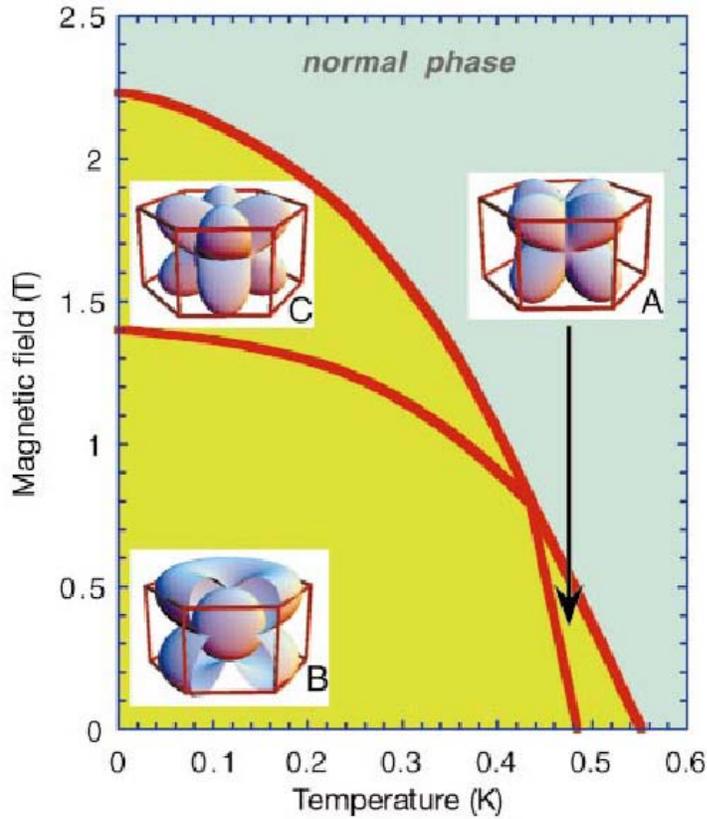

Fig. 10 (color online): Graph from Huxley et al. [61]. Magnetic field along c-axis. Nodes exist (Joynt and Taillefer [115]) in all three superconducting phases. The B phase has a line node in the basal plane and point nodes along the hexagonal *c* axis. In the A phase, according to Huxley et al., antinodes (maxima in the gap function) occur along bisectors of the a, a* directions (45 ° rotated from the a direction) while antinodes occur along the a direction and the direction perpendicular to the a direction. A, B, and C come together in an apparent tetracritical point. UPt$_3$ is unusual in that the superconducting gap does not have the same rotational symmetry as the crystalline lattice. The graphical pictures for the Fermi surfaces shown in the white squares for the A and B phases correspond to the E$_{2u}$ model (f-wave), see Strand et al. [170] for further depictions. Reprinted by permission from Macmillan Publishers Ltd: Nature [61], copyright (2000).

$\gamma_{residual}$ ≈ 150 mJ/molK$^2$ for their two samples. The Ott et al. sample was an adjacent piece to a sample in which dHvA oscillations were observed, i. e. presumably of high quality. The Sulpice sample was the first one to clearly show two transitions in the specific heat at T$_c$ (i. e., the sample was well ordered) – later recognized by Fisher et al. [60] and Hasselbach et al. [80] (Fig. 9) to represent transitions into the A and B phases. This strong sample dependence of $\gamma_{residual}$ in *a priori* well ordered and low-impurity samples is an indication of strong sensitivity of the UcS to low level amounts of impurities/defects (✓9).

There have been various further measurements consistent with the triplet E$_{2u}$ model and contrary to, e. g., the singlet E$_{1g}$ model. These include small angle neutron scattering [61] which studied

the alignment of the superconducting flux line lattice in field applied along the hexagonal c-axis (Fig. 4), Josephson tunneling experiments as a function of angle (Strand et al. [171]), and anisotropy of the thermal conductivity at low temperatures in the B phase (see the review [115] by Joynt and Taillefer). Polar Kerr effect measurements have been recently performed (Schemm et al. [65]) in UPt$_3$ and show the breaking of time reversal symmetry (✓8) in the B phase. These measurements are consistent with any one of the four 2D representations possible for hexagonal UPt$_3$, i. e. are also consistent with, but not proof of, the E$_{2u}$ representation. Kycia et al. [172] plot T$_c$ vs ρ$_0$ (inversely proportional to the mean free path) for various annealed single crystals of UPt$_3$ (see discussion in section 2.1.16.) and get good agreement (✓16) with the Abrikosov-Gorkov theory modified for UcS with anisotropy.

4.2.5. *UNi$_2$Al$_3$ / UPd$_2$Al$_3$:*

After the discoveries of heavy fermion superconductivity in CeCu$_2$Si$_2$ (1979), UBe$_{13}$ (1983), and UPt$_3$ (1984) there was a long period with no further discoveries (URu$_2$Si$_2$, discovered in 1986, has a γ value only as large as β-Mn and will be discussed below in the γ < 100 mJ/molK$^2$ section, section 4.3.) until 1991, when the group of Frank Steglich found the heavy fermion antiferromagnets UNi$_2$Al$_3$ and UPd$_2$Al$_3$, T$_c$ = 1.1 and 2 K respectively. (Geibel et al. [173-174]). These hexagonal materials, with coexistent antiferromagnetism (T$_N$ = 4.6 and 14 K respectively), have γ values extrapolated from just above T$_c$ (i. e. in the antiferromagnetic state) of 120 [173] and 140 [79] mJ/molK$^2$ respectively. Due to the greater availability of quality samples of UPd$_2$Al$_3$, research has tended to focus primarily on this compound. As well, the lower T$_c$ in UNi$_2$Al$_3$ has resulted in power law investigations over much too limited a range (e. g. between 0.5 and 1.0 K for 1/T$_1$ Tou et al. [175]) to allow any conclusions about nodal behavior (section 2.1.12.).

    Neutron scattering experiments (Geibel et al. [176]) established the local moment in UPd$_2$Al$_3$ to be 0.85 μ$_B$, while that in UNi$_2$Al$_3$ is (Schroeder et al. [177]) is much smaller, 0.24 μ$_B$. These were the first heavy fermion superconductors coexistent with a static local moment. (The very small moment (~0.02±0.01 μ$_B$) starting below 5 K in UPt$_3$ is thought to be dynamic, as discussed above.) Elastic neutron scattering (Krimmel et al. [178]) on a single crystal was measured to study the magnetic order in UPd$_2$Al$_3$. The data show a "remarkable" dip in the integrated scattering intensity (for the (0 0 ½) magnetic Bragg peak) as temperature is lowered through T$_c$, followed by a full recovery upon further cooling. This 10% 'bite' out of the scattering intensity, about 1 K wide and centered on T$_c$, indicates an interaction between superconductivity and magnetism different from the cessation of order parameter evolvement seen at T$_c$ in UPt$_3$ and remains unexplained.

    More recent neutron scattering experiments have been used to explore the connection between the superconducting pairing mechanism and the observed magnetic response of the antiferromagnetic UPd$_2$Al$_3$. Bernhoeft et al. [179] propose a superconducting interaction between quasiparticles strongly renormalized by low frequency spin fluctuations in the antiferromagnetic state. A more unique-to-UPd$_2$Al$_3$ mechanism for superconductivity was proposed by Sato et al. [33] based on their neutron scattering results, and further discussed theoretically by McHale et al. [180] and Chang et al. [181]. In this model, one f-electron is itinerant and experiences an effective attractive interaction produced by crystalline electric field excitations of the remaining two localized f-electrons. Sato discovered a resonance peak in the

neutron scattering intensity below $T_c$. They identify this as evidence for collective modes (bosons) which are local crystalline electric field excitations that propagate because of exchange between local moments. These 'magnetic exciton' bosons are then theorized ([33], [180-181]) to cause an effective attractive interaction between the itinerant (heavy) f-electrons, causing superconductivity.

Other evidence for UcS in UPd$_2$Al$_3$ includes $\kappa(\Theta, H)$ (Watanabe et al. [182]) (✓11). Specific heat has been measured (Caspary et al. [79]) down to 0.4 K, but the $T^3$ variation noted between 0.4 and 1 K is over too limited a temperature range to draw conclusions about nodal structure.

Josephson tunneling on heteroepitaxial thin films of UPd$_2$Al$_3$ on LaAlO$_3$ was carried out, where the film quality was somewhat less than for bulk specimens ($T_c$ decreased by 20%, $T_N$ decreased by 15%, residual resistivity increased a factor of seven (Huth et al. [183]; Geibel et al. [174]). The authors concluded that a feature in the tunneling conductivity data at 1.2 meV was a spin fluctuation mode that was coupled to the superconducting order parameter. Despite some sample quality issues, this tunneling work (✓7), coupled with the neutron scattering data (Sato et al. [184]) and Bernhoeft et al. [179]) that show a feature in the scattering intensity around 1.2-1.5 meV, is one of the more experimentally-grounded proposals for spin-fluctuation- mediated superconductivity in a heavy fermion system.

4.2.6. *PrOs$_4$Sb$_{12}$:*

PrOs$_4$Sb$_{12}$, $T_c$=1.8 K, in the skutterudite family, has a low temperature specific heat above $T_c$ (from where the normal state $\gamma$ is extrapolated) that is complicated by a low lying first excited crystal field level that is only 7 K above the ground state. Various extrapolations/estimations for $\gamma$ exist (Bauer et al. [185]; Maple et al. [186]) that range from 350 to 500 mJ/molK$^2$. Since PrOs$_4$Sb$_{12}$ has a nonmagnetic ground state, it has been conjectured (Maple et al. [187]) that the superconducting electron pairing may be unconventional and mediated by quadrupolar fluctuations.

Although the discovery work [185] did not see a double peak structure in the specific heat at $T_c$, numerous later works (Maple et al. [186]; Vollmer et al. [188]; Measson et al. [189-190]; McBriarity et al. [191]) resolved two anomalies. These anomalies, with respect to their relative amplitudes, are much like in UPt$_3$ discussed above with Fig. 9. However, the upper transition in the specific heat of PrOs$_4$Sb$_{12}$ is significantly (~factor of two) broader whereas in UPt$_3$ both transitions are equally narrow. The current consensus appears to be (see references and discussion in McBriarity et al. [191]) that the upper transition is due to inhomogeneous superconductivity and not due to an intrinsic second superconducting phase as convincingly proven to exist in UPt$_3$.

The superconducting upper critical field, $H_{c2}(0)$, is approximately 2 T. At fields starting ~ 4.5 T, there is (Aoki et al. [192]; Ho et al. [193]) a field induced ordered phase (non-superconducting) characterized by neutron scattering (Kohgi et al. [194]) as antiferro-quadrupolar ordering.

Concerning indications of UcS in PrOs$_4$Sb$_{12}$, muon spin resonance (Aoki et al. [195]) data show unambiguous evidence (spontaneous appearance of static internal magnetic fields below $T_c$) for time reversal symmetry breaking in the superconducting state (✓8). However, thermal conductivity (Izawa et al. [196]), $\kappa$, measurements as a function of field and angle are less clear an indication for UcS. These data indicated a transition between two superconducting

phases as a function of field at 0.52 K, with the higher field phase having four fold symmetry and the low field (<0.6 T) having two fold symmetry. This apparent change of the gap symmetry has been addressed theoretically by Curnoe et al. [197] and by Alrub and Curnoe [198-199]. Since there are no other measurements which support this second superconducting phase at low fields in PrOs$_4$Sb$_{12}$ (for example the small angle neutron scattering results up to 1 T of Huxley et al. [200] discussed in the next paragraph see no evidence of this gap symmetry change; similarly high resolution magnetization measurements (Tayama et al. [201]) down to 0.4 K and up to 1.4 T see no "appreciable anomaly" corresponding to the result of Izawa et al.), this $\kappa(H,\Theta)$ result must be viewed with some caution.

      In terms of temperature dependence of various measured parameters in the superconducting state, these measurements give mixed indications. Penetration depth $\lambda$ determined by $\mu$SR indicates (MacLaughlin et al. [202]) no nodal behavior, i. e. consistent with conventional superconductivity. The rather slow depression (Frederick et al. [203]) of T$_c$ with the doping of the non-magnetic Ru in PrOs$_{4-x}$Ru$_x$Sb$_{12}$ is also consistent with conventional superconductivity (section 2.1.9.). Scanning Tunneling Microscopy (STM) measurements (Suderow et al. [204]) indicate no nodal behavior, but a variation of the gap over the Fermi surface consistent with a multiband superconductor. This somewhat general conclusion of more than one gap value on the Fermi surface has since been refined. For example, Seyfarth et al. [205], using thermal conductivity, identified two fully open gaps, with a size ratio of ~3. In contrast to these measurements indicating fully gapped, nodeless behavior, several other measurements were interpreted as evidence for nodal behavior. Thermal conductivity measurements [206] down to 0.05 K saw a significant value of $\kappa$/T as T→0, strong evidence for nodal behavior. Penetration depth $\lambda$ determined using a tunnel diode oscillator (Chia et al. [207]) $\Rightarrow$ two point nodes on the Fermi surface like in the $^3$He A-phase, although their sample showed the upper superconducting transition thought to be due to inhomogeneous superconductivity, i. e. samples were not of the best quality. Finally, small-angle neutron scattering experiments (Huxley et al. [200]) measured distortions in the flux-line lattice consistent with nodes in the gap.

      In summary, despite improvements in sample quality the arguments for UcS in PrOs$_4$Sb$_{12}$ are not as uniform as for the other heavy fermion superconductors in section 4. Perhaps this is inherent in the proposed different superconducting pairing mechanism – exchange of quadrupolar rather than magnetic fluctuations.

### 4.2.7. *CeMIn$_5$, M=Co,Ir,Rh:*

These tetragonal 115 compounds are a subset of the more general series Ce$_n$M$_m$In$_{3n+2m}$, with M=Co, Ir, Rh, Pd and Pt. The basic building block of these compounds is CeIn$_3$, called the 'infinite layer' compound since, when n gets very large, Ce$_n$M$_m$In$_{3n+2m}$ approaches CeIn$_3$. These very interesting superconductors, along with the PuMGa$_5$, M=Co, Rh, have been frequently reviewed, see, e. g., Setai et al. [208], Sarrao and Thompson [209], Pfleiderer [70] and White et al. [210]. Ce$_2$RhIn$_8$ has T$_c$=2.0 K under 2.3 GPa pressure and Ce$_2$CoIn$_8$ has T$_c$=0.4 K at ambient pressure. For n=m=1, CeCoIn$_5$ (Petrovic et al. [211]) has T$_c$=2.3 K and $\gamma\approx$1150 mJ/molK$^2$, CeIrIn$_5$ (Petrovic et al. [212]) has T$_c$=0.4 K and $\gamma\approx$625 mJ/molK$^2$ (where $\gamma$ in this review is always defined as C$_{normal}$/T extrapolated to T→0 and not C/T at T$_c^+$ as is found in, e. g., Pfleiderer [70]), while CeRhIn$_5$ is an antiferromagnet, T$_N$=3.8 K, where $\gamma$=380 mJ/molK$^2$ (Fisher

et al. [213]). CeRhIn$_5$ becomes superconducting at 2.1 K under an applied pressure of 1.6 GPa. (For a detailed description of the structure, see Pfleiderer [70]) Although band structure calculations combined with dHvA measurements (see, e. g., Hall et al. [214]) give an approximately 2D Fermi surface with "undulating cylinders" for these compounds, the anisotropy in, e. g., the electrical resistivity, the magnetic susceptibility, or the upper critical magnetic field, is only about a factor of two – i. e. they are much more isotropic than the cuprates.

4.2.7.1 *CeCoIn$_5$:* Grown from an In flux, crystals of CeCoIn$_5$ – the highest T$_c$ heavy fermion compound known - are flat platelets, allowing easy determination of properties both in the basal a-b plane as well as along the c-axis. There is a peak in the resistivity vs temperature curve at around 50 K, so that the residual resistivity ratio (RRR), defined as $\rho(300 K)/\rho_{normal}(T\rightarrow0)$, of ≈14 underestimates the high quality of the crystals, in which dHvA oscillations can be seen. (Shishido et al. [215]). CeCoIn$_5$ has been identified as being in the 'superclean' limit, with the superconducting quasiparticle mean free path, ℓ, being over 1 μm and a short (~50 Å) coherence length, ξ (Kasahara et al. [216]).
.      As mentioned above in section 2.1.2., the normal state specific heat (extended down to low temperature by an applied field of 5 T) of CeCoIn$_5$ follows the non-Fermi liquid temperature dependence $C_{electronic}/T \propto -\log T$ (Kim et al. [25]), indicating the presence of quantum critical fluctuations (✓2). This is reinforced by a normal state resistivity $\rho \sim T$ from T$_c$ up to 20 K, again consistent with non-Fermi liquid behavior. (Petrovic et al. [211]) The NMR 1/T$_1$ in the normal state varies as $T^{1/4}$, consistent with spin fluctuation theory's prediction near an antiferromagnetic instability. (Kohori et al. [217]).

In addition, CeCoIn$_5$ possesses a large number of non-exponential temperature dependences in various superconducting state properties (✓5), all consistent with line nodes as listed in Table 1. These properties include: thermal conductivity $\kappa/T \sim T^2$ (Movshovich et al. [218]) – but only in the limited temperature range 0.033 – 0.100 K; superconducting state specific heat $C_{sc} \sim T^2$ [219] – but only between 0.1 and 0.4 K; NMR 1/T$_1 \sim T^3$ (Y. Kohori, et al. [217]) (again – common to heavy fermions) from T$_c$ down to 0.3 K; penetration depth $\lambda \sim T$ along the c-axis (Chia et al. [220]) interpreted as consistent with d-wave with line nodes along the c-axis.

Tunneling measurements (Zhou et al. [221] and Allan et al. [219]) imply nodal $d_{x2-y2}$ pairing symmetry with line nodes in the a-b plane. Angle resolved measurement of κ (Izawa et al. [222]) and of C/T (An et al. [223]) in field for T<0.1 T$_c$ – both very powerful techniques applied successfully in only a few UcS – also imply nodal $d_{x2-y2}$ pairing symmetry (11✓) with line nodes in the a-b plane.   The difference between An et al.'s C/T work's conclusions about the correct nodal direction (in the [110] direction) and previous somewhat higher temperature C/T (Θ,H) work (Aoki et al. [224]) which incorrectly assigned $d_{xy}$ pairing with nodes in the [100] direction is the inversion of the nodal/antinodal directions with lowering temperature in the superconducting state as theoretically explained by Vorontsov and Vekhter [77]. See also Das et al. [225].

The question of a possible pseudogap in CeCoIn$_5$ is at present the subject of conflicting measurments. The resistivity at 1.35 GPa (where T$_c$ as a function of pressure is a maximum at 2.6 K) shows a slight decrease ~0.4 K *above* T$_c$ that has been called (Sidorov et al. [226]) consistent with a pseudogap, although no such feature in ρ at lower pressure (Nicklas et al. [227]) had been remarked upon previously.   STM [221] and STS (Wirth et al. [228]) tunneling

studies support a pseudogap interpretation, with the work by Zhou et al. [221] showing a pseudogap feature 3 K above $T_c$. However, Van Dyke, Davis, and Morr [229] explain the tunneling STM data of Zhou et al. as not being due to a pseudogap, but rather that their measured dI/dV line shape can be explained by the electronic band structure. A further possible support for there being a pseudogap in $CeCoIn_5$ is the $\kappa(\Theta,H)$ work by Izawa et al. [222], where the oscillations seen with angle appear to survive 0.9 K above $T_c$, although a factor of 10 smaller in amplitude than the low temperature data. Whether these higher temperature remanent $\kappa(\Theta,H)$ oscillations can be explained by an anisotropy in the in-plane electrical resistivity is not clear. Works using Andreev reflection (L. Greene et al. [230] and W. K. Park et al. [231]) see, in contrast, no evidence of a pseudogap in $CeCoIn_5$.

In summary, there appears to be conflicting evidence, pro and con, for a pseudogap in $CeCoIn_5$, so at present this remains an open question. We will discuss the well established pseudogap behavior in both the electron and the hole doped cuprates below. Once that discussion is complete, it will become clear how, in comparison, uncertain the existence of a pseudogap in $CeCoIn_5$ is.

Joining the list of strong indications for UcS in $CeCoIn_5$ is the presence of a magnetic resonance in inelastic neutron scattering below $T_c$ (C. Stock et al. [32]). As discussed above in 2.1.3., such a resonance would normally be considered to be evidence for a sign change in the superconducting energy gap $\Delta$ on different parts of the Fermi surface, as would be consistent with the d-wave pairing symmetry already implied from numerous measurements discussed here. However, recent neutron scattering data (Song et al. [232]) on $Ce_{1-x}Yb_xCoIn_5$, x=0, 0.05, and 0.3, have called this interpretation of the observed resonance in $CeCoIn_5$ into question. Rather than being due to a spin exciton that implies a sign change of the gap function (d or s± symmetry), Song et al. describe the observed resonance as being due to a magnon-like excitation, with no implication about a sign change in $\Delta(k)$. This is still indicative of UcS (✓3), but now for a more strongly coupled version.

Another piece of evidence for UcS in $CeCoIn_5$, which is the result of many experiments leading up to what is now considered the final answer, is the presence of a second superconducting phase (✓6) – see Fig. 11 - as a function of field – just as already discussed in $U_{1-x}Th_xBe_{13}$ and $UPt_3$ above. This additional (so called Q-) phase (triplet p-wave) occurs just below $H_{c2}(0)$ and for T<0.3 K. This behavior was initially (H. Radovan et al. [233]; A. Bianchi, et al. [234]; H. A. Radovan et al. [235]) assigned to be a Fulde-Ferrell-Larkin-Ovchinnikov (FFLO) superconducting phase with a finite momentum caused by Zeeman (field) splitting of the bands. (As discussed also below in section 9, Organic Superconductivity, an FFLO phase

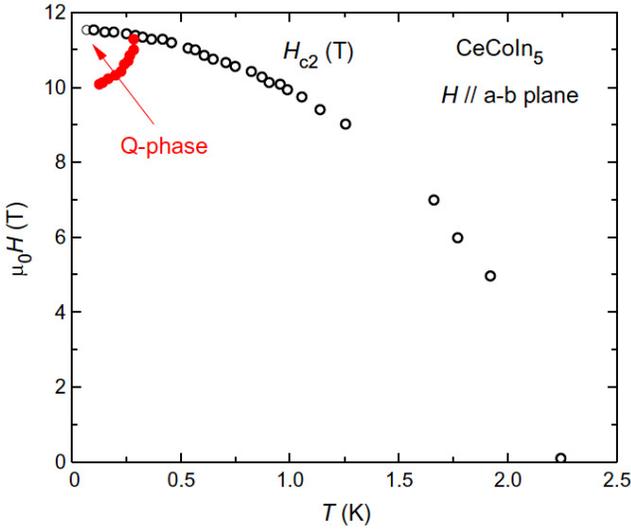

Fig. 11 (color online) The Q-phase exists in a narrow region of the H-T phase diagram (determined by specific heat (Radovan et al. [233]; Bianchi et al. [234]), T<~0.3 K, for CeCoIn$_5$, with the solid red circles representing second order phase transitions while the upper boundary to the Q-phase is first order. For field in the c-axis direction, the Q-phase exists in a much more restricted field region between 4.9 and 5 T. The Q-phase is extremely sensitive to doping, with less than 1% substitution of Cd, Hg, or Sn on the In site erasing any sign of a transition in the specific heat. (Y. Tokiwa et al. [236])

is by itself not proof of UcS, although it preferentially occurs in UcS.) Subsequent NMR results (B.-L. Young et al. [237]; Koutroulakis et al. [238]) revealed evidence for a local moment in this region of the phase diagram, rather than the long wavelength spin density modulation expected for the FFLO phase. Using sensitive high field neutron diffraction on a 155 mg single crystal of CeCoIn$_5$, Gerber et al. [239] concluded that modulated triplet *p*-wave superconductivity occurs in the Q-phase due to the interaction of the host's *d*-wave superconductivity with the spin-density-wave (SDW) order discovered by NMR.

A last piece of evidence for UcS in CeCoIn$_5$ is the response (J. P. Paglione et al. [240]) of its $T_c$ to substitution of various rare earths for Ce as a function of $\rho_0$. Qualitatively similar to the decrease of $T_c$ with increasing $\rho_0$ as discussed already for UPt$_3$ (see also section 2.1.16. and section 11.2. for URhGe below), although no fit to a modified Abrikosov-Gorkov model is made these data are also (✓~16) consistent with UcS.

To summarize the experimental measures that indicate UcS in CeCoIn$_5$, small characteristic temperature (✓1) superconductivity forms out of a non-Fermi liquid normal state (✓2), magnetic resonance below $T_c$ (✓3), some disputed evidence for a pseudogap (✓4), power law temperature dependences of various properties ($\kappa$, C/T, 1/T$_1$, and $\lambda$) in the superconducting state (✓5), presence of more than one superconducting phase (✓7), and $\kappa(\Theta, H)$ as well as $C(\Theta, H)$ show an angular dependence characteristic of nodal (in this case $d_{x^2-y^2}$) behavior (✓11) are all present. Concerning the metric for UcS of high sensitivity to impurities (2.1.9.), it is believed from band structure calculations that the M transition metal site in CeMIn$_5$ is relatively decoupled from the Ce site (Sarrao and Thompson [209]), thus decreasing the sensitivity of superconductivity to doping on the M site. However, doping of non-magnetic La on the Ce site in Ce$_{1-x}$La$_x$CoIn$_5$ does show a rapid, UcS-like depression of $T_c$ with x (50% suppression of $T_c$ for about x=0.1) (Petrovic et al. [241]).(✓9) Although this is less rapid than discussed above for CeCu$_2$Si$_2$, where $T_c$ falls by 50% for only approximately 6% doping by La on the Ce-site, the magnitude of this effect in CeCoIn$_5$ is still entirely consistent with UcS.

To summarize the theoretical and experimental insights into the UcS in CeCoIn$_5$, the large set of experimental results have not yet lent themselves to a comprehensive theoretical

understanding. Magnetic fluctuations are strongly believed to be important, see for example the magnetic interaction model proposed by Monthoux et al. [101]. In fact, using quasiparticle interference imaging, van Dyke et al. [242] present evidence that indeed the Cooper pairing in CeCoIn$_5$ is mediated by antiferromagnetic f-electron interactions. The normal ground state, with its non-Fermi liquid behavior in specific heat (C/T ~ -logT) – evidence for quantum criticality, is unusually robust against suppression by magnetic field with 5 T having apparently no effect. (Kim et al. [25]). The theoretical understanding of the field induced second superconducting Q-phase shown in Fig. 11 (modulated triplet p-wave symmetry due (Gerber et al. [239]) to the host's $d_{x2-y2}$ superconductivity interacting with the field induced spin density wave with a moment of 0.15 $\mu_B$ (Kenzelmann et al. [243]) was a tour de force.

4.2.7.2 *CeIrIn$_5$:* Due to its lower $T_c$ (0.40 K), CeIrIn$_5$ (Petrovic et al. [212]) is less thoroughly characterized than CeCoIn$_5$, e. g. there is no measure of a resonance in neutron scattering below the 0.4 K $T_c$. Power laws of superconducting properties indeed show non-exponential behavior, but suffer from a typical lowest-temperature-of-measurement of 0.05 K. For example, Movshovich et al. [218] report C/T $\propto T^2$ (indicative just as in CeCoIn$_5$ of line nodes in the gap) between 0.05 and 0.2 K, as well as $\kappa/T \propto a + bT$ (consistent with line nodes) between 0.05 and 0.2 K. The normal state specific heat C/T follows the non-Fermi liquid temperature dependence $\gamma_0-AT^{1/2}$ between 0.6 and 6 K (✓2) (Kim et al. [25]). C/T as a function of field and angle has been done in CeIrIn$_5$ under pressure to raise the superconducting temperature and to allow reaching sufficiently low temperature to obtain correct measurement of the nodal directions following the Vorontsov and Vekhter [77] discussion. For P=0.9, 1.5, and 2.05 GPa, $T_c$ for CeIrIn$_5$ is 0.50, 0.75 and 0.85 K respectively. (X. Lu et al. [244]). The result of these C(H, Θ) measurements were the same as in CeCoIn$_5$: 4 four fold rotation consistent with $d_{x2-y2}$ symmetry pairing with lines of nodes in the a-b plane (✓11).

No resistive or tunneling evidence for a pseudogap just above the bulk $T_c$ of 0.4 K exists, presumably because of the electrical short at 1.2 K due to a resistive transition, $\rho \rightarrow 0$, proposed by Bianchi et al. [245] to be due to filamentary superconductivity caused by crystalline defects.

By doping Hg on the In site, Bauer et al. [246] found long range antiferromagnetism in CeIr(In$_{1-x}$Hg$_x$)$_5$ for x>0.05. Shang et al. [247] doped both Hg and Sn on the In site as well as Pt on the Ir site. They find a maximum in the specific heat γ (proportional to the effective mass, m*) at about CeIr(In$_{0.97}$Hg$_{0.03}$)$_5$ and suggest that pure CeIrIn$_5$ is therefore "in proximity" to an antiferromagnetic quantum critical point. They further postulate that the superconductivity in CeIrIn$_5$ is mediated by magnetic quantum fluctuations.

4.2.7.3 *CeRhIn$_5$:* Hegger et al. [248] demonstrated superconductivity at 2.1 K in CeRhIn$_5$ via resistive measurements under an applied pressure of 1.63 GPa. At zero pressure, CeRhIn$_5$ is an antiferromagnet with $T_N$=3.8 K. There is coexistent antiferromagnetism (AFM) (local moment of 0.37 $\mu_B$) and superconductivity below 1.8 GPa demonstrated by ac specific heat and NMR, with the peak in $T_c$ vs pressure approximately at where pressure suppress $T_N \rightarrow 0$, i. e. at a quantum critical point (✓2) (T. Park et al., [249]). As shown below in Fig. 12, $T_N \rightarrow 0$ (causing quantum critical behavior and its possible influence on UcS – section 2.1.2) can also be reached by substituting either Co or Ir for Rh. In the case of CeRh$_{1-x}$Ir$_x$In$_5$, the specific heat above $T_c$ shows [250] non-Fermi liquid behavior (C/T rising as T is lowered), consistent with a magnetic quantum critical point at $T_N \rightarrow 0$ causing UcS. In contrast, the specific heat [251] above $T_c$ in CeRh$_{1-x}$Co$_x$In$_5$ where $T_N \rightarrow 0$ (around x=0.6-0.7, see Fig. 12) does not show the non-Fermi

liquid behavior characteristic of pure CeCoIn$_5$. (This specific heat evidence of lack of quantum criticality near where T$_N$→0 in CeRh$_{1-x}$Co$_x$In$_5$ is consistent with dHvA measurements, which show [252] no divergence in the effective mass m* in this region of the phase diagram.) In the coexistence region, neutron scattering (Aso et al. [253]) implies direct coupling of the antiferromagnetic and superconducting order parameters. Specific heat under pressure (Fisher, et al. [213]) in the superconducting state give the non-exponential power law C ~ T$^2$ from 0.5 K (lowest temperature of measurement) to 2 K, too limited a temperature range to strongly infer UcS. Also in the superconducting state, NQR measurements (T. Mito et al. [254]) (as for all the heavy fermions) find 1/T$_1$ ~ T$^3$, although only between 0.35 and 2 K. In another NQR under pressure work (S Kawasaki et al. [255]), a claim is made that a small anomaly in the 1/T$_1$ data in 1.6 GPa at 4 K (T$_c$ at that pressure is 1 K, with T$_N$=2.7 K) corresponds to the opening of a pseudogap. The possibility of a pseudogap in CeRhIn$_5$ under pressure has not been confirmed by any other measurement. Angle resolved specific heat in a field, C(H, Θ), reveals 4 fold oscillations (✓11) implying d-wave pairing symmetry in both the coexistent AFM and superconductivity regime (P<1.8 GPa) *and* in the regime P>2.3 GPa, where superconductivity exists *after* the AFM is suppressed (T. Park et al. [256]).

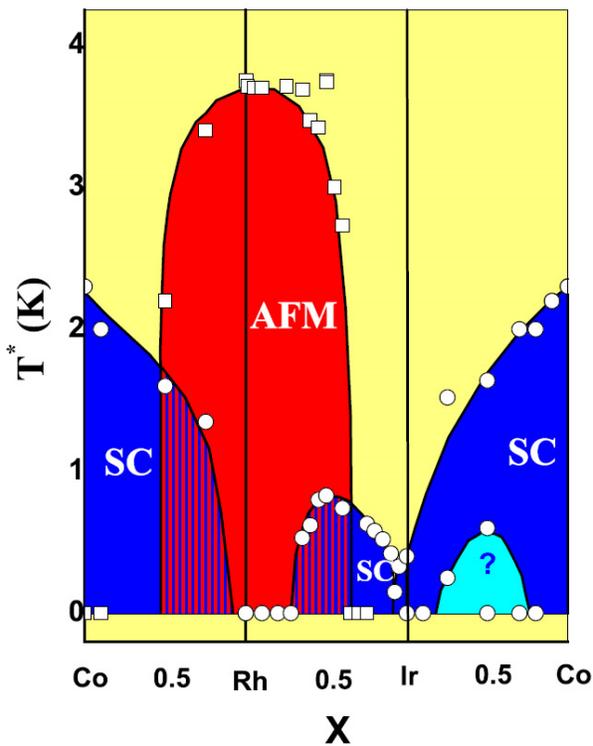

Fig. 12 (color online) Phase diagram [209] of CeMIn$_5$, where M is varied between Co, Rh, and Ir as well as between Ir and Co. The two regions highlighted with narrowly spaced parallel vertical lines exhibit true microscopic coexistent antiferromagnetism and superconductivity. As of the time of the Sarrao and Thompson [209] review in 2007, microscopic measurments to determine the nature of the light blue phase region in CeIr$_{1-x}$Co$_x$In$_5$, in which a second superconducting transition had previously been observed (Pagliuso et al. [257]) had not been completed. Figure reproduced with permission from J. L. Sarrao and J. D. Thompson, J. Phys. Soc. Japan 76 (2007), p. 051013 [209].

In Fig. 12 is displayed the tuning of the electronic properties in CeMIn$_5$ by doping rather than via pressure. Even though all three of the CeMIn$_5$, M=Co, Ir, Rh (all isovalent), are UcS, Fig. 12 shows clearly that superconductivity is relatively insensitive to doping on the M site, even when the doping is between the non-superconducting endpoint CeRhIn$_5$ and either CeCoIn$_5$ or CeIrIn$_5$. This aspect, that the M site appears to be rather decoupled from the Ce site in terms of the effect of impurities on T$_c$, was already discussed above for CeCoIn$_5$ but Fig. 12 reinforces this point.

Thus, in summary, all three CeMIn$_5$, M=Co, Ir, Rh, appear to exhibit UcS, with the most complete case being made for M=Co, where the results are particularly rich in scope.

### 4.3. "Heavy" Fermion Superconductors, $\gamma < 100$ mJ/molK$^2$

Below are listed a few of the less strongly correlated compounds which are often lumped under the "heavy fermion" label. This is meant to be a representative, not exhaustive, list. The ferromagnetic compounds UGe$_2$ and URhGe are discussed in section 11. (ferromagnetic UcS).

#### 4.3.1 *PuCoGa$_5$:*

Also occurring in the tetragonal 115 structure as do the CeMIn$_5$ just discussed, PuCoGa$_5$ has T$_c$=18.5 K and was, surprisingly enough, the *first* Pu compound found to superconduct at any temperature (Sarrao et al. [108]). (PuRhGa$_5$, T$_c$=8.5 K, was the obvious follow up, with Rh isoelectronic to Co, and was characterized shortly thereafter.) As discussed in the Introduction to section 2, PuCoGa$_5$ is on the border of being a heavy fermion system, with a $\gamma$ value of approximately 70/95 mJ/molK$^2$ (Sarrao et al. [108]/Bauer et al. [258]). Also, of course, the T$_c$ of 18.5 K makes PuCoGa$_5$ clearly different from heavy fermion systems, where the highest T$_c$ is 2.3 K for CeCoIn$_5$. However, as a possible UcS it clearly belongs together with the discussion of the three 115's just discussed, where the $\gamma$ values (defined as C$_{normal}$/T as T$\to$0) indeed indicate large effective mass, 'heavy' electrons for CeMIn$_5$, M=Co, Ir, Rh, with $\gamma$ = 1150, 625, and 380 mJ/molK$^2$ respectively. An additional argument for considering the Pu 115 superconductors together with the CeMIn$_5$ materials is that Bauer et al. [258] pointed out (see Fig. 13) that T$_c$ in both the PuMGa$_5$ and the CeMIn$_5$ compounds has essentially the same variation with the c/a tetragonal axes ratio. This commonality of T$_c$ vs c/a, combined with the calculated/measured-by-dHvA approximately two dimensional Fermi surfaces for both sets of superconductors, led Bauer et al. to propose a common mechanism of magnetically mediated UcS that is strongly linked to the structural c/a ratio. A theoretical understanding of the origin of the T$_c$ scaling with c/a was declared (Sarrao and Thompson [209]) an "important open question" and seems of fundamental interest, thus joining the 'to-be-understood'/puzzle list of this review.

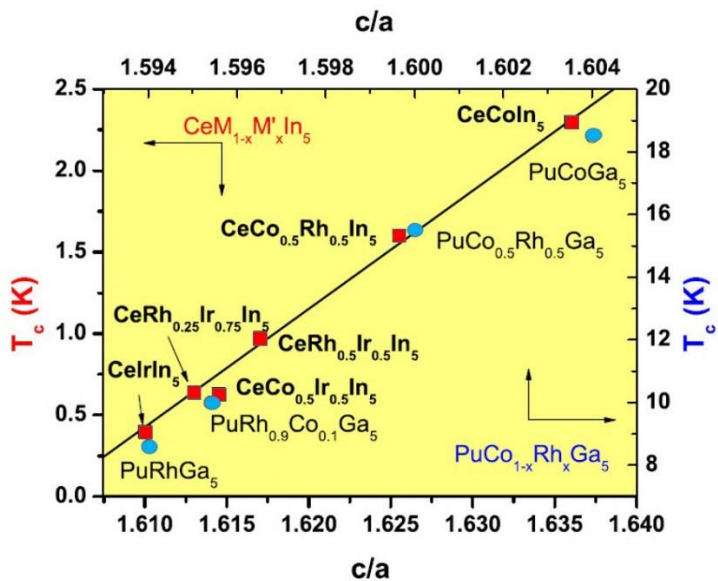

Fig. 13 (color online) T$_c$ for the CeMIn$_5$ compounds has the same linear behavior vs the ratio of the c-axis to the a-axis, along with the same slope, as do the plutonium based PuMGa$_5$ superconductors. (Figure based on Bauer et al. [258] and Sarrao and Thompson [209].) Reprinted figure with permission from Bauer et al., Phys. Rev. Lett. 93 (2004), p. 147005. [258] Copyright (2004) by the American Physical Society.

The low temperature non-exponential temperature dependences of superconducting $PuCoGa_5$ are as follows. For specific heat on $PuCoGa_5$ (using the rare $^{242}Pu$ isotope to decrease the self heating due to radioactive decay by a factor of 15 compared to $^{239}Pu$) down to 1 K, $C \propto T^2$ up to 7 K (Bauer et al. [258]). Measurement of the penetration depth, $\lambda$, between 3 and 12 K results in $\lambda \propto T$, consistent with d-wave pairing (Morris et al. [259]). As is common for heavy fermion systems, $1/T_1 \propto T^3$, but only down to ~ $T_c/2$, below which $1/T_1 \propto T$ (Curro et al. [260]) which of course as seen in Table 1 does not fit theoretical predictions for either line or point nodes. Thus, although these non-exponential temperature dependences are consistent with UcS, they are – due to the internal heat generated in the sample by the rapidly decaying Pu nuclei – known only over limited temperature ranges. The NMR normal state $1/T_1$ data [260], $T>T_c$, were analyzed as consistent with antiferromagnetic spin fluctuations.

The normal state temperature dependence of the resistivity in $PuCoGa_5$ is $\rho \propto T^{1.35}$ from $T_c$ to 50 K, consistent with non-Fermi liquid behavior although there is no sign of a magnetic phase transition or a quantum critical point to cause this behavior. Perhaps more interesting, the normal state magnetic susceptibility published in the discovery paper (Sarrao et al. [108]) follows a Curie-Weiss temperature dependence, $\chi = \chi_0 + C/(T - \Theta)$ with a fairly large effective moment of 0.68 $\mu_B$, all the way from room temperature downwards until it is interrupted by the diamagnetic superconducting transition at 18.5 K. This is unlike any known heavy fermion system (superconducting, magnetic, or non-ordering) (Stewart [106]), for example $\chi$ of $UBe_{13}$ deviates from its high temperature Curie-Weiss behavior already below 100 K. This evidence that the superconductivity forms directly out of a normal state with a 0.68 $\mu_B$ effective moment is in and of itself evidence for UcS, as discussed in a theory by Flint et al. [109]. Polarized neutron scattering data (Hiess et al. [261]) – albeit on a sample of $PuCoGa_5$ with a somewhat depressed $T_c$ (15 vs 18.5 K) due to Sb impurities from the preparation - finds that the magnetic behavior in $PuCoGa_5$ is dominated by orbital contributions at q=0 and is inconsistent with the localized Pu 5f electron picture suggested by the Curie Weiss behavior reported by Sarrao et al. [108].

Additionally, the measured $\chi$ just above $T_c$ in the Hiess et al. work is ~3-4 times smaller than is the Sarrao et al. sample. Further unpublished work by the Los Alamos group (J. D. Thompson [262]) showed an array of results in six different samples of $PuCoGa_5$ for $\chi$ - all still able to be fit to $\chi = \chi_0 + C/(T - \Theta)$ (although with varying $\chi_0$ and $\Theta$ values and – with one exception - only able to be fit down to 25-125 K depending on sample) – but with magnitudes indeed varying from 3.15 memu/mol just above $T_c$ (the Sarrao et al. discovery result), to 2.0 memu/mol in a $^{242}Pu$ sample (Curie-Weiss behavior down to $T_c$), and all the way down to ~0.9 memu/mol just above $T_c$ like in the Hiess et al. sample. This range of results for $\chi$ remains unexplained, and may at least partly be due to the large background subtraction of the encapsulation required. Thus, this important point of the intrinsic behavior of $\chi$ down to $T_c$ in $PuCoGa_5$ – the subject of a concerted (but inconclusive) examination at Los Alamos National Laboratory - is likely to remain an unsolved puzzle.

### 4.3.2 *CePd$_2$Si$_2$:*

The study of $CePd_2Si_2$ represents a significant subfield of UcS, that where antiferromagnetism is suppressed by applied pressure and a dome of superconductivity appears in the phase diagram,

see Fig. 2 for an overview. Specifically in CePd$_2$Si$_2$, the peak of the dome is at T$_c$=0.43 K at 2.8 GPa. Thus, CePd$_2$Si$_2$ fulfills condition ✓2 discussed above in section 2 where quantum critical fluctuations are thought to be important for the superconducting pairing. CePd$_2$Si$_2$ is an antiferromagnet with T$_N$=10 K, and the specific heat above T$_N$ in zero pressure extrapolates to give a $\gamma$=65 mJ/molK$^2$. At p$_c$ = 2.8 GPa, the normal state resistivity follows $\rho \propto T^{1.2}$ up to ~ 40 K (Grosche et al. [263]), consistent with quantum critical spin fluctuations. Mathur et al. [26] argue (also for the similar (see also Pfleiderer [70]) CeIn$_3$, where T$_N$=10.2 K, $\gamma$=140 mJ/molK$^2$, at p$_c$=2.5 GPa, T$_c$=0.19 K, and $\rho \propto T^{1.6}$) for superconducting pairing mediated by magnetic spin-spin interactions. Due to the high pressures, very low temperatures, and small volume of sample involved, the set of measurements on these compounds is quite restricted. They serve as important, but understudied in comparison even to PuCoGa$_5$, examples of UcS.

### 4.3.3 URu$_2$Si$_2$:

URu$_2$Si$_2$ was discovered to be a superconductor, T$_c$=1.5 K, by W. Schlabitz et al. [264] (see also Palstra et al. [265] and Maple et al. [266]. The specific heat $\gamma$ value extrapolated to T=0 from the data in the normal state above T$_c$ is 65-75 mJ/molK$^2$ ([264], [266].) At 17.5 K there is an anomaly in the specific heat which, in the initial discovery papers, was assigned to be due to antiferromagnetism. The entropy associated with this transition was found by Schlabitz et al. to be about 0.15 Rln2, certainly a large enough amount for neutron scattering to find the putative local moment. Thirty years later, it is fair to say that the majority of work on URu$_2$Si$_2$ has been focused on this 17.5 K anomaly, rather than on the superconductivity (the subject of the present review.) The reason for the effort expended is that the measured order U moment is only 0.04 $\mu_B$ (neutron scattering Broholm et al. [267] or 0.02 $\mu_B$ (magnetic xray scattering, Isaacs et al. [268]), i. e. the anomaly in the specific heat is not predominantly due to magnetic ordering. What exactly this order (dubbed 'hidden order' as its origin remains hidden from our understanding) is due to remains the subject of vigorous investigation, both experimental and theoretical, and debate (for a review, see Mydosh and Oppeneer [269]).

Concerning the superconducting properties of URu$_2$Si$_2$, the temperature dependences in the superconducting state that imply line nodes (C $\propto$ T$^2$ Fisher et al. [270]; 1/T$_1$~T$^3$ Kohori et al. [271]) are less than conclusive because of limited temperature dependence (C/T, 0.2-1.0 K) and the possibility of a point node 1/T$_1$ $\propto$ T$^5$ (Table 1) behavior being masked (Heffner and Norman [48]). However, $\kappa(\Theta, H)$ (Kasahara et al. [272]) and C($\Theta$, H) (Yano, et al. [273]) measurements give definite evidence (✓11) for horizontal line nodes in the hole band and point nodes in the electron band – an interesting difference in nodal topology in one material. These two works build on the earlier $\rho$ and $\kappa$ work of Kasahara et al. [274], who argued for unconventional d-wave superconductivity with two (a light hole band and a heavy electron band) superconducting energy gaps. Schemm et al. [66] have recently presented evidence for breaking of time reversal symmetry (✓8) in URu$_2$Si$_2$.

Point-contact spectroscopy measurements on a URu$_2$Si$_2$ single crystal with the tip applied along the a-axis direction (on the 0.5 mm thin edge of the crystal) yielded evidence for a pseudogap (✓4) that extended above Tc = 1.37 K by 0.6 K (Morales and Escudero [275]). Since the point contact is expected to apply significant pressure (estimated to be ~1 GPa in another experiment (Rodrigo et al. [276]) on URu$_2$Si$_2$ to investigate the 17.5 K transition), the alignment of the tip on the single crystal is of critical importance, since pressure increases T$_c$ (~0.1 K for 0.2 GPa) when applied in the c-axis direction, and reduces T$_c$ when applied in the a-axis

direction (Bakker et al. [277]). Confirmation of these important results – only the second indication of a pseudogap in a heavy fermion superconductor, after CeCoIn$_5$ - by another measurement technique would be desirable.

4.3.4 *CeIn$_3$:*

CeIn$_3$ is an antiferromagnet, $T_N$=10.2 K, $\gamma$=130 mJ/molK$^2$, which superconducts (the superconducting dome, see Fig. 2, is peaked at $T_c$=0.22 K) under 2.5 GPa pressure at the point in the phase diagram where pressure suppresses $T_N$ to T=0 (Walker et al. [278]; Knebel et al. [279]). There is evidence [279] in the resistivity, $\rho = \rho_0 + aT^{\sim 1.6}$, for antiferromagnetic quantum critical fluctuations (✓2) near the critical 2.5 GPa pressure. The other indication for UcS is a sensitivity [70] to non-magnetic impurities (✓9). For a theory of the superconductivity in CeIn$_3$, see Fukazawa and Yamada [280].

4.3.5 *PrPt$_4$Ge$_{12}$*

This skutterudite is included here for completeness. With its high $T_c$ (7.9 K), it is an unlikely candidate for triplet superconductivity, where $T_c$ should be highly sensitive to defects. With a $\gamma$ value of 87 mJ/molK$^2$ (Gumeniuk et al. [281]), it is further quite unlike PrOs$_4$Sb$_{12}$, $T_c$=1.8 K, with a $\gamma \geq 350$ mJ/molK$^2$. However, according to muon spin resonance measurements (Maisuradze et al. [282]), PrPt$_4$Ge$_{12}$ breaks TRS below its $T_c$. Maisuradze et al. speculate that the nature of the pairing (rather than triplet) could be a complex $\ell$=0 (spin singlet) function with an internal phase, which would [282] also break TRS. In terms of power laws, Maisuradze et al. [283] find C ~ T$^3$ (consistent with point nodes, see Table 1) over the limited temperature range of 0.4 to 1.6 K and fit their penetration depth data also to a model with point nodes.

4.4. *Summary:*

Having discussed a number of heavy fermion superconductors as pertains to the question of whether their superconductivity is unconventional, we have seen a wide variety of properties and proposed explanations. As clear from the discussion, this question is sufficiently complex that the answer is not always straightforward. For example, the prototype heavy fermion superconductor, CeCu$_2$Si$_2$, with its several properties consistent with UcS (including strongly enhanced effective electron mass, strong antiferromagnetic fluctuations, sensitivity to non-magnetic defects, nearness in the phase diagram to a magnetic phase transition) has recently had its specific heat down to 0.04 K analyzed as a fully gapped superconductor (Kittaka et al. [50]). Such a two gap model involves a relatively large number (four) of fitting parameters, and awaits confirmation of the two gap nature from other measurements – such as ARPES, as was important in substantiating the two gap nature in MgB$_2$ and the IBS. Further understanding of the ~0.5 K anomaly below $T_c$ in C/T in pure UBe$_{13}$ and its possible connection to the second superconducting transition in U$_{1-x}$Th$_x$Be$_{13}$ might aid in understanding the precise nature of the UcS lower transition and its pairing symmetry, as well as its cause, in U$_{1-x}$Th$_x$Be$_{13}$.

As we have discussed, the proposed mechanisms for superconductivity in heavy fermion systems, although spin-fluctuation mediation is a strong favorite, include other possibilities: antiferro-quadrupolar (PrOs$_4$Sb$_{12}$) and magnetic exciton (UPd$_2$Al$_3$). Of course, some heavy

fermion systems are inarguably UcS, e. g. UPt$_3$, U$_{1-x}$Th$_x$Be$_{13}$, and CeCoIn$_5$ all have more than one superconducting phase; CeCoIn$_5$ has a narrow spin resonance below T$_c$ in the neutron scattering, indicative of a sign change in the order parameter as well as tunneling evidence for d$_{x2-y2}$ pairing symmetry; U$_{1-x}$Th$_x$Be$_{13}$ and UPt$_3$ both – in the best quality samples – have large (sample dependent) residual specific heat γ values in the superconducting state (30% and from 0 up to 45% respectively of γ$_{normal}$).

As will be seen when the other superconducting classes are discussed in the following sections, the heavy fermion superconductors, diverse as their properties are (see Table 2 for a summary), do not present the full breadth of complexity to be found in UcS.

Table 2: Presence of Strongly Suggestive Evidence Indicating UcS in Heavy Fermion Superconductors. The symbol '✘' means that that property was specifically looked for but not found

|  | 1 | 2 | 3 | 4 | 5 | 6 | 7 | 8 | 9 | 11 | 14 | 16 |
|---|---|---|---|---|---|---|---|---|---|---|---|---|
|  | T$_F$ << Θ$_D$ | QC | Spin res | PG | T$^α$ | >1 sc phase | Phase sens. tunneling ⇒ non-BCS | TRSB | T$_c$/non-mag imp | C(H,Θ) κ(H,Θ) | γ$_{residual}$ large | T$_c$ vs ρ$_0$ fits AG model |
| CeCu$_2$Si$_2$ | ✓ | ✓ | broad |  | ~✓ |  |  |  | ✓ | ✘ |  |  |
| UBe$_{13}$ | ✓ | ✓ |  | ✓ |  |  |  |  | ✓ |  |  |  |
| U$_{1-x}$Th$_x$Be$_{13}$ | ✓ | ✓ |  |  | ✓ |  |  |  |  |  | ✓ |  |
| UPt$_3$ | ✓ |  | ✘ |  | ✓ |  |  | ✓ |  |  | ✓ | ✓ |
| U(Ni/Pd)$_2$Al$_3$ | ✓ |  | ✓Pd |  |  |  | ✓ for Pd |  |  | ✓ for Pd |  |  |
| PrOs$_4$Sb$_{12}$ | ✓ |  |  | mix |  |  | ✘ | ✓ |  |  |  |  |
| CeCo/Ir/RhIn$_5$ | ✓ | ✓Co,Ir | ✓Co | ?Co |  | ✓Co |  |  | ✓ for Co | ✓Co,Rh |  | ~✓ Co |
| PuCoGa$_5$ |  |  |  |  |  |  |  |  |  |  |  |  |
| CePd$_2$Si$_2$ |  | ✓ |  |  |  |  |  |  |  |  |  |  |
| URu$_2$Si$_2$ |  |  |  | ✓ |  |  |  | ✓ |  | ✓ |  |  |

## 5. Cuprates (Hole and Electron Doped)

Superconductivity was discovered in Ba (hole)-doped $La_2CuO_4$, $T_c \sim 35$ K (tetragonal structure with planes of $CuO_2$), in 1986 by Bednorz and Mueller. Measurements of $T_c$ vs pressure (Chu et al. [284]) showed an increase in $T_c$ of 25% at 1.3 GPa, leading to doping the compound with the smaller Y ion on the La site to exert "chemical" pressure. This led to superconductivity at 93 K (Wu et al. [285]) in what was later determined to be $YBa_2Cu_3O_{7-\delta}$ (a different composition than the original Bednorz and Mueller compound and with an orthorhombic structure with chains of Cu-O as well as planes of $CuO_2$). Based on these discoveries in two different copper perovskite structures, the race for higher $T_c$, and for applications, was on.

Although the search for higher $T_c$ in the 'cuprate' superconductors has reached a $T_c$ as high as 134 K (for $HgBa_2Ca_2Cu_3O_{8+\delta}$), which increases to 164 K at 31 GPa (Gao et al. [286]), the search for understanding many fundamental issues, including even the pairing mechanism ("The Secret", Anderson [102], see section 2 above), remains unfinished. Both experimental (see, e. g., the recent laser ARPES measurements (J. M. Bok et al. [287]) that suggest quantum critical fluctuations as the "glue") and theoretical work continues apace. As Abrahams [288] observes, it is difficult "to give a comprehensive survey" of the theory of the cuprates. A partial list of theoretical reviews includes Orenstein and Millis [289], Norman and Pepin [290], Sachdev [291], Lee, Nagaosa and Wen [292], Anderson [293], Norman [294], Ogata and Fukuyama [295], Abrahams [288], Norman [103], Scalapino [14], and Das, Markiewicz and Bansil [296].

There are several collections of experimental reviews of the cuprates, e. g. the five volumes of The Physical Properties of High Temperature Superconductors [#1-5 references] ed. by D. M. Ginsberg, the two volumes (30 and 31) of Handbook on the Physics and Chemistry of the Rare Earths (ed. by K. A. Gschneidner, Jr., L. Eyring, and M. B. Maple ([#1,2 references] devoted to High Temperature Superconductors, here are also specialized experimental reviews, or even multiple reviews, on just one measurement technique or property in the cuprates (e. g. <u>SQUID devices:</u> Koelle et al. [297], <u>phase sensitive detection of the pairing symmetry:</u> Van Harlingen [298], <u>Andreev reflection spectroscopy:</u> Deutscher [299], <u>scanning tunneling microscopy (STM) and spectroscopy (STS):</u> Fischer et al. [300], <u>photoemission measurements</u>: Damascelli et al. [301], <u>pseudogap behavior</u>: Timusk and Statt [40] *and* Rice et al. [39] *and* Norman, Pines and Kallin [302], and <u>isotope effect</u>: Zhao et al. [303] *and* Franck [304].

Thus, although not the first discovered UcS (which were heavy fermions) nor even the second (organic superconductors), the amount of theoretical and experimental work on the cuprate superconductors probably exceeds that on all the other classes of UcS combined. Distilling this body of theoretical and experimental work down to just one link in the chain of understanding UcS in general will necessitate brevity on subjects worthy of the entire length of this review.

An overview of a representative cuprate phase diagram (both the more heavily studied hole doped side and the less studied electron doped side - represented by $R_{2-x}Ce_xCuO_{4-\delta}$, R=Pr, Nd, Sm, discovered by Tokura, Takagi and Uchida [305] - is given in Fig. 14. Like sometimes found in the heavy fermion superconductors already discussed and in some organic, IBS, non-centrosymmetric, and perhaps the cobalt oxide hydrate superconductors discussed below, UcS in

the cuprates forms near to (sometimes coexistent with) antiferromagnetism in the phase diagram. In addition to the ordered antiferromagnetic phase, in the cuprates there are a variety of incommensurate spin-density-wave striped phases (see, e. g., Lake et al. [306]) that compete and interact with the superconducting phase. The nearness of magnetism in the phase diagram leads to proposals of spin excitation mediated pairing just like in the heavy fermions (see, e. g., Chubukov, Pines and Schmalian [307] and Scalapino [14]). However, the differences between the cuprate and the heavy fermion superconductors, as will be discussed, are numerous.

We begin with a brief outline of the electron doped $R_{2-x}Ce_xCuO_{4-\delta}$, then discuss their superconducting properties, followed by similar treatment for the more-completely-studied hole doped cuprates.

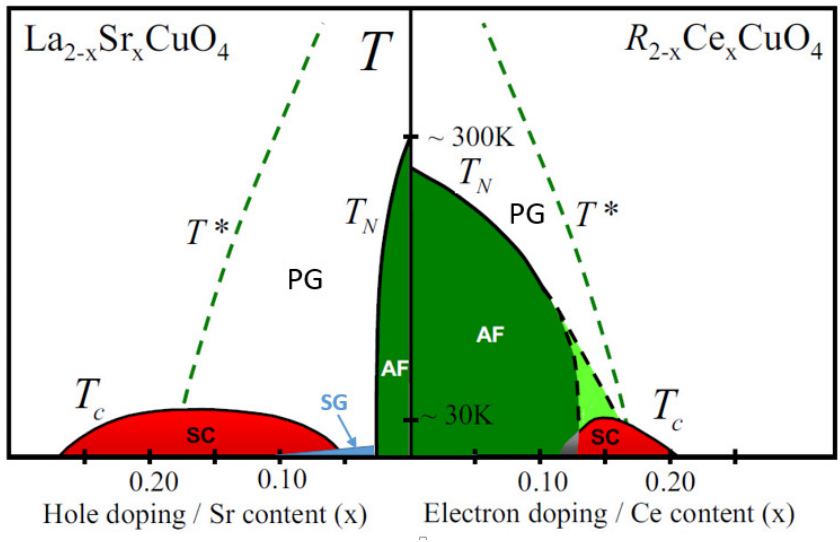

Fig. 14 Phase diagram of the high temperature superconducting cuprates with the hole doped cuprates represented by $La_{2-x}Sr_xCuO_4$, and the electron doped cuprates represented by $R_{2-x}Ce_xCuO_4$, after Armitage, Fournier and Greene [41]. Both sides show rapid suppression of $T_N$ with doping. On the hole doped side, there is a region of spin glass behavior connecting the antiferromagnetic regime and the superconducting dome, and actually extending into the dome (Julien [308]). Below T*, pseudogap behavior is observed as discussed below in the text, although the origin of the behavior on the two sides of the phase diagram appears to be different [41], [309]. As will be discussed, this difference is likely tied to the greater extent of the antiferromagnetic region with doping as well as its coexistence with the superconducting dome on the electron doped side of the phase diagram. The light green region represents uncertainty in the extent of the antiferromagnetic region. At zero doping, both $La_2CuO_4$ and $R_2CuO_4$, R=Pr, Nd, Sm, are antiferromagnetic Mott insulators. Despite the specific compounds listed in the figure, this phase diagram roughly fits the 200+ known cuprate superconductors, hole and electron doped. Reprinted figure with permission from Armitage et al., Rev. Mod. Phys. 82 (2010), p. 2421. [41] Copyright (2010) by the American Physical Society.

5.1. *Electron Doped:*

Two recent reviews exist for the electron doped cuprates, Armitage, Fournier, and Greene [41] and Fournier [42]. The focus of much of the former is on normal state properties. There are more than just the $R_{2-x}Ce_xCuO_4$ (R=Pr, Nd, Sm, Eu) electron doped compounds, there are also the $Sr_{1-x}Ln_xCuO_2$, Ln=La, Nd, Sm, Eu, as well as $R_{2-x}Th_xCuO_4$, R=Pr [310], Nd [311]. The $Sr_{1-x}Ln_xCuO_2$ are the so called 'infinite layer' compounds, which consists of $CuO_2$ planes alternating

with alkaline earth planes, the limit with large n of the perovskite structure stacking, see Siegrist et al. [312]. This is similar, as discussed above, to $CeIn_3$ being the infinite layer, n→∞, compound for $Ce_nM_mIn_{3n+2m}$. Some examples are $Sr_{0.84}Nd_{0.16}CuO_2$, $T_c$=40 K (Smith et al. [313]) and $Sr_{0.9}La_{0.1}CuO_2$, $T_c$=42 K (Kikkawa et al. [314]) vs $T_c$=24 (22, 20, 13) K for $R_{2-x}Ce_xCuO_4$, R=Nd (Pr, Sm, Eu) and $T_c$=20 and 23 K for $R_{1.85}Th_{0.15}CuO_4$, R=Nd, Pr respectively. For a fourth kind of electron doped cuprate, there is also F doping to add electrons to the undoped parent compound, $Nd_2CuO_{4-y}F_y$, $T_c$=27 K (James et al. [315]).

The focus of the current review is on the superconducting properties which, unavoidably, are influenced by the properties of the normal state out of which the condensation into the superconducting state occurs. Thus, we offer here a brief synopsis of the complex normal state on the right side of the phase diagram in Fig. 14 to serve as a background. In the above discussion for the heavy fermion superconductors, individual compounds could be extremely different from one another and were discussed individually. In contrast, the electron doped cuprates ($R_{2-x}Ce_xCuO_4$, R=Nd (Pr, Sm, Eu) and the infinite layer compounds, e. g. $Sr_{0.84}Nd_{0.16}CuO_2$ and $Sr_{0.9}La_{0.1}CuO_2$) lend themselves for the most part to broad generalities. For a more in-depth discussion thereof, see Armitage, Fournier and Greene [41].

Although not of particular import for the superconducting properties, the antiferromagnetism for the electron doped $R_{2-x}Ce_xCuO_4$, R=Nd (Pr, Sm, Eu), is not just the simple ordering of Cu spins (for R=Nd, $T_N$=255 K) in the $CuO_2$ layers like on the hole-doped side, despite the similar ordering temperatures. The rare earth R atoms give rise to a "rich set" [41] of magnetic properties due to coupling between the Cu and R atom spins, including, e. g., further ordering transitions of the Cu spins. These further transitions ($T_{N2}$=75 K, $T_{N3}$=30 K for R=Nd (Lynn and Skanthakumar [316]) are followed by ordering of the R atoms at low temperatures. For example, for R=Nd, there is a further antiferromagnetic transition at 1.7 K (Ghamaty et al. [317]) with an ordered moment on the Nd of 1.3 $\mu_B$ (Matsuda et al. [318]).

The normal state resistivity for 'optimally doped' (meaning at the peak of the superconducting dome) electron doped cuprates follows a $T^2$ (Fermi liquid) temperature dependence from $T_c$ up to 200 K (e. g. in $Nd_{1.85}Ce_{0.15}CuO_4$ ('NCCO') Tsuei, Gupta and Koren [319]) in contrast to the non-Fermi liquid $\rho=\rho_0+AT$ dependence reported for some of the heavy fermion superconductors and for the hole doped cuprates.

However, the electron doped cuprates display a broad arrange of behavior just as do the hole doped cuprates, and several samples display linear-with-temperature resistivities, thought to be consistent with spin fluctuations and quantum criticality (Jin et al. [320]). Resistivity in thin film $La_{2-x}Ce_xCuO_4$ shows a band of $\rho=\rho_0+AT$ (✓2) vs composition behavior above the superconducting dome, with the maximum in temperature extent (up to 50 K) at optimal doping x=0.10 [320]. In thin films of $Pr_{2-x}Ce_xCuO_4$ at the particular concentration of x=0.17 (just past optimal doping), $\rho=\rho_0+AT$ is found between 0.04 and 30 K, in this case in applied fields up to 12 T to suppress $T_c$ (~ 19 K at zero field).

As was mentioned with Fig. 14, pseudogap behavior appears to be different in the electron-doped cuprates, with the majority of the pseudogap behavior seemingly related to the antiferromagnetic spin correlations. (Armitage, Fournier and Greene [41]; Motoyama et al. [321]. In contrast, in the hole-doped cuprates, T* (as will be discussed below for these materials) possibly signals a phase transition that extrapolates to a quantum critical point near a hole doping p=0.19.

Turning now to investigation of the superconducting state properties which can provide evidence for UcS (Unconventional Superconductivity), we begin with tunneling into the electron doped cuprates (which has been quite revealing as discussed below for the hole doped cuprates).

5.1.1. *Tunneling*

This work has had a number of materials issues, as well as puzzling results. In addition, early work (e. g. point contact spectroscopy, Huang et al. [322]) was interpreted as consistent with predominantly electron-phonon mediated coupling, which in the more recent work is no longer the case.
**Unlike in the hole doped materials discussed below**, point contact tunneling in zero field shows no sign of a gap above $T_c$ (the pseudogap), although other measurement techniques (e. g. optical conductivity, ARPES, photoemission) do as depicted by the dashed T* line in the right side of Fig. 14 above, see the review by Armitage, Fournier and Greene [41]. One puzzling tunneling result still under recent discussion (J. Yuan et al. [323]) is the existence (Biswas et al. [324] and Kleefisch et al. [325]) – known since 2001 - of a so-called 'normal state energy gap' (NSG) of ~5 meV in $M_{2-x}Ce_xCuO_4$, M=Pr,Nd, at 2 K in magnetic field > $H_{c2}$. Work in 2003 by L. Alff et al. [326] added LCCO to the list of electron doped cuprates that show an NSG. This NSG is distinct from the large (~100 meV) 'pseudogap,' which is observed via ARPES (discussed next below.) The explanations under discussion for the NSG include preformed superconducting singlet pairs (Y. Dagan et al. [327]), hidden order parameter (viz. discussion of $URu_2Si_2$ above) under the superconducting dome [326], and a Coulomb gap caused [324] by electron-electron interactions. See ref. [41] for a complete discussion. Although not a focus of recent work in understanding the superconductivity in electron doped cuprates, this NSG joins our list of interesting and unresolved puzzles.

In zero field and below $T_c$, point contact spectroscopy on NCCO, $T_c$=25.1 K, gives (Shan et al. [310]) a superconducting energy gap of ~4 meV. (The early work of Huang et al. [322] on NCCO, $T_c$=22 K, gave $\Delta$=3.7 meV.) STM measurements (Niestemski et al. [329]) on $Pr_{0.88}LaCe_{0.12}CuO_4$ (PLCCO), $T_c$=24 K, resulted in a superconducting energy gap of 7.2±1.2 meV. Obviously, the ratios for $2\Delta/k_BT_c$ obtained from these two measurement methods (3.7 and 7.0 respectively, vs 3.52 from BCS theory) differ significantly. Such variation in $\Delta$ is also seen in other measurements, as will be discussed below.

More interesting than this (long standing in the hole doped cuprates) disagreement on $\Delta$, is Niestemski et al.'s analysis of their STM data on PLCCO. They conclude that their tunneling spectra can be related to an electron-bosonic (where the bosons are spin fluctuations) mode coupling at energies of 10.5±2.5 meV. This energy is consistent with a (inferred from inelastic neutron scattering and discussed below) magnetic resonance ($\checkmark$3) mode energy (~11 meV) in PLCCO (S. D. Wilson et al. [330]).

Another use of tunneling data is via model fits to determine s- vs d-wave symmetry. In the electron doped cuprates, these fits are not definitive. Most n-type cuprate tunneling data are on epitaxial thin films or crystals. The lack of a zero bias conduction peak (ZBCP) (normally if such a peak is present it implies a sign change in the phase of the order parameter, i. e. consistent with d-wave or s± as found in the IBS) can be due to inherent disorder. The critical comparison of the mean free path and the coherence length, $\xi_0 \sim \ell$, for disorder to suppress the ZBCP (Aprili et al. [331]) is easily exceeded (Fournier [42]) in the electron doped cuprates due to the relatively long $\xi_0 \sim$ 50Å.

### 5.1.2. ARPES

Although of great utility for the hole doped cuprate superconductors and the IBS, because of those materials' larger superconducting gaps, ARPES results for the superconducting gap in electron doped superconductors, with $\Delta \sim 4$ meV, are less conclusive. Armitage et al. [332] interpret their ARPES data on single crystals of NCCO as consistent with an anisotropic energy gap, consistent with d-wave pairing. However, within their measurement uncertainty, they cannot rule out an anisotropic s-wave ordering parameter with a small gap below their detection limit. Matsui et al. [333] present ARPES data on PLCCO that are described as "basically consistent" with $d_{x2-y2}$ pairing symmetry.

When it comes to a high energy gap-like structure (the 'pseudogap'), ARPES see such a structure (✓4), e. g. in single crystal PLCCO, $T_c$=26 K, without difficulty [333].

### 5.1.3. Specific Heat:

Yu et al. [334] measured the low temperature C/T of PCCO, $T_c$=22 K, and found that the field dependence was proportional to $H^{1/2}$, consistent with d-wave pairing (as first done by K. Moler et al. [81] for YBCO.) However, as is now understood and discussed below in IBS, such behavior in the two band electron doped cuprates can be mimicked by two fully gapped s-wave bands.

### 5.1.4. Penetration Depth

As discussed above in section 2.1.5., the power law dependence of the penetration depth can be affected by impurities and therefore be difficult to interpret definitively. In addition, early measurements of $\lambda$ in NCCO returned results (exponential temperature dependence) consistent with s-wave pairing symmetry. Cooper [335] reanalyzed these penetration depth data using the strong Nd magnetic contribution to arrive at $\lambda \propto T^1$ or $T^2$. In the case of PCCO, where the magnetic contribution of Pr is weaker, Prozorov et al. [336] found that their data on single crystals for $\Delta\lambda$ varied as $T^2$ between 0.025 K up to 0.3 $T_c$, consistent with d-wave pairing with impurities.

### 5.1.5. NMR

Despite the lack of a pure power law, the $1/T_1$ data for single crystal PLCCO, $T_c$=24 K, of Zheng et al. [45] can be well fit to a model for disordered d-wave superconductors. No evidence of a pseudogap in the penetration depth data is observed.

### 5.1.6. Raman

Blumberg et al. [337] interpret their Raman data on single crystals of NCCO, $T_c$=22 K, as consistent with $d_{x2-y2}$ order parameter symmetry. They further analyze their data as indicating that the maximum in $\Delta$ is near where the antiferromagnetic fluctuations are at a maximum in the Brillouin zone (a so-called 'hotspot'), emphasizing the role of such fluctuations in the superconductivity.

### 5.1.7. *Isotope Effect*

Batlogg et al. [338] prepared powdered samples of NCCO, $T_c$=24.5 K, and exchanged approximately 85% of the $^{16}O$ for $^{18}O$. In contrast with results for hole doped $La_{2-x}M_xCuO_4$ (see section 5.2.4.), they found no isotope effect with a limit on $\alpha$ ($T_c \propto M^{-\alpha}$) of ≤0.05.

### 5.1.8. *Phase sensitive experiments*

Phase sensitive measurements (✓7) were performed using patterned thin film SQUID devices with Josephson junctions. The first results, on the half-flux quantum effect (discussed below, see also Figs. 17 and 18, with the hole doped cuprates where the technique was first employed), were performed by Tsuei and Kirtley [339] on NCCO, $T_c$=22-25 K, and $Pr_{1.85}Ce_{0.15}CuO_{4-\delta}$, $T_c$=22-23 K. Later work has been performed in $La_{1.895}Ce_{0.105}CuO_4$, $T_c$=29 K (Chesca et al. [340] and in the infinite layer $Sr_{0.85}La_{0.15}CuO_2$, $T_c$=18 K (Tomaschko et al. [341]). These works offer convincing proof of $d_{x2-y2}$ pairing symmetry. Obviously, these tour-de-force efforts in the electron doped cuprates benefited from the years of effort achieving the same results in the hole doped cuprates (for reviews, see Tsuei and Kirtley [4] and Kirtley [342]).

### 5.1.9. *Neutron Spin Resonance*

Wilson et al. [330] reported a sharp magnetic excitation (resonance) in the inelastic neutron scattering of PLCCO, $T_c$=24 K, in the superconducting state, where the energy of the resonance, $E_r$, was approximately 5.3 $k_BT_c$. This result further linked the electron doped cuprates to the hole doped, where such a resonance (with $E_r \approx 5.8$ $k_BT_c$) had been seen (see also below) since 1991 (Rossat-Mignod et al. [343]). In PLCCO, the integrated scattered intensity around Q=(1/2,1/2,0) in the reciprocal lattice of the $CuO_2$ planes (the antiferromagnetic Bragg peak position in the undoped compound) increases dramatically below $T_c$ – just as seen in YBCO [343] and other hole doped cuprates (see below). Although not understood, this resonance is theorized to be fundamentally linked to the superconducting pairing mechanism.

### 5.1.10. *Summary for Electron Doped Cuprates*

Although not as mature a materials class as the hole doped cuprates, these materials – thanks to the definitive phase sensitive experiments which build on prior experience in the hole doped superconductors – appear to be clearly UcS, with nodal $d_{x2-y2}$ pairing symmetry. Besides the phase sensitive, tricrystal experiments, the fit to the NMR data [45] and the inelastic neutron scattering resonance mode [330] seem quite clear evidence for UcS. The STM data [329] on PLCCO and the Raman data [337] on NCCO provide indications of the importance of spin fluctuations for the (unconventional) superconductivity. The underinvestigated apparent gap (the NSG) in the electronic energy spectrum of approximately the same size as the superconducting gap, but found in the normal state for H>$H_{c2}$ remains an interesting puzzle.

Table 3: Strong Indications of UcS in the electron-doped cuprates

|   | 1 | 2 | 3 | 4 | 5 | 6 | 7 | 8 | 9 | 11 | 13 | 14 |
|---|---|---|---|---|---|---|---|---|---|----|----|----|

| | $T_F \ll \Theta_D$ | QC | Spin res | PG | $T^\alpha$ | >1 sc phase | Phase sens. tunneling $\Rightarrow$ non-BCS | TRSB | $T_c$/non-mag imp | $C(H,\Theta)$ $\kappa(H,\Theta)$ | Lack of isotope effect | $\gamma_{residual}$ large |
|---|---|---|---|---|---|---|---|---|---|---|---|---|
| electron-doped cuprates | ✗ | ✓ | ✓ | ~✓ | ~✓ | | ✓ | | | ✗ | ✓ | ✗ |

## 5.2. Hole Doped

### 5.2.1. Introduction

In terms of the phase diagram sketched above in Fig. 14, the hole doped cuprates are roughly similar to the just-discussed electron doped materials. The major differences are the broader-in-composition superconducting dome and the lack of coexistence of long range antiferromagnetism and superconductivity. After a brief introduction, we will discuss the detailed differences as they pertain to UcS.

Discovered in 1986 by Bednorz and Mueller, $T_c$ above 30 K in the low electronic density of states Ba-doped $La_2CuO_4$ (the tetragonal '214' perovskite structure, with two layers containing La followed by a $CuO_2$ layer, then repeat – see Fig. 15) was a definite surprise. Until then, higher $T_c$ had been sought in the high density of states A15 superconducting class (for a review, see Stewart [2], where the pairing is mediated by electron-phonon coupling. The pathway to higher $T_c$ in the cuprates, which are believed to have a different primary pairing mechanism (with some contribution from electron-phonon coupling), led initially to the $YBa_2Cu_3O_{7-\delta}$ (orthorhombic '123' perovskite structure with both Cu-O chains and $CuO_2$ planes – see Fig. 16) but soon diverged along numerous paths.

Rather than digress into the fascinating history of these discoveries, we simply list the resultant main cuprate superconductor 'classes' (the Bi-, Tl-, and Hg-based families: $Bi_2Sr_2Ca_{n-1}Cu_nO_{2n+4}$, $Tl_2Ba_2Ca_{n-1}Cu_nO_{2n+4}$ $HgBa_2Ca_{n-1}Cu_nO_{2n+3-\delta}$, n=1, 2, 3, …) - with some representative $T_c$ values given in Table 4 – that were discovered after the two discovery compounds ($La_{2-x}M_xCuO_4$, M=Ba,Sr,Ca and $YBa_2Cu_3O_{7-\delta}$) and refer the reader interested in more detail to the recent history of cuprate research by Chu et al. [344]. Below, after a discussion of $La_{2-x}M_xCuO_4$ to start the discussion of the hole doped cuprates, we discuss the 'strongly suggestive' pieces of evidence of section 2 for determining unconventional superconductivity across all the hole doped cuprate materials, since the properties therein are much more uniform across the classes than in section 4, where the heavy fermion superconductors were often entirely different from each other.

Table 4: $T_c$ and number of $CuO_2$ planes in the unit cell (all structures are tetragonal except for YBCO, which is orthorhombic) for some representative (data from [344]) cuprate superconductors. Note the peak in $T_c$ at n=3, noticed early by P. W. Anderson [345].

| Compound | Abbreviation | $T_c$(K) | n | Year of discovery |
|---|---|---|---|---|
| $La_{1.85}Sr_{0.15}CuO_4$ | 214 | 30 | 1 | 1986 |
| $YBa_2Cu_3O_7$ | 123 or YBCO | 93 | 2 | 1987 |

| | | | | |
|---|---|---|---|---|
| $Bi_2Sr_2CuO_6$ | Bi-2201 | 9 | 1 | 1987 |
| $Bi_2Sr_2CaCu_2O_8$ | Bi-2212 | 84 | 2 | 1988 |
| $Bi_2Sr_2Ca_2Cu_3O_{10}$ | Bi-2223 | 110 | 3 | 1988 |
| $Bi_2Sr_2Ca_3Cu_4O_{12}$ | Bi-2234 | 90 | 4 | 1988 |
| $Tl_2Ba_2CuO_6$ | Tl-2201 | 90 | 1 | 1987 |
| $Tl_2Ba_2CaCu_2O_8$ | Tl-2212 | 110 | 2 | 1988 |
| $Tl_2Ba_2Ca_2Cu_3O_{10}$ | Tl-2223 | 125 | 3 | 1988 |
| $TlBa_2Ca_3Cu_4O_{11}$ | Tl-1234 | 104 | 4 | 1988 |
| $HgBa_2CuO_4$ | Hg-1201 | 95 | 1 | 1993 |
| $HgBa_2CaCu_2O_6$ | Hg-1212 | 114 | 2 | 1993 |
| $HgBa_2Ca_2Cu_3O_8$ | Hg-1223 | 133 | 3 | 1993 |
| $HgBa_2Ca_3Cu_4O_{10}$ | Hg-1234 | 125 | 4 | 1993 |

5.2.2. *$La_{2-x}M_xCuO_4$, M=Ba,Sr,Ca*

First presented as a mixed phase $Ba_xLa_{5-x}Cu_5O_{5(3-\delta)}$ sample by Bednorz and Mueller [8], Cava et al. [346] refined the superconducting phase to be $La_{2-x}M_xCuO_4$, doped with M=alkaline earth ions on the La site (for Ba, $T_c$=35 K, for Sr, $T_c$=30 K).  Another method to hole dope $La_2CuO_4$ is to use high pressure (0.3 GPa) at 600 °C to drive additional oxygen into the lattice, to form $La_2CuO_{4+\delta}$, $\delta\approx0.13$, with $T_c$=28 K.  (Schirber et al. [347])

As discussed above already, $La_{2-x}M_xCuO_4$ has layers of $CuO_2$ (see Fig. 15), as do all the cuprate superconductors, but no CuO chains, as does only the $YBa_2Cu_3O_7$ class (see Fig. 16). Doping by alkaline earths introduces holes into the $CuO_2$ planes.   Both the $CuO_2$ planes, and the CuO chains in YBCO, are important for superconductivity.

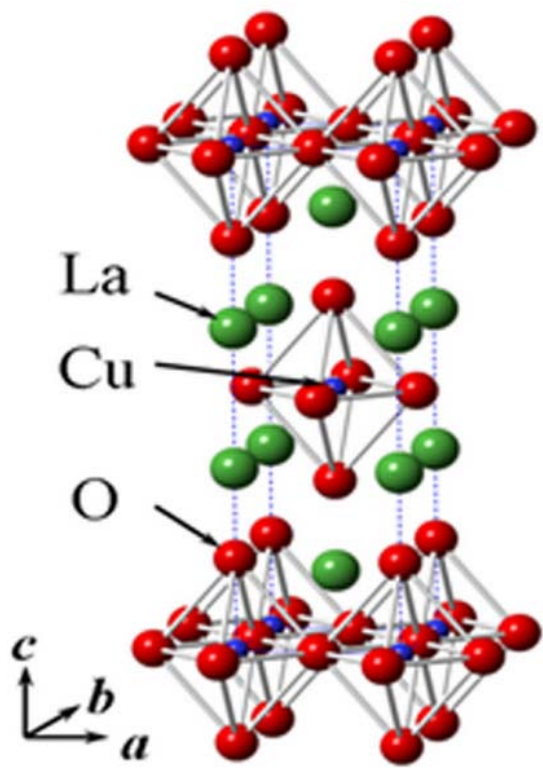

Fig. 15 (color online) (Hosono et al. [348]). This $La_2CuO_4$ tetragonal structure (called the T214 phase, and also sometimes called a one layer cuprate structure) has two apical oxygens above and below the Cu (blue atom) in the center of the structure, and the four oxygens above and below these two apical oxygens form a square and are on the unit cell corners. In the electron doped $R_{2-x}Ce_xCuO_4$ structure (called the T'214 phase, not shown) discussed above in the text, the apical oxygens are missing and the four oxygens above and below the planes of La atoms (shown in the figure at the left for the T214 and in the same place for the T'214) are centered on the faces and not at the corners. Figure reproduced with permission from Hosono et al., ref. [348] according to the license at
https://creativecommons.org/licenses/by/3.0/

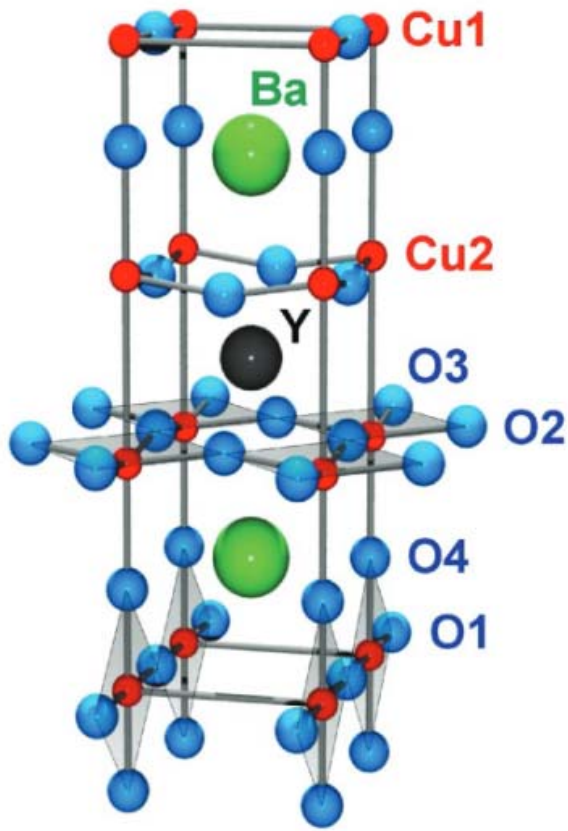

Fig. 16 (color online) YBCO (from S. Thiess et al. [349]). The direction of the a, b, and c axes is the same as in Fig. 15. There are two sheets of $CuO_2$ in the a-b plane around the central Y atom involving the Cu2 atoms, with chains of Cu-O in the b-axis direction above and below the Ba atoms involving the Cu1 atoms. Reprinted figure with permission from Thiess et al., Phys. Rev. B 92 (2015), p. 075117. [349] Copyright (2015) by the American Physical Society.

Since $La_{2-x}M_xCuO_4$, M=Ba, Sr, or Ca, was the first discovered cuprate high temperature superconductor, this material rapidly attracted an enormous amount of experimental and theoretical attention. For example, from the publication of the discovery by Bednorz and Mueller in June, 1986, the resonating valence bond theory [350] of Anderson appeared already in March, 1987. As an index to the amount of work that has transpired in trying to understand the cuprates, the discovery paper has been cited ~9400 times, while the seminal theory paper of Anderson has been cited ~5700 times, as of the writing of this review.

In the theory of Anderson, in order to capture the essential physics of superconductivity in $La_{2-x}M_xCuO_4$, the focus is on the Cu d-electron orbital which, after hybridization with the O anions, results in a single d-band. With the addition of on-site Coulomb repulsion with strength U, Anderson proposes a nearly half filled single band 2D Hubbard model (although in reality there is coupling between the $CuO_2$ layers). As reviewed by Scalapino [14], there have since been a large number of theoretical approaches (e. g. random phase approximations [RPA], renormalized mean field theory, self consistent renormalization, slave boson approximations, various numerical Monte Carlo approaches, and many others) to arrive at the predicted physical properties of such Hubbard models.

The fundamental question in the cuprates, and indeed in all UcS, is what is the interaction that causes the pairing of the electrons? One facet that *is* known is that the pairing symmetry in the cuprate superconductors is d-wave, from phase sensitive tunneling, ARPES, and several

other measurement techniques as will be discussed. At present, the consensus in the cuprates is that the cause of the pairing interaction remains "disputed" (Anderson [102]) and that "diverging views" remain (Chu et al. [344]). There is however a pairing mechanism often mentioned in the cuprates (as well as in the IBS) that is certainly a strong candidate, namely the exchange of antiferromagnetic spin fluctuations (see the review by Scalapino [14]) rather than of virtual phonons as in the BCS theory.

In this primarily experimental review, our task vis-à-vis theory is to inform the reader of the current state of theoretical understanding where consensus has been reached, and – after an introductory discussion - to provide appropriate recent references to introduce the ongoing debatable issues. We now turn to the rather thorough experimental characterization of the superconducting properties as they shed light on UcS in the five sub-classes ($La_{2-x}M_xCuO_4$, $YBa_2Cu_3O_{7-\delta}$, and the Bi, Tl, and Hg based compounds) representative of hole doped cuprates.

5.2.3. *Discussion of phase sensitive experiments/the half-flux quantum effect*

Since much of the following discussion of the hole doped cuprates and their unconventional superconductivity focuses on their pairing symmetry being d-wave, we present the proof of d-wave pairing by phase sensitive tunneling experiments first, somewhat out of historical order. As will be discussed, numerous other measurement techniques also strongly support the presence of d-wave pairing in the hole doped cuprates.

In 1994, Tsuei et al. [64] developed a clever technique to determine if a superconductor has $d_{x^2-y^2}$ pairing symmetry. They realized that the symmetry of a pair wave function can be probed at a weak link junction where the Cooper pairs tunnel through. This is because the sign of the Josephson tunneling current between two putative d-wave superconducting regions joined by the weak link is a function of how their relative order parameters are aligned at the interface.

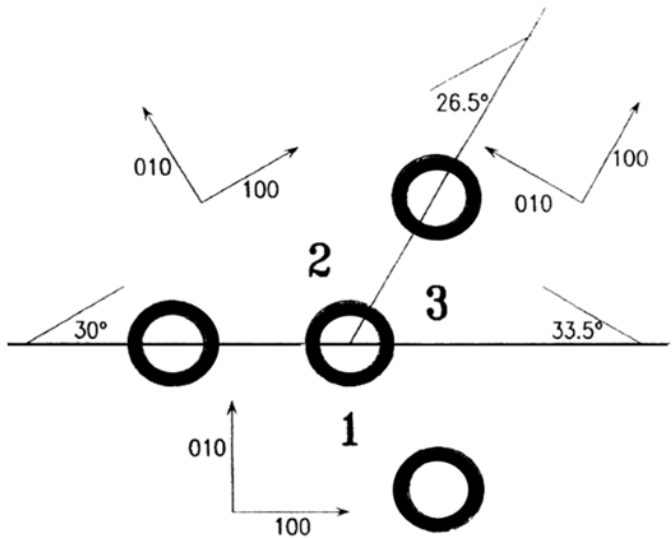

Fig. 17: A sketch from Tsuei et al. showing four YBCO rings epitaxially deposited on SrTiO$_3$, where the SrTiO$_3$ substrate has three crystals (1, 2, and 3) coming together at the center of the diagram with the crystalline orientations as shown. Reprinted figure with permission from Tsuei et al., Phys. Rev. Lett. 73 (1994), p. 593. [64] Copyright (1994) by the American Physical Society.

In the figure is a schematic of four rings of YBCO, deposited as an epitaxial thin film on a $SrTiO_3$ (perovskite structure) crystal specially manufactured such that the three adjacent grains (1, 2, and 3) shown had their x and y axes oriented as shown. Using ion-milling photolithography techniques, four rings (ID 48 μm, width 10μm) were made as depicted. One of the rings (over grain 1) serves as a control with no weak links. Two of the rings (across the 1-2 grain boundary and across the 2-3 boundary) have two weak links each where, with the epitaxy, the covering YBCO superconductor crystalline orientation in the ring changes at the grain boundary to form a weak link. These two rings also serve as controls. The fourth ring, at the "tricrystal" point where 1, 2, and 3 come together, in zero field (Tsuei et al. achieved B<0.4μT) is designed (for details, see the review by Tsuei and Kirtley [4] as well as Tsuei et al. [64]) with its *three* weak links, each one generating a phase shift of π in the superconducting current, to be 'frustrated' (where one circuit around the ring does not cause a phase shift equal to an integer multiple of 2π) for $d_{x2-y2}$ pairing symmetry. For all other pairing symmetries, including g-wave, Tsuei et al. were able to show different, non-frustrated behavior would obtain.

The most measureable effect caused by a superconducting ring containing an odd number of π phase shifts is that – under the conditions in the Tsuei et al. design – one half of a flux quantum, $\Phi_0/2$, will be spontaneously generated. This generation of $\Phi_0/2$ in Josephson junctions in a frustrated geometry of superconductors with unconventional pairing was first theoretically proposed for the heavy fermion superconductors by Geshkenbein, Larkin and Barone [351]. Using a high resolution scanning SQUID microscope, Tsuei et al. were able to show that a magnetic field equivalent to one half of a flux quantum was present in the three junction central ring, proving conclusively that YBCO has $d_{x2-y2}$ pairing symmetry. For a scan of the fields at the four rings, see Fig. 18.

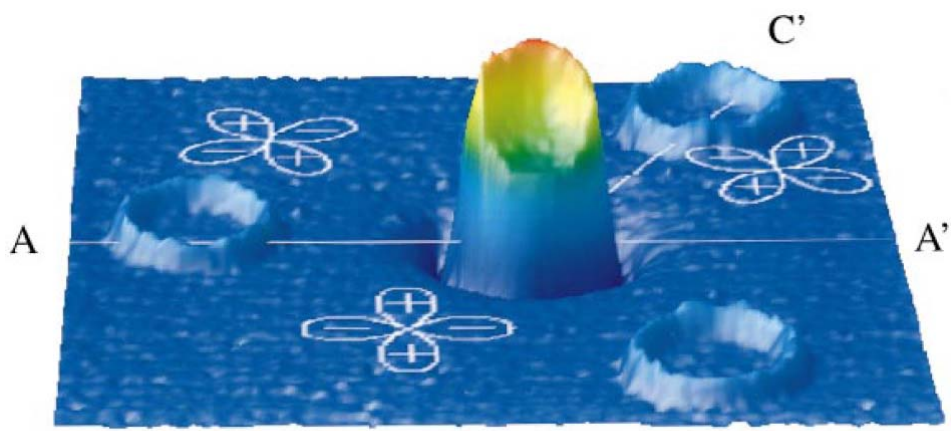

Fig. 18 (color online) (from Tsuei Kirtley [4]) The tricrystal, half flux quantum experiment of Tsuei et al. [64] that showed conclusively that YBCO has $d_{x2-y2}$ pairing symmetry. The three outer rings, in this scanning SQUID microscope image of the sketched arrangement of Fig. 17, show no magnetic field, while the center ring, with three Josephson junctions, i. e. with frustration, shows one half of a superconducting quantum of flux. The grain boundaries are along the line A-A', and along the line connecting the center of the picture to C'. Reprinted figure with permission from Tsuei and Kirtley, Rev. Mod. Phys. 72 (2000), p. 969. [4] Copyright (2000) by the American Physical Society.

5.2.4. *Isotope Effect*

In BCS theory, the dependence of $T_c$ on the molar mass of a superconductor is $T_c \propto M^{-1/2}$. As discussed in section 2.1.13., there are well known exceptions even in the elemental superconductors, e. g. the $T_c$ of Ru is essentially independent of M (T. H. Geballe et al. [78]) due to renormalized Coulomb interaction effects. Thus, if there *is* an isotope effect, this implies that lattice vibrations (phonons) play some role in the microscopic mechanism for superconductivity. As discussed in section 2.1.13., this does not necessarily imply that electron phonon coupling is responsible for the pairing interaction. While it is clear that electron phonon coupling alone is not sufficient to explain the high $T_c$'s in the cuprates (references: R. Heid et al. [352], Bohnen et al. [353], Giustino, Cohen and Louie [354]), as we will discuss there are models that suggest that such coupling can *enhance* the pairing correlations from another dominant interaction, e. g. spin fluctuations. For reviews of the isotope effect in cuprates, see J. Franck [304] and Zhao, Keller and Conder [303].

In the case of the *optimally doped* cuprates, only a small isotope shift (where $^{16}O$ is replaced by $^{18}O$ or $^{63}Cu$ is replaced by $^{65}Cu$), $\alpha$ significantly less than 0.5, is observed. Thus, in general, the role of phonons for the <u>peak</u> high temperature superconductivity in the cuprates is not considered to be of primary importance [303]. As well, even for the under hole-doped cuprates, $\alpha$ behaves monotonically (except for the special case of $La_{2-x}M_xCuO_4$) and in general remains too small to explain the high $T_c$ values observed. Despite these statements, the isotope effect in the hole doped cuprates has been extremely well characterized and can serve as a guide for efforts to characterize this effect in other UcS classes. Thus, we describe the results in some detail.

5.2.4.1. *Optimally hole-doped cuprates:* In optimally doped $La_{1.85}Sr_{0.15}CuO_4$, Franck, Harker and Brewer [355] report the unsubstituted compound had $T_c$=37.94 K. For isotopic substitution of $^{16}O$ by $^{18}O$, $T_c$ decreased by 0.41 K, to give $T_c=M^{-\alpha}$, $\alpha$=0.10, while for substitution of $^{63}Cu$ by $^{65}Cu$, $T_c$ decreased by 0.14 K to give $\alpha$=0.12, i. e. the two values for $\alpha$ are comparable. This equality of $\alpha$'s for both O and Cu isotopic substitution is important (J. P Franck [304]) to rule out a special role of phonons involving the oxygen atoms. (For the case of the isotope effect being especially influenced by phonons involving the Fe atoms in IBS, see section 6.) The oxygen substitution result was confirmed by Zhao et al. [356], with a $T_c$ decrease of 0.52 K.

In $YBa_2Cu_3O_{7-\delta}$, for an oxygen content of 6.94 and $T_c$=93 K (i. e. optimally doped), the isotope exponent $\alpha$ for substitution of $^{63}Cu$ by $^{65}Cu$ is (Zhao et al [357]) slightly negative, -0.017. For oxygen substitution in $YBa_2Cu_3O_{7-\delta}$ at optimal doping, $T_c$~92 K and $\delta$~0.1, $\alpha$=0.025 (Zech et al. [358]). Results for oxygen substitution in $Bi_2Sr_2Ca_2Cu_3O_{x\sim10}$ ("2223") (Katayama-Yoshida et al. [359]), $T_c$=110 K and in Pb stabilized 2223 (Bornemann, Morris and Liu [360]), $T_c$=108 K gave $\alpha$=0.03 and -0.013 respectively. Thus, although the $\alpha$ exponents for optimally doped $La_{1.85}Sr_{0.15}CuO_4$ seem small, they are four to five times larger than in optimally doped $YBa_2Cu_3O_{7-\delta}$ and Bi 2223.

5.2.4.2. *Under hole-doped cuprates:* The above is not to say that the isotope effect is small for *all* levels of hole doping in the cuprates. In fact, for an oxygen content that gives $T_c$~50 K in YBCO, $\alpha$ exceeds 0.2 (Zech et al. [358])/0.3 (Kamiya et al. [361]) (although the variation of $\alpha$ with doping remains monotonic). A careful reexamination (Soerensen and Gygax [362]) of $^{18}O$

doping experiments in YBCO, substituted with Pr and Ca on the Y site and Zn on the Cu site, called earlier results into question but concluded that the Zn-doped results were accurate. Thus, α for $^{18}$O substituted $YBa_2(Cu_{1-x}Zn_x)_3O_{7-\delta}$, with $T_c$ suppressed by the Zn doping below 10 K is found to be over 0.3.

As an interesting separate case, in $La_{2-x}Sr_xCuO_4$ as a function of Sr content α can exceed 1 (Zhao et al. [363]) in the deeply underdoped regime, see Fig. 19.

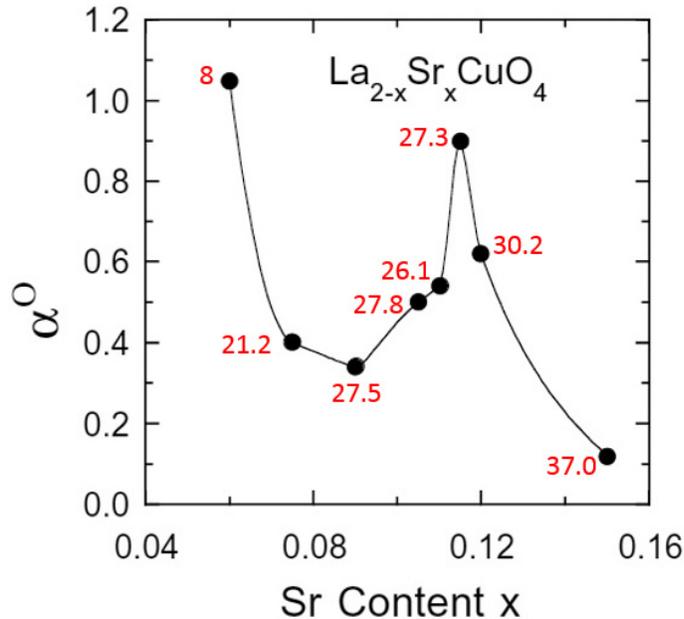

Fig. 19: (color online) Isotope effect α, from the shift in $T_c$ from replacement of $^{16}$O by $^{18}$O, in $La_{2-x}Sr_xCuO_4$ as a function of x. (Zhao et al. [363]). $T_c$ in units of K for the $^{16}$O samples at each Sr composition where α was measured shown in red. The non-monotonicity of $T_c$ vs x (vs the smooth $T_c$ vs x dome shown in the phase diagram in Fig. 14) is partly explained by the fact that the data for x=0.105, 0.11, and 0.115 are from a different work (Zhao et al. [364]). Zhao et al. [363] – in the face of skepticism about the validity of their reported strong dependences of $T_c$ on oxygen mass depicted in Fig. 19 – give convincing experimental verifications for the accuracy of their data. They explain their data for x≤0.09 based on a polaron model of superconductivity. This figure is from Zhao et al. [363]. © IOP Publishing. Reproduced with permission. All rights reserved.

5.2.4.3. *Theory:* As an example of the development over time of theory to correctly explain the varied isotope results for the hole doped cuprates, the enhancement of α in $La_{2-x}Sr_xCuO_4$ around x=0.115 (where $T_c$≈27 K) was argued by Pickett, Cohen and Krakauer [365] to be due to strong anharmonicity in the phonon modes that is sensitive to the O mass – thus the peak in α. This explanation, which is based on a drop in the electronic density of states at the Fermi energy, N(0), due to an anharmonicty-driven phase transformation, assumed an electron-phonon coupling mechanism for superconductivity.

More recently, Johnston et al. [366] performed calculations in a five band model for electron phonon coupling in the cuprates which tracked the electron phonon coupling to oxygen phonons. They found that cuprates with the largest electron coupling to phonons in the $B_{1g}$ branch (the out of phase bond-buckling planar oxygen branch), which unlike isotropic phonon modes primarily provide d-wave (known to be dominant in the cuprates) pairing, have the largest $T_c$ values. They argued that electron phonon coupling enhances the pairing correlations provided by another, more dominant interaction. Note that [366] to first order there is no electron phonon couping to the $B_{1g}$ branch phonon modes in single layer (e. g. $La_{2-x}M_xCuO_4$) cuprates, consistent with (see Table 4 and Fig. 19) their lower $T_c$ values.

Bang [367], in results similar to Johnston et al., (see similar approaches also by Honerkamp, Fu and Lee, [368]; Nunner, Schmalian and Bennemann [369]; Nazarenko and Dagotto [370]) found that anisotropic phonon interaction (the $B_{1g}$ buckling phonon mode of the in-plane oxygen motion) can enhance d-wave pairing and therefore can enhance $T_c$ together with a dominant coupling interaction that Bang assumes to be antiferromagnetic (AFM) spin fluctuations. With the assumption that the electron phonon coupling strength, $\lambda_{ph}$, is unchanged with doping and that the $\lambda_{AFM}$ increases with increasing doping, Bang put forward a simple formula for the isotope effect $\alpha = \frac{1}{2} \lambda_{ph}/(\lambda_{AFM} + \lambda_{ph})$ which gives a plausible explanation of the behavior of the isotope effect (decreasing as doping approaches the optimal concentration/$T_c$) in the n≥2 cuprates (e. g. YBCO but excluding single layer $La_{2-x}M_xCuO_4$ with its absent-to-first-order $B_{1g}$ mode coupling.)

Greco and Zeyher [371] calculate in a t-J model that the large measured $\alpha$ values in the underdoped YBCO- and Bi-based cuprates come from a shift of the electronic density of states from low to higher energies. This shift is caused by a ground state (e. g. a charge density wave state) that *competes* with (rather than enhancing) the superconductivity, with the competing state also the source of the pseudogap. In this approach, the large values of $\alpha$ for under-hole-doped cuprates are not evidence for strong electron phonon coupling. When they apply their model to the data for $\alpha$ shown in Fig. 19, they propose a *second* pseudogap around the 1/8 doping level in $La_{2-x}(Ba/Sr)_xCuO_4$ that is due to static (Ba) or fluctuating (Sr) stripes (for a discussion of charge and spin stripes in the cuprates, see Tranquada [372].) This second pseudogap indeed competes with the superconductivity, with $T_c$ falling to ~5 K (Huecker et al. [373]) for Ba doping at x=0.125 (from above ~30 K for optimally doped, x=0.155), and dipping several Kelvin (see Fig 19) to a local minimum in $T_c$ vs x for Sr doping at x=0.125 (Suryadijaya, Sasagawa and Takagi [374]). Thus, in the model of Greco and Zeyher, the large isotope effect $\alpha$ in $La_{2-x}(Ba/Sr)_xCuO_4$, rather than enhancing $T_c$, suppresses it.

In summary, clearly the isotope effect in single layer $La_{2-x}M_xCuO_4$ remains the exception amongst the cuprates. The behavior of $\alpha$ for optimally hole-doped materials, n≥2, appears to be understood by numerous authors to provide evidence for electron coupling to anisotropic phonon ($B_{1g}$ oxygen) modes providing at least an *enhancement* in $T_c$ over an alternative d-wave pairing interaction such as exchange of spin fluctuations.

5.2.5. *Pseudogap*

A 'pseudogap' refers to particular observed properties above $T_c$ in both the electron and hole doped cuprates (although there are differences between the two, e. g. point contact spectroscopy in the electron doped cuprates does not show a pseudogap). This term was first used for the cuprates by J. Friedel [375]. A pseudogap is nothing more or less than a partial gap – admittedly that manifests itself differently in different measurements – where parts of the Fermi surface become gapped *above* the superconducting transition temperature. Part of the justification for tying the pseudogap at $T^*>T_c$ with the actual superconducting gap at $T_c$ is that the pseudogap, at $T^*>T_c$, in the under doped cuprates (e. g. in under-hole-doped BiSCCO, 2212) has (Harris et al. [376]), with some extra smearing ("dirty d-wave" scenario) near the nodal angle, the *same* $d_{x2-y2}$ angular dependence as the superconducting, $T<T_c$, gap. NMR data were the first (Warren et al. [377]), see the discussion in the review [40] by Timusk and Statt, to show evidence of this partial gap in the normal state. Other, later measurements including ARPES (see review by Damascelli, Hussain and Shen [301]), optical conductivity (see review by Basov and Timusk [378]),

resistivity (see Fig. 20), electronic specific heat, thermoelectric power (Tallon et al. [379]), electronic Raman scattering (F. Slakey et al. [380] and Opel et al. [381]), and magnetic neutron scattering (Rossat-Mignod et al. [343]) also gave clear indications.

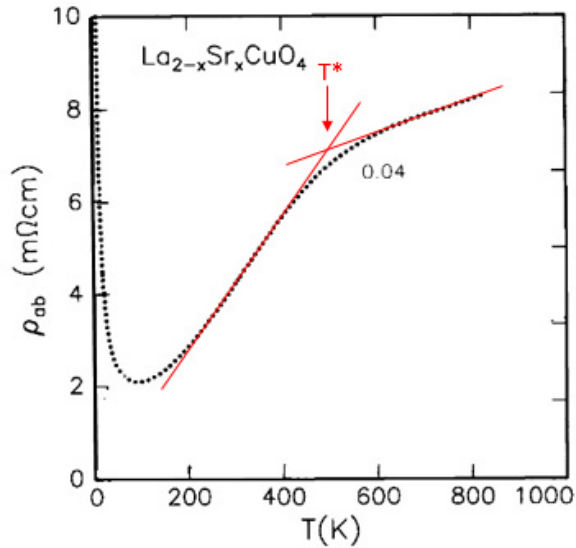

Fig. 20 (color online) (Takagi et al. [382]). Resistivity vs temperature of an underdoped single crystal of $La_{1.96}Sr_{0.04}CuO_4$, showing the decrease in the rate of rise of the resistivity with temperature at the onset of the pseudogap at $T^*\approx 500$ K. Data are black points; red lines are guides to the eye. Reprinted figure with permission from Takagi et al., Phys. Rev. Lett. 69 (1992), p. 2975. [382] Copyright (1992) by the American Physical Society.

Although, as discussed above for the heavy fermion superconductors $CeCoIn_5$ and $URu_2Si_2$ (and below for IBS and the cobaltates), there are other superconductors where evidence for a partial gap above $T_c$ has been observed, the reviews that exist on the pseudogap (see, e. g., Timusk and Statt[40]; B. Batlogg et al. [383]; Norman, Pines and Kallin [302]; Lee [384]; the latter half of the article by Abrahams [288]; Rice, Yang and Zhang [39]; Fujita et al. [385]; Tallon and Loram [386]) all focus on the hole-doped cuprates (see the schematic sketch of the phase diagram in Fig. 14).

The pseudogap in the cuprates remains a central focus of much research today, and should be named as one of the outstanding puzzles in UcS. As Norman says in his 2012 overview of the theory of cuprates, "the nature of the pseudogap phase is key." There are several important open questions concerning the pseudogap in the cuprates: what is its cause, how does it relate to/interact with the superconducting gap, and – seemingly a straightforward, easily answereable question - where does $T^*$ go to zero (a potential quantum critical point) as a function of composition? Concerning this last question, despite an enormous amount of data, there exists no consensus. One set of experiments and analysis states that $T^*\rightarrow 0$ at an approximate hole doping of p=0.19 (e. g., see Vishik et al. [387]), somewhat past optimal doping (see Fig. 14). Another set of experiments and analysis states that $T^*$ intersects the edge of the superconducting dome on the overdoped side (see Fig. 21). This seeming important contradiction has developed with time, and relies on the evolution of analysis of data taken with surface techniques such as ARPES and STM. We will now give a short review of some of the many various observational techniques (far more than for any other class of UcS) that demonstrate pseudogap behavior in the cuprates, some of which agree with Fig. 14 and some of

which with Fig. 21. This question of where T* intersects the x-axis in a temperature vs composition phase diagram has long been a point of contention, although current thinking seems to trend toward the scheme depicted in Fig. 14 or, more precisely, around p.=0.19.

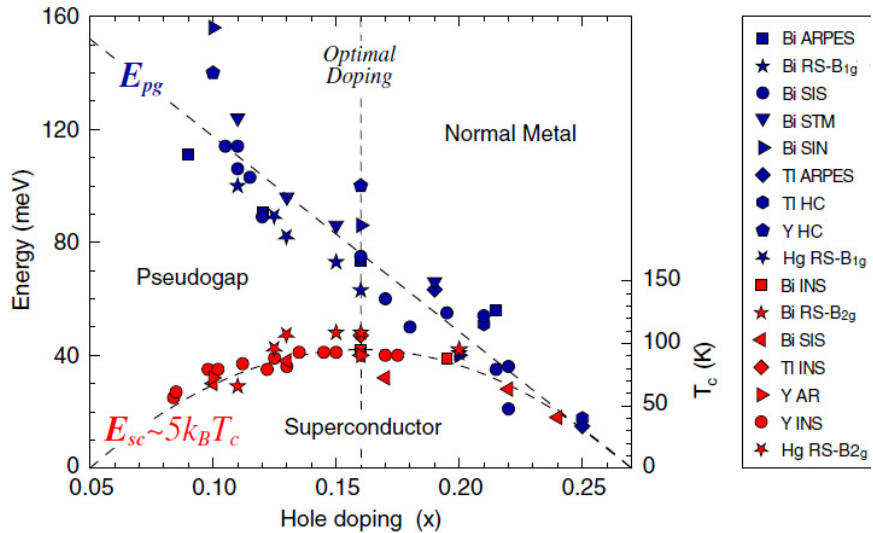

Fig. 21 (color online) Pseudogap ($E_{pg}=2\Delta_{pg}$) and superconducting energy gap, $E_{sc}$, (in red) for a number of hole doped cuprates with $T_c^{max}$ ~ 95 K (Bi2212, YBCO, Tl2201 and Hg1201, see Table 4.) (Huefner et al. [38]). The measurement methods (for refs. see Huefner et al.) are given in the righthand key: ARPES, Raman Spectroscopy (which can give two spectra: $B_{1g}$ (antinodal) and $B_{2g}$ (nodal), each representing a different average over the Fermi surface), various tunneling spectroscopies (SIS, STM, and SIN), and Heat Conductivity (HC). In addition, the energy $\Omega_r$ of the magnetic resonance (see next section) - as determined by Inelastic Neutron Scattering (INS) – which scales with $T_c$ just as the superconducting energy gap, $E_{sc}$, does is also shown (in red). (figure is reproduced from ref. [38]. © IOP Publishing. Reproduced with permission. All rights reserved.)

5.2.5.1 *ARPES:* ARPES is a very powerful tool for characterizing the partial gap in the normal state in the cuprates (the pseudogap), as reviewed by Damascelli, Hussain and Shen [301] and Vishik et al. [388]. ARPES first succeeded in detecting the cuprate pseudogap in 1996 (Marshall et al. [389]; Loeser et al. [390]; Ding et al. [391].) Marshall et al., using the easily cleavable hole doped $Bi_2Sr_2CaCu_2O_{8+\delta}$ (2212), (as did both Loeser et al. and Ding et al.) discovered portions of missing Fermi surface (i. e. where a gap had opened) in underdoped samples, $T_c$=65-67 K, related to the opening of a pseudogap *above* $T_c$. Measurements (Marshall et al.) were taken at 110 K/100 K by Loeser et al. For an extreme underdoped 2212 sample, $T_c$=10 K, Ding et al. found that the pseudogap extended up to at least 301 K. In their review of ARPES in cuprates, Damascelli, Hussain and Shen [301] show in a plot (their Fig. 62) of T* vs composition as determined by various groups in $Bi_2Sr_2CaCu_2O_{8+\delta}$ (the cleanest and most cleavable cuprate) that T* extrapolates approximately to the far (overdoped) side of the superconducting dome, consistent with Fig. 21. Tallon and Loram [386] argue however that ARPES shows two gaps, the larger one of which (the leading edge gap) is the pseudogap (which extrapolates to the composition axis, T*→0, at approximately p=0.19) and the smaller one of which is associated with the superconducting gap (which joins together with the superconducting dome in the overdoped regime). In a more recent comprehensive review of ARPES, tunneling, resistivity, NMR and other measurement techniques on $Bi_2Sr_2CaCu_2O_{8+\delta}$, Vishik et al. [387] show a plot of

T* determined by these measurements. From a consideration of the higher temperature data, T* extrapolates to the overdoped end of the superconducting dome as shown in Fig. 21. However, the low temperature determined T* data bend backwards under the superconducting dome (similar to $T_{Structural}$ in Co-doped $BaFe_2As_2$ discussed below for IBS in section 6.1) and approach 0 at p=0.19.

    It is fair to say that ARPES measurements, along with tunneling discussed below, have been at the forefront at trying to unravel both the important pseudogap-related questions: the correct, intrinsic behavior of the pseudogap and the relationship between the pseudogap and the superconducting gap. Although this effort is far from complete, recent STM (discussed below in section 5.2.5.3) work is providing important clues.

    To finish the section on ARPES investigation of the pseudogap in hole-doped cuprates, Fig. 22a shows, in one quadrant of the Brillouin zone, ARPES data for a partially (pseudo-) gapped Fermi surface ("Fermi arc") at T>$T_c$, while Fig. 22b shows a continuous Fermi surface measured at T<$T_c$.

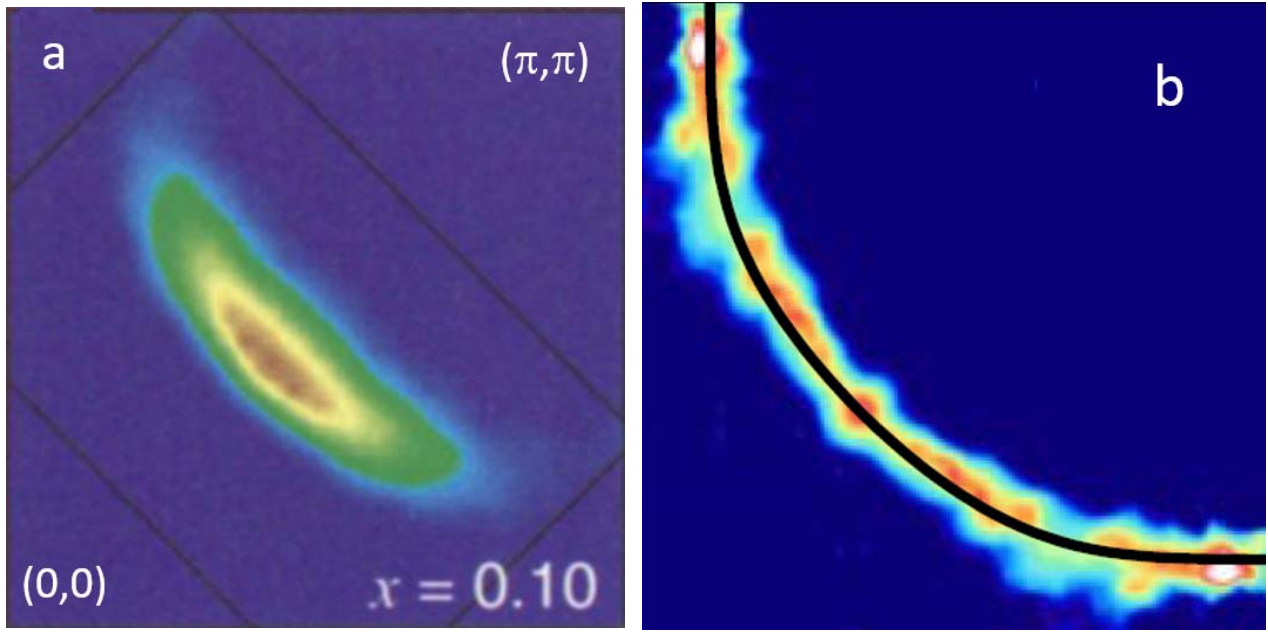

Fig 22: (color online) a.) A graph of the ARPES spectral intensity (color coded from low [blue], through green then yellow and high is represented by brown) in one quadrant of the first Brillouin zone (lower left 0,0 to upper right π,π is the nodal direction) showing the missing portions of the T>T* Fermi surface, where the pseudogap has opened along the anti-nodal direction 0,π to π,0. The sample is $Ca_{1.9}Na_{0.1}CuO_2Cl_2$, $T_c$=13 K, data taken (Shen and Davis [392] at 15 K. The black outline shows the area in k-space where measurements were made. b.) ARPES spectral intensity (from blue [low] to yellow then red and finally white) for $Tl_2Ba_2CuO_{6+\delta}$, with δ adjusted to give $T_c$=30 K, also in one quadrant of the first Brillouin zone showing a complete presence of electron density at the Fermi surface. The quarter circle black line is a tight binding fit of the data. Data were taken (Platé et al. [393]) at 10 K. The lower left corner in both plates is the 0,0 zone origin; the upper right is the π,π point. Fig. 22 a: Reprinted from Materials Today, vol. 11, K. M. Shen and J. C. Davis, p. 14, Copyright (2008), with

permission from Elsevier. Fig. 22b: Reprinted figure with permission from Platé et al., Phys. Rev. Lett. 95 (2005), p. 077001. [393] Copyright (2005) by the American Physical Society.

5.2.5.2. *NMR:* While ARPES provides evidence for the opening of a pseudogap in the charge channel, NMR is sensitive to behavior in the spin channel. Warren et al. [377] found a sharp decrease in the spin lattice relaxation rate $1/T_1$ (called a 'spin' gap) at T*=100 K in the underdoped, 60 K superconductor $YBa_2Cu_3O_{6.7}$. This first indication of a pseudogap was called a 'precursor' effect, which seems quite descriptive as a harbinger of the opening of the full superconducting gap. For a short review of NMR in the cuprates, with a table of T* values for various cuprates determined from the point where the spin relaxation rate drops precipitously upon cooling, see Kotegawa et al. [394]. Timusk and Statt [40], in their Fig. 22, summarize Knight shift and $1/T_1$ data for a broad range of hole doped cuprates, and state that the line through the determined pseudogap T* values merges into the $T_c$ dome "slightly into the overdoped region of the phase diagram." This is essentially consistent with Fig. 14, and also agrees with Knight shift data on eight different hole concentrations in $Y_{0.8}Ca_{0.2}Ba_2Cu_3O_{7-\delta}$ presented by Tallon and Loram [386].

5.2.5.3. *Tunneling Spectroscopy:* Because of the short coherence length in cuprate superconductors (e. g. $\xi$~25 Å in the a-b plane, 2-3 Å in the c-direction, in YBCO – similar in BiSCCO), the need for clean surfaces restricted the application of this technique for a number of years to the easily cleavable 2212 BiSCCO kept under ultra high vacuum conditions. Tao, Lu and Wolf [395] reported a gap in the tunneling conductance in the normal state, T*=110-150 K, in 2212 BiSCCO, $T_c$=85-90 K. Qualitatively, the dI/dV vs V data show a dip in the conductivity at zero bias, with two symmetrically placed conductance peaks at about ±40 mV, for T<$T_c$. In the normal state, the peaks disappear but a reduced dip of the conductance remains. The width of this conductance dip (or the size of the pseudogap) is temperature independent from T* down to $T_c$, unlike the behavior of a real superconducting gap.

The cause of this pseudogap, both in the tunneling conductance and in the other measurement techniques discussed here, has been a question of key theoretical interest. One proposal is that the pseudogap was a precursor (the nomenclature used by Warren et al. in discussing their NMR data) to the full superconductivity that occurs at lower temperature, with 'preformed' (incoherent) Cooper pairs at T* but no phase coherence/no diamagnetic fluctuations above $T_c$ ( see the review by M. Norman [396]). Gomes et al. [397], using STM on several dopings of BiSCCO 2212, find that pairing gaps nucleate in nanoscale regions above $T_c$. For underdoped samples, they find that such pairing occurs well below T* and that the broad dip in the STM-measured conductance starting at T* is a second energy scale unrelated to superconducting pairing. Nernst effect measurements (Wang, Li and Ong [398]) corroborate this point of view, with the onset of the Nernst signal (indicating vortex excitations $\Rightarrow$ superconducting pairing) lying between the superconducting dome and the T* line in both 2212 BiSSCO and in LSCO. As a related piece of information for what the second energy scale could be, Valla et al. [399], using photoemission and STM, find that the pseudogap phase in $La_{2-x}Ba_xCuO_4$, x=1/8 (where static spin and charge orders – 'stripes' – form and $T_c$ is suppressed to T=0) has an energy gap at the Fermi energy consistent with a phase incoherent d-wave superconductor, but that the Cooper pairs form spin-charge ordered structures instead of superconducting pairs.

As seen in Fig. 21, T* determined from tunneling (and other) determinations of the pseudogap from 2008 and before extrapolates to the high end of the superconducting dome. Tallon and Loram [386], similarly to the discussion above (section 5.2.5.1) of the ARPES data, argue for two gaps in the tunneling data, the higher one of which in their view corresponds to the pseudogap and which extrapolates with composition somewhere around the middle of the superconducting dome. More recent STM work tends to argue that T* of the pseudogap more likely joins the superconducting dome just past optimal doping, see, e. g., [400].

In addition, recent work (K. Fujita et al. [401]) has revealed further information: that the cuprate pseudogap phase involves a new charge density state with a d-symmetry form factor. (These charge modulations were first seen by STM in 2002 (Hoffman et al. [402].) Recent work (S. Badoux et al. [403]) suggests, in the hole doped cuprates, that this pseudogap phase is distinct from the charge ordering.
Finding how this fluctuating charge ordering relates to the pseudogap and the superconductivity (see Huxley [404] for a discussion) is clearly one of the future foci in investigating pseudogap physics in the cuprates.

5.2.5.4. *Specific Heat* The specific heat displays evidence for the cuprate pseudogap in several ways. When the doping level is at or above optimal doping, the jump in the specific heat at $T_c$, $\Delta C$, is approximately 1 times the normal state electronic specific heat, as expected (D. Einzel [405]) for a d-wave superconductor. When the doping falls below optimal, this $\Delta C$ starts to fall precipitously (see Fig. 23). Second, the extrapolation of the normal state electronic specific heat from above $T_c$ to T=0, which is defined as $\gamma$ ($\equiv C_{en}(T \to 0)$) also starts to decrease in size (see Fig. 23) below optimal doping, where $\gamma$ is proportional to the electronic density of states at the Fermi energy. Third, the normal state entropy from above $T_c$ should extrapolate to 0, which it does until the doping decreases below optimal in the cuprates. At this point $S_n$ begins to extrapolate to a negative (unphysical) value – indicating a decrease in low energy spectral weight.

Tallon and Loram [386] plot T* determined from the normal state specific heat vs hole doping p for $Y_{0.8}Ca_{0.2}Ba_2Cu_3O_{7-\delta}$. The values go all the way to zero at p=0.19.

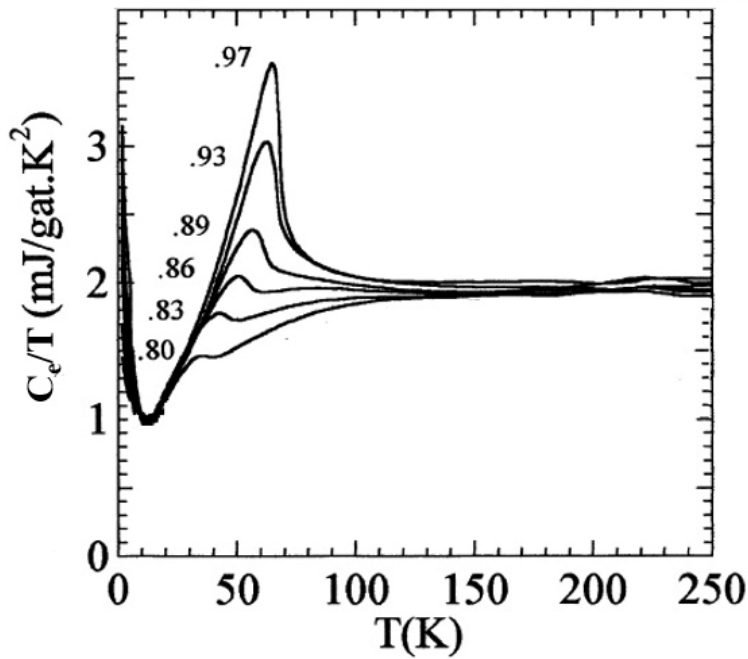

Fig. 23: Normal and superconducting state electronic specific heat, $C_e$, for $YBa_2(Cu_{0.98}Zn_{0.02})_3O_{6+x}$, $0.80 \leq x \leq 0.97$ (J.W. Loram et al [43]). Note how the extrapolation of the normal state, $T>T_c$, $C_e/T$ to $T=0$ decreases for $x=0.83$ and $0.80$ compared to the higher oxygen doped samples. Also, $\Delta C/T_c$ clearly shrinks very rapidly with decreasing x for the underdoped samples. Both properties are consistent with a pseudogap which, as we have seen from the ARPES data, is accompanied by a decrease in the electronic density of states at the Fermi energy. Reprinted from J. Phys. Chem. Solids, vol 62, J. W. Loram et al., p. 59, copyright (2001) [43], with permission from Elsevier.

5.2.5.5. *Optical Measurements* Evidence for a pseudogap in, e. g., $YBa_2Cu_4O_8$ ($T_c$=82 K, naturally underdoped) and in underdoped $YBa_2Cu_3O_{6.6}$ ($T_c$=59 K) is found in the scattering rate and far infrared conductivity along the c-axis direction determined from the optical reflectance (Basov et al. [406]). For example, there is a gaplike depression (sharp decrease from linear with frequency) in $1/\tau$ at around 750 cm$^{-1}$ in the normal state, T=85 K, for $YBa_2Cu_4O_8$ (Timusk and Statt [40]). Tallon and Loram [386] discuss infrared conductivity data for $Y_{0.8}Ca_{0.2}Ba_2Cu_3O_{7-\delta}$ and point out that such data combine two gaps, that from the pseudogap and that from the superconducting gap. At higher hole doping, they say that the IR conductivity data is dominated by the superconducting gap, and thus a plot of T* joins the overdoped side of the superconducting dome and goes to zero at the high doping end of the dome when $T_c \rightarrow 0$.

5.2.5.6. *Pseudogap Summary:* The phenomenon of a pseudogap in the underdoped cuprates is observed by a number of different types of measurements (in both spin – e. g. NMR – and charge – e. g. ARPES – channels) – a far richer set of observations of this phenomenon that in any other class of UcS. See Fig. 21 for a summary of values measured for T* and $T_c$ in a select subset of the hole doped cuprates where T* merges with the superconducting gap *in the strongly overdoped regime.*

As is apparent, some of the above set of data (e. g. NMR) for the pseudogap and the data shown in Fig. 21 are contrary to the T* line extrapolating to zero near p=0.19 as shown in Fig. 14. Measurements that give the Fig. 14 result, implying the possibility of a quantum critical point under the superconducting dome, in addition to the data discussed above, include polarized elastic neutron diffraction in $(Y,Ca)Ba_2Cu_3O_{6+x}$ (B. Fauqué et al. [407]) and $HgBa_2CuO_{4+\delta}$ (Li et al. [408]), polar Kerr effect in YBCO, where T* is found *below* $T_c$ (J. Xia et al. [409]), and

resonant ultrasound spectroscopy in YBCO (Shekhter et al. [410])  For a review of these pseudogap measurements consistent with the plot in Fig. 14, see K. Fujita et al. [385].

To make clear that the question of where T*→0 is still open, there are ARPES data (Kondo et al. [411]) on $(Bi;Pb)_2(Sr;La)_2CuO_{6+\delta}$ (Bi2201) which appear to show conclusively that T* in this cuprate joins the horizontal axis *above* where the superconducting dome goes to zero on the overdoped side.

The proposed theories for explaining the cuprate pseudogap are still under discussion. (for a review, see Scalapino [14])  In any case, the pseudogap is related to the real superconducting gap, having the same d-wave symmetry, however with the difference that the size of the pseudogap is temperature independent.

5.2.6. *Neutron Spin Resonance*

As discussed above for the electron doped cuprates, a narrow-in-energy (resolution limited) resonance that is magnetic in nature has been observed in the superconducting state in the neutron scattering data for the hole doped cuprates.  This resonance - peaked at the antiferromagnetic wavevector - was first discovered in bilayer YBCO, $T_c$=91 K by Rossat Mignot et al. [321].  Such a resonance, where the resonance energy scales with $T_c$, was later reported for 2212 BiSCCO, also bilayer, (Fong et al.[412]), and single layer $Tl_2Ba_2CuO_{6+\delta}$ (He et al. [413]), - but not in single layer $La_{2-x}Sr_xCuO_4$ (Eschrig [29]). (For a discussion of the difference between the neutron scattering data in $La_{2-x}Sr_xCuO_4$ and the other hole doped cuprates, see Birgeneau et al. [414] and Eschrig [29]).   This neutron spin resonance in YBCO, 2212 BiSCCO, and $Tl_2Ba_2CuO_{6+\delta}$ is thought to be associated with the pairing boson – magnetic instead of phononic - for the high temperature superconductivity because the energy of the resonance $\Omega_r$ (see Fig. 21) scales with $T_c$ just as the superconducting gap $E_{sc}$ does.   For discussion of the implications of the neutron spin resonances seen in the hole doped cuprates, see the aforementioned reviews by Eschrig [29] and Birgeneau et al. [414].

5.2.7. *Power Law*

Most experimental characterization of cuprates does not focus on non-exponential temperature dependences of measured quantities as evidence for UcS nearly as much as in the heavy fermions, instead focusing on the properties just discussed above.  However, some data do exist in the literature, and will be discussed here.  For a review, see Tsuei and Kirtley [4].

Penetration depth data on optimally doped YBCO (Hardy et al. [415]) show that $\lambda \sim T$ (indicative of line nodes in a clean d-wave superconductor, Table 1) between 3 and 20 K.  The thermal conductivity of the hole doped cuprates (e. g. for optimally doped and Zn-doped YBCO Taillefer et al. [416]) finds $\kappa/T = a + bT^2$.  Originally the size of 'a' was thought to be universal for the hole doped cuprates, consistent with calculations for a gap with d-wave symmetry.  More recent work (Sun et al. [417]) has shown a breakdown of this universality, stressing the importance of electronic inhomogeneities in understanding $\kappa$.

In some classes of superconductors, the temperature dependence of the NMR $1/T_1$ spin lattice relaxation time deep in the superconducting state can be used (see Table 1 above) to

distinguish point vs line nodes (see discussion for the heavy fermions, where $1/T_1$ behaves typically as $T^3$, or indicative of line nodes). In the cuprates (see also discussion for electron doped PLCCO above), however, the temperature dependences of NMR and NQR measurements are not so simple and are typically (see, e. g., data by Corey et al. [418] on $YBa_2Cu_4O_8$) fit to model calculations over a broad temperature range (see, e. g., the model by Pines and Wróbel [419]) for d-wave superconductivity.

Determinations of power laws in the electronic component of the specific heat in the cuprates are also not of the same utility as for some other classes of UcS. This is because of impurity effects (Schottky anomaly) dominating the low temperature C/T (specific heat divided by temperature) in all but the purest samples of, e. g., YBCO (Revaz et al. [420]). $C_{electronic} \propto \alpha T^2$, with a coefficient $\alpha$ of ~0.2 mJ/molK$^3$ (quite small!), was found in YBCO (Junod et al. [421]) in the low field limit. In the high field, low temperature limit $C \propto TB^{1/2}$ [421] – see theory papers on the specific heat of d-wave superconductors with line nodes by Volovik [422] and Kuebert and Hirschfeld [423]. (We will revisit this subject when discussing the IBS below.)

### 5.2.8. *Breaking of Time Reversal Symmetry: Polar Kerr Effect or μSR measured spontaneous appearance of an internal magnetic field below $T_c$*

Xia et al. [409] report a polar Kerr effect (and breaking of time reveral symmetry) *above* $T_c$ near the pseudogap temperature T* in underdoped YBCO. At optimally doped YBCO, the onset of the polar Kerr effect is below $T_c$. Since the time reveral symmetry is not broken *at* $T_c$, this is not evidence for UcS. Thus, the behavior of the polar Kerr effect (occurring above $T_c$) is different in YBCO than in the heavy fermion superconductors.

### 5.2.9. *Superconductivity is extremely sensitive to impurities, non-magnetic as well as magnetic*

YBCO has been shown to be extremely sensitive to Zn substitution on the Cu site, e. g. the $T_c$ for pure YBCO is suppressed from 92 K to ~50 K for $YBa_2Cu_{2.9}Zn_{0.1}O_{7-\delta}$. (Chien, Wang and Ong [424]) However, even though Zn is <u>a priori</u> a spinless dopant, it has been shown (Bobroff et al. [425]) that Zn (as well as Li) substituted on the Cu site causes local moments on its near neighbor Cu atoms. This is also the case for Al substitution for Cu in $La_{1.85}Sr_{0.15}CuO_4$, where $T_c$ is suppressed at the rate of 23 K/%Al. (Ishida et al. [426]). Thus, just as in the heavy fermion superconductors, it is difficult to achieve non-magnetic doping in the highly correlated cuprates. However, when the $T_c$ suppression from the induced moments is calculated [426], it is almost a factor of 40 smaller than the measured suppression, and the anisotropy of the d-wave order parameter is understood theoretically as the dominant cause of the strong scattering effect on $T_c$.

### 5.2.10. *C(H,theta)*

As will be discussed for IBS, and as has been discussed above for CeCoIn$_5$, angle resolved specific heat in field can graphically display the nodal structure of UcS. Although many measurements for the cuprates give clear evidence for $d_{x^2-y^2}$ pairing symmetry, with the accompanying nodal structure, $C(H,\Theta)$ measurements in the cuprates have been less successful – presumably due to the low temperature magnetic impurities (see section 5.2.7.) above). For

example, measurements by Park and Salamon [427] were unable to resolve the expected 4 fold variation with $\Theta$ in $C(H,\Theta)$ for YBCO.

5.2.11. *Specific heat $\gamma$ varies as $H^{1/2}$ in the Superconducting State:*

As already discussed in section 2.1.15., Moler et al. [81] did the original 'proof' of lines of nodes in the gap (theory by Volovik) in YBCO, showing that $\gamma \propto H^{1/2}$. The problem with that as an indication of UcS is that a field dependence close to the Volovik $H^{1/2}$ prediction can be obtained for superconductors with two fully gapped bands [83]. The original work by Moler et al. was improved upon later in better samples of YBCO by Wang et al. [428], who also used the powerful technique of scaling to convincingly show the presence of line nodes.

In addition to the superconducting properties of the hole-doped cuprates as they pertain to UcS, the question of quantum criticality (section 2.1.2.) is of interest. Due to the high critical fields of the hole-doped cuprates, measurement of non-Fermi liquid behavior in the normal state resistivity down to low temperatures is rare.

5.2.12. *Superconductivity forms out of a non-Fermi liquid normal state, implying that quantum critical fluctuations are involved in the Cooper pairing.*

Cooper et al. [429] found $\rho = \rho_0 + AT^\alpha$, $\alpha$ within 0.1 of 1.0, in $La_{2-x}Sr_xCuO_4$ around the optimal doping at x=0.19 that extends above $T_c$ up to 200 K. Measurements in 48 T to suppress superconductivity were performed in one sample, x=0.23, and found linear-with-T behavior down to below 2 K (lowest temperature of measurement). For various reasons, including that the T-linear behavior of the resistivity extends over a fairly broad composition range (0.18 ≤ x ≤ 0.28), this non-Fermi liquid behavior is not typical of the known behavior (see Stewart [24]) around a quantum critical point. (~✓2) Similar results (R. Daou et al. [430]) in a lower $T_c$ (~20 K) doped sample of $La_{1.6-x}Nd_{0.4}Sr_xCuO_4$ in a 35 T applied field found linear in T resistivity between 2 (lowest temperature of measurement) and 25 K.

5.2.13. *Summary*

Now that the cuprates (electron- and hole-doped) have been (briefly in relation to the extant body of knowledge) discussed, we see that indeed the cuprates offer an important additional set of properties bearing on the overall question of UcS beyond what is seen with heavy fermion superconductors. Most importantly, the broad range of properties in the cuprates that show pseudogap behavior (from ARPES to specific heat to even the polar Kerr effect) makes the rather sparse indications of a pseudogap in just two heavy Fermion compounds ($CeCoIn_5$ and $URu_2Si_2$) seem – in comparison - both qualitatively and quantitatively uncharacteristic of that class of materials. Secondly, the proof of d-wave pairing in the cuprates by phase sensitive measurements – despite $CeCoIn_5$, $URu_2Si_2$, and others heavy fermion compounds having convincing evidence for d-wave symmetry – is lacking for the heavy fermion superconducting class. Thirdly, the large body of knowledge on the isotope effect (or sometimes the lack thereof) in the cuprates provides important insight for the theorists into the superconducting pairing mechanisms that is totally lacking in the heavy fermion superconductors with their lower (≤ 2.3 K in $CeCoIn_5$) $T_c$'s. (Although $PuCoGa_5$, $T_c$=18.5 K, and $PuRhGa_5$, $T_c$=8.7 K, have higher $T_c$

values, their γ values do not really qualify them as 'heavy' Fermion systems. Plus, the problems of substitution using the extremely radioactive Pu isotopes and the attendant $T_c$ decay with time ($\Delta T_c \sim -0.5$ K/month) due to self-irradiation damage effects [431] seem insurmountable obstacles to determining the isotope effect accurately enough.)

In terms of puzzles that remain unsolved in the cuprates, certainly that of the provenance of the pseudogap, about which so much experimentally is known, - together with convincingly explaining the pairing mechanism for the superconductivity – must be rated number one. The many differences between $La_{x-2}(Sr,Ba)_xCuO_4$ and the other hole doped cuprates also certainly need to be further understood.

Table 5: Strong Indications of UcS in the hole-doped cuprates

|  | 1 | 2 | 3 | 4 | 5 | 6 | 7 | 8 | 9 | 11 | 13 | 14 |
|---|---|---|---|---|---|---|---|---|---|---|---|---|
|  | $T_F \ll \Theta_D$ | QC | Spin res | PG | $T^\alpha$ | >1 sc phase | Phase sens. tunneling $\Rightarrow$ non-BCS | TRSB | $T_c$/non-mag imp | $C(H,\Theta)$ $\kappa(H,\Theta)$ | Lack of isotope effect | $\gamma_{resi dual}$ large |
| hole-doped cuprates | ✗ | ~✓ | ✓ | ✓ | ✓ |  | ✓ | ✓ | ~✓ | ✗ | Opt. doped ✓ | ✗ |

## 6. Iron Based Superconductors

### 6.1. *Introduction*

What the community considers to be iron based superconductors (IBS) involves superconductivity in iron 3d electrons, not just superconductivity in a compound that contains iron such as $Th_7Fe_3$, $T_c$=1.8 K discovered in 1961 (Matthias, Compton and Corenzwit [432]) or $Lu_2Fe_3Si_5$ discovered in 1980 ($T_c$=6.1 K) (H. F. Braun [433]). Although the modern IBS are commonly traced back to the discovery by Kamihara et al. [434] in 2008 of $T_c$=26 K in $LaFeAsO_{1-x}F_x$, in fact the first discovery of superconductivity in an iron pnictide was at the less exciting $T_c$=5 K in LaFePO [435]. Both these compounds belong to the so-called '1111' family (see Table 6 below), named after the stoichiometry of the principle constituents. (And, of course, the first example of superconductivity in iron 3d electrons is iron itself under pressure, with $T_c$ at the even lower value of 2 K at 21 GPa (Shimizu et al. [436]).

IBS since their rapid emergence starting in 2008 have continued to bring new interesting physics, with some of the important problems being related to unusual charge and spin ordering (stripes). Like in the cuprates (see Fig. 14), there is often magnetism in the phase diagram in the various families (listed below in Table 6) of IBS. In fact, the IBS may be even more influenced by magnetism and magnetic fluctuations since – unlike in the cuprates - there are families (the

122, the doped 11's, and the 122* (also called 245's)) where magnetism is *coexistent* with superconductivity (see the representative phase diagram for Co-doped 122 BaFe$_2$As$_2$ in Fig. 24.)

There are numerous overviews on both experiment and theory for the IBS, for example see reviews by Johnston [437], Paglione and Greene [438], Stewart [30], Hirschfeld, Mazin and Korshunov [439], Hosono and Kuroki [27], Chang et al. [440], Hosono et al. [348].

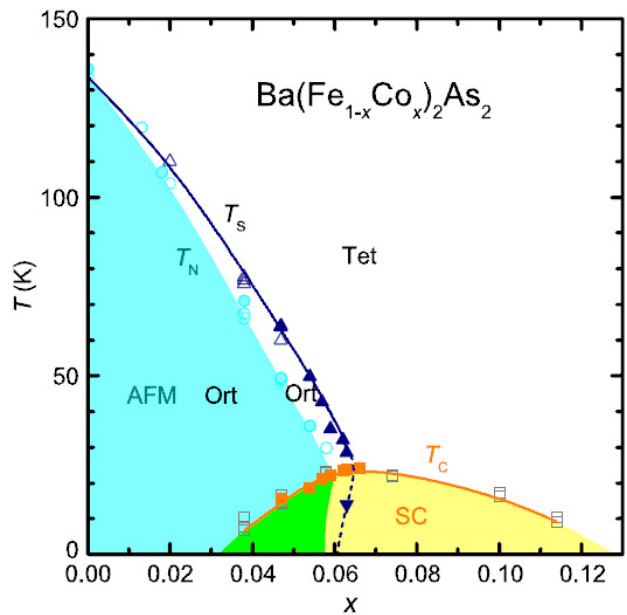

Fig. 24 (color online) Nandi et al. [441]. As Co is doped on the Fe site in BaFe$_2$As$_2$, the structural (tetragonal →orthorhombic) and antiferromagnetic phase transitions at $T_S$ and $T_N$ respectively split apart (both are about 138 K at x=0) and decrease in temperature until they intersect the superconducting dome at around the maximum $T_c$ (currently ≈27 K, J S Kim et al. [442] at 'optimal doping'). The retrograde behavior of $T_S$ with x below $T_c$ is caused [441] by magnetoelastic coupling between nematic magnetic fluctuations and the lattice. (For a discussion of the theory of nematic order in IBS, see ref. [443]. The superconductivity competes with the time-dependent magnetic order (Pratt et al. [444]), which allows the lattice to transform back to tetragonal as the sample at optimal doping is cooled below $T_c$. Reprinted figure with permission from Nandi et al., Phys. Rev. Lett. 104 (2010), p. 057006. [441] Copyright (2010) by the American Physical Society.

In addition, just like in the cuprates there are numerous reviews on just one aspect of IBS, see, e. g., the review on neutron scattering (Dai [445]), the recent collection of topical reviews in Comptes Rendus Physique [446] and the earlier collection of topical reviews in Reports on Progress in Physics [447]. Each review presents a snapshot of the current status at that time of IBS. Unfortunately for the timelessness of the reviews, this status keeps evolving almost yearly. For example, several new families (see Table 6 below) of IBS have appeared since the earliest reviews in 2010, i. e. the 122* (defect structure of the 122 structural family – sometimes called the 245 family) was discovered in early 2011, the 10-3-8 and 10-4-8 families in late 2011 and the 112 family (monoclinic Ca$_{1-x}$La$_x$FeAs$_2$, T$_c$=34 K, Katayama et al. [448]) was discovered in 2013. Even since the middle of 2015, the statement of Hosono and Kuroki [27] that T$_c$ in the 1111 family remains below 10 K except for the "iron oxy-arsenides" has been at least questioned in one paper in early 2016 by an

Table 6: T$_c$ and structure for eight prototypical IBS. Hosono and Kuroki [27] note two additional IBS families achieved by intercalation: NH$_3$ and alkali metal co-intercalation in FeSe,

and 11-derived 1111. Note that $T_c$ for bulk IBS maxed out already at 56 K in less than a year after the discovery of $T_c$=26 K by Kamihara et al. [434]. According to band structure calculations, the five Fe 3-d orbits dominate the electronic density of states at the Fermi energy of this class of superconductor, thus the name 'iron based' is appropriate.

| Compound | Abbreviation/Structure | $T_c$(K) | Discovery | references |
|---|---|---|---|---|
| $LaFeAsO_{1-x}F_x$ | 1111/tetragonal | 26 | 2008 | [434] |
| $Gd_{0.8}Th_{0.2}FeAsO$ | | 56 | 2008 | [449] |
| $Ba_{1-x}K_xFe_2As_2$ | 122/tetragonal | 38 | 2008 | [450] |
| LiFeAs | 111/tetragonal | 18 | 2008 | [451] |
| FeSe | 11/tetragonal | 8 | 2008 | [452] |
| $Sr_2ScO_3FeP$ | 21311,42622/tetragonal | 17 | 2009 | [453] |
| $Sr_2VO_3FeAs$ | | 37.2 | 2009 | [454] |
| $K_{0.8}Fe_{1.6}Se_2$ | 122*,245/tetragonal | 32 | 2011 | [455] |
| $(Ca_{0.85}La_{0.15}FeAs)_{10}Pt_3As_8$ | 10-3-8/triclinic | 34.6 | 2012 | [456] |
| $(CaFeAs)_{10}Pt_4As_8$ | 10-4-8/tetragonal | 38 | 2011 | [457-458] |
| $Ca_{1-x}La_xFeAs_2$ | 112/monoclinic | 34 | 2013 | [448] |
| $Ca_{1-x}La_xFeAs_{2-y}Sb_y$ | | 47 | 2014 | [459] |

unconfirmed report of a new sub-family of the 1111's involving nitrogen, ThFeAsN, $T_c$=30 K (C. Wang et al. [460]). The recent theoretical review by Hirschfeld [461] talks about "many . . . fascinating problems" still to be solved in the field of IBS. Even some problems thought to be solved (just as discussed above for the heavy Fermion $CeCu_2Si_2$) (Kittaka et al. [50]), e. g. the quantum critical point and accompanying divergent effective mass (Walmsley et al. [462]) in $BaFe_2As_{2-x}P_x$ at x=0.3 have been called into question [445] by the recent result that the magnetic order being suppressed there in the phase diagram is weakly first order, i. e. inconsistent with quantum criticality.

In addition to the continuing discoveries in *bulk* IBS, much recent work [463] starting in 2012 has been spent on interfacial superconductivity involving monolayer FeSe ($T_c$=8 K in bulk form) on $SrTiO_3$ (see section 13.3. below). The $T_c$ has been pushed as high as 109 K in single layer FeSe on $SrTiO_3$ (Ge et al. [464]). Although the measurement tools discussed above in section 2 mostly do not function in monolayer films, ARPES (Lee et al. [465]), STM and other measurements have generated conflicting theoretical explanations for the high $T_c$ in FeSe on $SrTiO_3$, including a simple s-wave pairing symmetry (Fan et al. [466]). In addition to the monolayer FeSe interfacial superconductivity work, 2 dimensional techniques have been used on epitaxial 10 nm insulating films of FeSe, where a gating voltage causes superconductivity starting at 35 K. (Hanzawa et al. [467])

Thus, cognizant of the continuing vibrant development of the IBS, the current review focuses on a broad overview of the unconventionality of these superconductors – which definitely need to be discussed in any review of UcS. We present the evidence for UcS in the bulk IBS, which join the HFS and the cuprates as clearly unconventional in their superconductivity. We will organize this section by the measurement techniques implying UcS – as done above for the cuprates – rather than individual materials as done for the HFS.

### 6.2. *Gap Structure*

Presenting a finalized overview of UcS in the IBS is somewhat hampered by the ongoing discussion of the gap structure. Early on, Mazin and co-workers (Mazin et al. [468]) proposed s± symmetry, based on the exchange of antiferromagnetic spin fluctuations (a repulsive interaction) between different parts of the Fermi surface with opposite signs for the gap function $\Delta(k)$ (see the simple depiction in Fig. 25, see also Chubukov [469]) in momentum space. Neutron

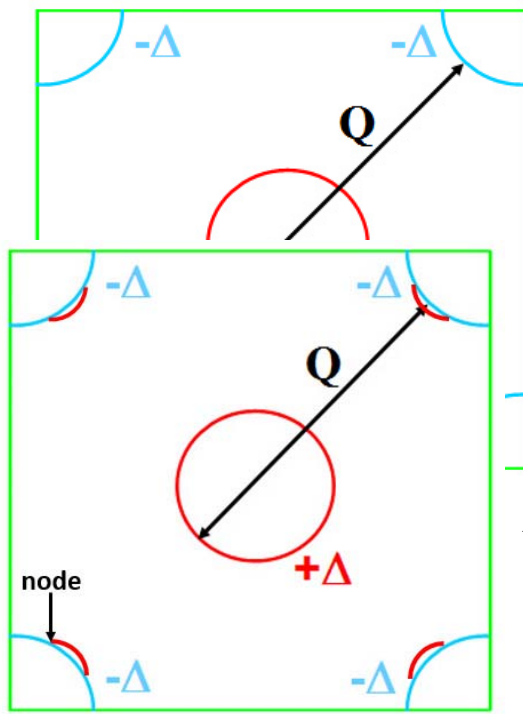

Fig. 25 (color online) On the left, schematic of the nodeless s± gap symmetry proposed for IBS by Mazin et al. [468]. The hole pocket is in red at the center Γ point (0,0) with energy gap +Δ, while the electron pockets are in blue at the corner M (π,π) points with energy gap -Δ. The spin density wave momentum wave vector Q spans the two nested pockets. This repulsive interaction is theorized to mediate attractive pairing in IBS. On the right, s± gap symmetry with nodes (where the sign of the gap changes from – to + at the intersecton of the red and blue lines) on the corner M electron pockets. One of the 8 'accidental' nodes is indicated.

scattering experiments have established the presence of strong magnetic fluctuations in most of the IBS, even in compounds such as FeSe (Q. Wang et al. [470]) which have no long range order.

As shown, the s± gap symmetry on the left in Fig. 25 has no nodes, which are not required by symmetry to be present in an s± superconductor. Nodes (shown in the right diagram) *may* be present in the structure of one or more of the pockets in s±, but come from details of the pairing interaction. Because such nodes in an s± superconductor are not dictated by symmetry, theorists call these nodes 'accidental.'

There are some exceptions where experimental evidence suggests d-wave gap symmetry in the IBS. For example, see Tafti et al. [471] and Reid et al. [472] in heavily overdoped 122 $Ba_{1-x}K_xFe_2As_2$, x>0.8, Tafti et al. and Reid et al. However, a recent work (Kim et al. [473]), with a painstaking analysis of low temperature specific heat for x up to 0.91 in $Ba_{1-x}K_xFe_2As_2$, indicates still the possibility of deep minima – 0.2 to 0.4 K in temperature units - rather than true nodal behavior. In any case, it is probably fair to say that the consensus is that spin-fluctuations mediate the pairing and are the cause of superconductivity, i. e. s± gap structure obtains in the IBS. Some of the first data consistent with this were NMR measurements in all the IBS families which show a lack of a peak (the Hebel-Schlichter coherence peak consistent with simple s-wave) in $1/T_1$ (the spin lattice relaxation time) just below $T_c$. Parker et al. [474] offer a theoretical discussion of why the lack of this coherence peak is consistent with nodeless s± gap structure (Fig. 25).

For spin singlet (s- or d-wave) pairing, the spin susceptibility part of the NMR Knight shift should decrease to zero below $T_c$ in all crystalline directions - thus ruling out triplet (p- or f-wave, i. e. spin triplet pairing.) As discussed in the review by Stewart [30], such a decrease in the NMR Knight shift below $T_c$ has been measured in the IBS families. *However*, it is also the case that the view that all the evidence for sign changes in the gap structure (see discussion below) has another explanation and that s++ pairing mediated by orbital fluctuations (Kontani and Onari [475], Onari and Kontani [476]) is the correct gap structure/pairing mechansim for IBS continues to be espoused by some. As concluded in the review of Hosono and Kuroki [27], "existing theories all confront some kind of difficulties." Thus, we begin our detailed discussion of UcS in the IBS below with two experimental indications of sign changes in the gap (INS and tunneling), and then consider the other 'strongly suggestive evidences' of UcS outlined above in section 2. We perforce leave the resolution of the interminable debate of the pairing mechanism (unconventional in *either* the s± or the s++ case) to the future.

6.2.1. *Inelastic neutron scattering (INS) – consistent with s±*

Inosov, in his review [477], says that neutron spin resonance measurements are not straightforward proof of a sign change in the gap, and the two scenarios of s++ vs s± are "difficult to disentangle." Dai, in his review [445] states that this spin resonance "can arise" from sign reversed quasiparticle excitations in an s± superconductor. Observations of a narrow peak in the neutron scattering intensity in the superconducting state, which is typically taken as *consistent* with a sign change in the gap function somewhere on the Fermi surface, have been made (see Table 4 in Stewart [30] and references therein) in Co and Ni doped $BaFe_2As_2$,

FeSe$_{0.5}$Te$_{0.5}$, Ba$_{0.6}$K$_{0.4}$Fe$_2$As$_2$, LaFeAsO$_{1-x}$F$_x$ and BaFe$_2$(As$_{0.65}$P$_{0.35}$)$_2$, with the observed ratio of the resonance energy, E$_r$, to k$_B$T$_c$ about 5, just as in the cuprates, where this resonance is a sign of the sign change in the d-wave symmetry gap. More recent work (Q. Wang et al. [470]) has found a similar sharp resonance in FeSe (which has no long range magnetic order) below T$_c$ with E$_r$/k$_B$T$_c$ also about 5. As reported by Knolle et al. [478], their INS results for 111 LiFeAs are more "ambiguous," and do not display a true resonance mode – making the choice in LiFeAs between s± and s++ more difficult.

### 6.2.2. *Phase sensitive tunneling/STM/STS* $\Rightarrow$ *s±*

Phase sensitive tunneling was instrumental in proving that cuprates have UcS. A good quality of the junction is important, and difficult to achieve in the IBS. A clever work-around was carried out by Chen et al. [479]. Using a polycrystalline sample of NdFeAsO doped with fluorine, T$_c$=46 K, they formed a superconducting loop with Nb with multiple junctions. The observance of half-integer flux quantization in this loop for some junctions is convincing evidence for a sign change in the superconducting order parameter of the 1111 sample. Since Hicks et al. [480] performed STM measurements on a similar NdFeAsO$_{1-x}$F$_x$, T$_c$=48 K, sample and ruled out a strong d-wave component, Chen et al. argue that the sum of their and Hicks et al.'s results argue for a sign change s± gap structure.

Since the work of Chen et al., other researchers have done tunneling work in further IBS – all concluding that their data favor the s± gap structure: directional point contact Josephson effect in Ba$_{0.6}$K$_{0.4}$Fe$_2$As$_2$ (Tortello et al. [481]); STM in LiFeAs (Chi et al. [482]); STS in Co-doped BaFe$_2$As$_2$ (Teague et al. [483]); point contact Andreev reflection in Co-doped BaFe$_2$As$_2$ (Tortello et al. [484]); STM in FeSe$_{0.4}$Te$_{0.6}$ (Hoffman [485]) and Hanaguri et al. [486]).

### 6.3. *Superconductivity forms out of a non-Fermi liquid normal state, implying that quantum critical fluctuations are involved in the Cooper pairing.*

As apparent from the sample phase diagram in Fig. 24 for Co-doped BaFe$_2$As$_2$, there are examples of IBS where a second order antiferromagnetic transition is suppressed to T=0 by doping, raising the possibility of a quantum critical point.

As discussed first in section 2.1.2, one of the indications of non-Fermi liquid behavior is when the resistivity deviates from Fermi liquid behavior ($\rho = \rho_0 + AT^2$) and varies as, e. g., $\rho = \rho_0 + aT$, i. e. linear in temperature, over a significant temperature range above T$_c$. This was the case in the discussion above of, for example, CeCoIn$_5$. As detailed in Stewart [30], there are a number of IBS where such $\rho = \rho_0 + aT$ evidence for quantum criticality (and therefore for quantum critical fluctuations) has been found: Ba(Fe$_{0.9}$Co$_{0.1}$)$_2$As$_2$ (from T$_c$=22 K up to 100 K) (Ahilan et al. [487]), SrFe$_{1-x}$Ir$_x$As$_2$, x>0.4, up to 300 K (F. Han et al. [488]), BaFe$_2$As$_{1.4}$P$_{0.6}$ above T$_c$=30 K up to 150 K (Kamihara et al. [489]), for FeSe from T$_c$ ~ 8 K up to almost 50 K (Sidorov, Tsvyashchenko and Sadykov [490] and Masaki et al. [491]). Since there is no magnetic order in FeSe, the source of the non-Fermi liquid behavior in this compound may not be from quantum criticality where a second order (usually magnetic) phase diagram is suppressed to T=0.

Another possible evidence for quantum critical behavior is the expected divergence of the electronic effective mass, m*, at the quantum critical point. Corroborating the resistivity data, divergence of m* has also been seen in BaFe$_2$As$_{1.4}$P$_{0.6}$ (Walmsley et al. [462]). However, as

mentioned above in the Introduction to this section (6) on IBS, recent work (including NMR, high resolution x-ray, and neutron scattering) (Hu et al. [492]) indicates that the antiferromagnetic order in BaFe$_2$(As$_{1-x}$P$_x$)$_2$ vanishes in a weakly *first order* fashion, which would rule out quantum criticality from a second order phase transition being suppressed to T=0.

Thus, clearly some of the IBS show non-Fermi liquid behavior, mostly $\rho = \rho_0 + aT$, indicating at least the possibility of quantum critical fluctuations being involved in the (therefore unconventional) superconductivity.

### 6.4. *Pseudogap*

Until now in this review pseudogap behavior, where part of the Fermi surface becomes gapped above T$_c$ at T* for some unknown reason, has been primarily a property of the cuprates. Do the IBS show pseudogap behavior that can be identified as separate from remnant SDW fluctuations?

6.4.1. *ARPES*

One of the most direct measurements of a pseudogap – a gap opening in the electronic density of states at the Fermi energy above T$_c$ – is ARPES. ARPES measurements have been done on several IBS systems to date. (Due to the multiple bands at the Fermi surface in the IBS, not to mention the difficulties of obtaining clean surfaces, ARPES measurements are obviously more difficult/laborious in the IBS than in the cuprates.)

Shimojima et al. [493] report convincing ARPES data in all of the Fermi surface bands ($\varepsilon$, $\delta$, $\gamma$, and $\alpha/\beta$) for optimally doped BaFe$_2$(As$_{1-x}$P$_x$)$_2$, x=0.3, for a gap opening well above T$_c$=30 K (where at this concentration T$_{SDW}$ has already been suppressed below the superconducting dome, Stewart [30]): T*~90 K. For lower x, the pseudogap also forms above T$_S$/T$_N$ (e. g. T*~160 K for x=0.07, while T$_{SDW}$~110 K [30], and for larger x the pseudogap disappears around x=0.6.

On underdoped Ba$_{0.75}$K$_{0.25}$Fe$_2$As$_2$, T$_c$=26 K, Xu et al. [494] report ARPES data on the $\alpha$, $\beta$ and $\gamma$,$\delta$ Fermi surface sheets and show convincing data of a pseudogap that extends up to about 120 K. This is clearly above the SDW transition at this composition which, according to Johrendt and Pöttgen [495] is somewhere below 90 K. Xu et al. discuss extensively the possible insights to be gained from their data, and conclude that both the superconducting gap and the pseudogap could have their origins from antiferromagnetic correlations (the s± model). They further note the similarity between their underdoped data in Ba$_{1-x}$K$_x$Fe$_2$As$_2$ and data in underdoped cuprates, and argue for a possible "unifying picture" of high temperature superconductivity based on antiferromagnetic fluctuations. Contrary wise, Shimojima et al. interpret their pseudogap data in P-doped BaFe$_2$As$_2$ as evidence for the importance of orbital ordering, and that such a mechanism "is not applicable" to the cuprates with their single d$_{x2-y2}$ orbital.

Sato et al.[496] performed photoemission spectroscopy on LaFeO$_{1-x}$F$_x$, $0 \leq x \leq 0.14$, (the prototypical IBS, T$_c$~25 K for $0.07 \leq x \leq 0.11$) and show evidence for a very clear gap opening in the electronic density of states at the Fermi energy. T* is about 115 K/100 K for x=0.07/0.11 respectively, which as shown by red triangles in Fig. 26 is consistent with possible remnant SDW fluctuations (extrapolated from x≤0.045). The size of the pseudogap, $\Delta_{PG}$, was about 17 meV/14.5 meV respectively (195 K/170 K in temperature units.)

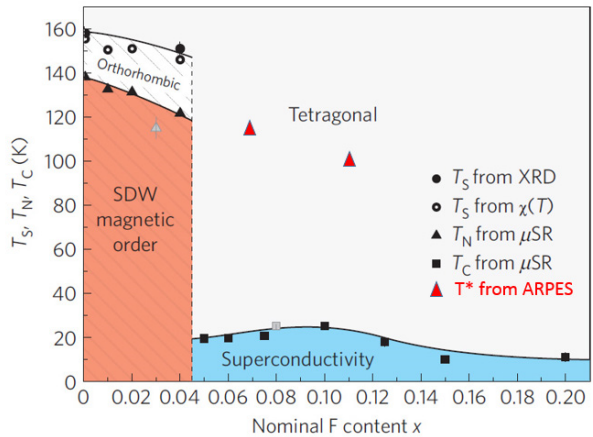

Fig. 26 (color online) (Luetkens et al. [497]) μSR determined phase diagram for LaFeAsO$_{1-x}$F$_x$. Red triangles are from the photoemission spectroscopy data of Sato et al. [496] which found a gap opening at those temperatures and compositions as discussed in the text. Reprinted by permission from Macmillan Publishers Ltd: Nature Materials [497], copyright (2009).

### 6.4.2 *NMR*

Nakai et al. [498] used a fit of their $1/T_1T$ spin lattice relaxation rate data on LaFeO$_{1-x}$F$_x$, x=0.11, to infer $\Delta_{PG} \sim 170$ K, in good agreement with the photoemission work of Sato et al. As pointed out in the $^{75}$As NMR work by Grafe et al. [499] on LaFeAsO$_{0.9}$F$_{0.1}$, measurements of $1/T_1T$ – unlike in the cuprates - do <u>not</u> (at least in the early data) show a fall off below a temperature identified as T* in the IBS. Grafe et al. offer as one speculation that perhaps As is insensitive to the antiferromagnetic fluctuations.

In a doped triclinic 10-3-8 (non-magnetic) IBS, T$_c$=13 K, Surmach et al. [500] using INS find an increase in scattered magnetic intensity (at the spin excitation peak energy) below T*=45 K. Using NMR, and contrary to most IBS results, Surmach et al. find a strong decrease in $1/T_1T$ starting below a T* of 31 K. In the related 10-4-8 IBS, T$_c$=33 K, Kobayashi et al. [501] and also Ikeuchi et al. [502] see no evidence of pseudogap behavior in their NMR data.

Note that the falloff in $1/T_1T$ observed in LaFeO$_{1-x}$F$_x$, $0.045 \leq x \leq 0.075$ – i. e. T$_c$ from 19 to 22 K – by Hammerath et al. [503] was identified by them as due to a glassy spin freezing (see Fig. 26), and not a pseudogap. Further, Ma et al. [504], upon analyzing their NMR data for the 122* IBS Tl$_{0.47}$Rb$_{0.34}$Fe$_{1.63}$Se$_2$, argue that NMR data in general in the IBS do not give evidence for a pseudogap. Thus, in summary, NMR data appear to be of limited utility for determining pseudogap behavior in the IBS, unlike in the cuprates where NMR data were the first indication of a pseudogap.

Based on NMR $1/T_1$ data, Cui et al. [505] report a nearly composition independent pseudogap-like phase in Ca(Fe$_{1-x}$Co$_x$)$_2$As$_2$, $0.023 \leq x \leq 0.059$.

### 6.4.3. *Optical/STS/Resistivity/STM*

Moon et al. [506], using infrared spectroscopy, find a decrease in the scattering rate below about 700 cm$^{-1}$ below 200 K (T*) in BaFe$_2$(As$_{0.67}$P$_{0.33}$)$_2$ and Ba(Fe$_{0.92}$Co$_{0.08}$)$_2$As$_2$ (both optimally doped). The first result in P-doped BaFe$_2$As$_2$ agrees in principle with the ARPES data of Shimojima et al. [493] discussed above, although the values of T* disagree (90 vs 200 K). In contradiction to the second result in Co-doped BaFe$_2$As$_2$, Massee et al. [507], using scanning

tunneling spectroscopy (STS), report no evidence for a pseudogap in Ba(Fe$_{1-x}$Co$_x$)$_2$As$_2$, x=0.04, 0.07, and 0.105.

Kwon et al. [508] measured the optical reflectivity of Ba$_{0.6}$K$_{0.4}$Fe$_2$As$_2$, T$_c$=36 K (where T$_{SDW}$ has been suppressed below the superconducting dome), and make a strong case for pseudogap behavior starting below 100 K. These data and their interpretation are not inconsistent with the ARPES data of Xu et al. [494] discussed above for Ba$_{0.75}$K$_{0.25}$Fe$_2$As$_2$, T$_c$=26 K and T*=120 K, considering (as discussed for the cuprates) that T* should fall with increasing doping. Kwon et al. also discuss the preformed pairing model (proposed by Emery and Kivelson [509]) (see Rice, Yang, and Zhang [39] for a discussion) of the pseudogap to explain their data.

Hess et al. [510] identify features (inflection point in ρ at ~150 K where the resistivity falls below the linear-with-T behavior at higher temperatures) in the resistivity in dopings of LaFeAsO$_{1-x}$F$_x$ (x≥0.05) and SmFeAsO$_{1-x}$F$_x$ (x≥0.06) as "pseudogap-like behavior." These resistivity features are much less distinct (see Fig. 20) than seen in the cuprates. The authors themselves point out the similarity to the SDW transition in the parent compounds (x≤0.04 in LaFeAsO$_{1-x}$F$_x$, see Fig. 26). In fact, according to μSR data (Drew et al. [511]), the SDW transition in SmFeAsO$_{1-x}$F$_x$ exists well into the region of the superconducting dome, which is an exception in the 1111 to the typical behavior shown in Fig. 26 where the magnetism ends before the start of superconductivity. Despite this, the article by Hess et al. is still often cited (see, e.g., Massee et al. [507] as evidence for pseudogap behavior in the IBS.

Hoffman [485] in a review of STM measurements in IBS states that there are no STM data supporting the existence of a gap above T$_c$ in IBS.

**Summary:**

Whether or not the IBS display pseudogap behavior as observed in the cuprates (a gap above T$_c$ due to unknown origins) must be judged at this time as an open question. The data until now are mixed – in some cases discussed above, there is indeed a gap opened at a temperature T* above T$_c$ and not associated with any possible SDW remnant fluctuations. In other cases this gap is identifiable with the SDW transition of the parent compound decreasing in temperature with increasing doping – e. g. in 1111 LaFeAsO$_{1-x}$F$_x$ (Fig. 26).

The measurements of possible pseudogap behavior are still incomplete vis-à-vis the full set of IBS families and the various measurement methods. There seems to be some consensus regarding pseudogap-like features – not associated with remnant SDW fluctuations - in P-doped and K-doped BaFe$_2$As$_2$. As clearly pointed out in the review on STM in the IBS by Hoffman [485], very often – particularly in the 1111 and 122 families – surface sensitive measurements in the IBS can be non-reproducible due to surface states caused by difficulty in obtaining well-cleaved surfaces. Further, many of the standard methods of determining pseudogap behavior - discussed above in section 2 and also in section 5 for the cuprates – seem not to show clear (or any) pseudogap-like behavior in the IBS, e. g. in NMR. Thus, the cuprates must still be judged as the only UcS with overwhelmingly clear pseudogap behavior, although the IBS are certainly the strongest alternative UcS class for pseudogap behavior.

### 6.5. Power law (instead of BCS exponential) temperature dependences in properties like penetration depth, $\lambda \propto T$, or NMR spin lattice relaxation time ($1/T_1$), indicating nodes in the superconducting gap function

$\Delta\lambda(T) \propto T$ is a clear indication of nodal, UcS behavior. This linear temperature dependence has been found (Fletcher et al. [512] and Hicks et al. [513]) in the 1111 superconductor LaFePO ($T_c$=6 K), and in the 122 IBS's KFe$_2$As$_2$ ($T_c$=3.4 K, RRR=1200) (Hashimoto et al. [514]) and BaFe$_2$(As$_{0.7}$P$_{0.3}$)$_2$ ($T_c$=30 K, Hashimoto et al. [515]).

$1/T_1 \propto T^3$ is normally taken as indication of nodal behavior and has been observed in 1111 LaFeAsO$_{0.9}$F$_{0.1}$ ($T_c$=26 K Grafe et al. [499], Nakai et al. [498] and Nakai et al. [516]). However, Chubukov, Efremov, and Eremin [517] argue that the $1/T_1 \sim T^3$ observed in this material is in the presence of sufficient impurities that the superconductor could in fact be nodeless.

### 6.6. $C(H, \Theta)$; angle resolved photoemission spectroscopy

The measurement of the angular dependence of the electronic density of states at the Fermi energy in the superconducting state (determined from specific heat data) in zero and applied magnetic fields, coupled with theory, can be used to indicate nodal, unconventional superconductivity – with the caveat that deep minima can mimic nodal behavior.

Malone et al. [518] measure $C(H, \Theta)$ in BaFe$_2$(As$_{0.67}$P$_{0.33}$)$_2$ ($T_c$=28.8 K) down to 0.5 K and infer line node behavior. Zeng et al. [519] measured $C(H, \Theta)$ in FeSe$_{0.4}$Te$_{0.6}$, $T_c$=14.5 K, and analyzed their data as consistent with a nodeless s± gap structure. Hoffman [485] points out that the STM data of Hanaguri et al. [486] on a crystal of Fe(Se,Te) with similar $T_c$ are also consistent with deep minima but still nodeless.

Zhang et al. [520] measured angle resolved photoemission (ARPES) data on the gap structure of BaFe$_2$(As$_{0.7}$P$_{0.3}$)$_2$ and analyze their data as consistent with nodes (see Fig. 25) in an s-wave pairing symmetry scenario, and inconsistent with d-wave pairing.

### 6.7. Time reversal symmetry breaking (TRSB)

TRSB has not been observed in IBS, although the work of Tafti et al. [471] suggests a transformation from d-wave to s± in KFe$_2$As$_2$ must proceed as a function of pressure through a TRSB s+id state. However, there is significant theoretical work suggesting its possibility – perhaps in Ba$_{1-x}$K$_x$Fe$_2$As$_2$ at strong overdoping. A short summary of these theory works would include Lee, Zhang and Wu [521], Platt et al. [522], and Håkansson, Löfwander and Fogelström [523]. The idea involves competition between s± and d$_{x2-y2}$ gap symmetries as a function of doping (or pressure), with the possibility that an s+id (or s+is) state could form that would have TRSB. For an overview of these arguments, see Hirschfeld [461].

### 6.8. Response to non-magnetic impurities

Such impurities are pair breaking only if they scatter electrons between parts of the Fermi surface with different signs. Before the advent of IBS and their proposed s± gap structure, the belief was that a strong influence of non-magnetic impurities on $T_c$ implied d-wave superconductivity.

The use of non-magnetic impurities as a way to distinguish s++ from s± gap symmetry has been thoroughly discussed in Hirschfeld's review [461]. The point raised is that claims of whether a putative non-magnetic impurity causes a 'slow' ($\Rightarrow$no gap sign change) or 'fast' ($\Rightarrow$gap sign change) depression of $T_c$ are "mostly not relevant." What is needed [461] to arrive at conclusions of s++ vs s± in the IBS, with their multiband character, is to examine $T_c$ suppression <u>together with</u> the accompanying change in the residual resistivity, $\Delta\rho_0$, which is proportional to the amount of pair breaking. The case to date where this has been done is the injection of non-magnetic (as checked (R. Prozorov et al. [524])) by low temperature measurements on $\Delta\lambda(T)$) impurities into Ba(Fe$_{0.76}$Ru$_{0.24}$)$_2$As$_2$ ($T_c$=17.8 K) via electron irradiation. Such irradiation, which creates local defects, avoids the problem of chemical impurities which can have the additional effect of doping carriers at the Fermi energy. Prozorov et al. conclude that their $T_c$ suppression vs $\Delta\rho_0$ is "fully compatible" with s± pairing.

### 6.9. *Isotope effect*

In the introductory section 2 above, we discussed the fact that there is an Fe isotope effect ($T_c \propto M^{-0.35}$), $^{54}$Fe replacing $^{56}$Fe, seen both in IBS (SmFeAsO$_{0.85}$F$_{0.15}$, $T_c$=41 K, and in Ba$_{0.6}$K$_{0.4}$Fe$_2$As$_2$, $T_c$=38 K, Liu et al. [525], with however no O isotope effect, $^{18}$O replacing $^{16}$O, in either compound. Due to the site selectivity of the observed isotope effect, this is interpreted to mean that in the IBS phonon modes than involve the Fe (a magnetoelastic effect) affect the spin fluctuations (theorized to provide the pairing mechanism) and therefore the superconductivity. Thus, the measured Fe isotope effect in the IBS is not an indication of conventional superconductivity.

### 6.10. *Specific heat $\gamma$ varies as $H^{1/2}$ in the Superconducting State*

In addition to the C(H, $\Theta$) measurements (Malone et al. [518]) on P-doped BaFe$_2$As$_2$ discussed in section 6.6., where line node behavior was inferred, Wang et al. [526] measured the specific heat to 35 T on BaFe$_2$(As$_{0.7}$P$_{0.3}$)$_2$. Their low temperature $\gamma$ data varied as $H^{1/2}$ only up to ~ 4 T, with linear behavior at higher fields. They analyzed their data in a two band model assuming nodes or deep minima on the small mass electron band, to explain the crossover from Volovik $\gamma \propto H^{1/2}$ behavior to a higher field, linear (non-nodal) behavior already at 4 T.

### 6.11. *Summary*

IBS are clearly unconventional. There are some data that strongly support the s± gap structure and exchange of antiferromagnetic spin fluctuations model first proposed for the IBS by Mazin et al. [468], including phase sensitive tunneling, STM, STS, and the electron-irradiation creation of non-magnetic impurities and their effect on $T_c$ and $\Delta\rho_0$. There are some data that support d-wave pairing in select materials, e. g. heavily overdoped Ba$_{1-x}$K$_x$Fe$_2$As$_2$. There are data which have been called inconclusive vis-à-vis distinguishing s++ and s± gap structure, including some INS (Inosov [477]). As proof of UcS of whatever gap structure (other than BCS simple s-wave), there is good agreement between ARPES and optical data for a pseudogap in near optimally P-doped and K-doped BaFe$_2$As$_2$, there is strong evidence for nodal behavior from penetration

depth, C(H, Θ), and ARPES data, and there is at least indicative evidence for a connection between magnetism (antiferromagnetic spin fluctuations) and superconductivity.

Table 7: Strong Indications of UcS in the iron based superconductors

|  | 1 | 2 | 3 | 4 | 5 | 6 | 7 | 8 | 9 | 11 | 13 | 14 |
|---|---|---|---|---|---|---|---|---|---|---|---|---|
|  | $T_F \ll \Theta_D$ | QC | Spin res | PG | $T^\alpha$ | >1 sc phase | Phase sens. tunneling $\Rightarrow$ non-BCS | TRSB | $T_c$/non-mag imp | $C(H,\Theta)$ $\kappa(H,\Theta)$ | Lack of isotope effect | $\gamma_{residual}$ large |
| IBS | ☒ | ✓ | ✓ | ~✓ | ✓ |  | ✓ |  | ~✓ | ✓ | ✓ | ~✓ |

## 7. Sr$_2$RuO$_4$

Sr$_2$RuO$_4$, with the same structure as the original 214 cuprate superconductor discovered by Bednorz and Mueller, was discovered to be superconducting, $T_c$=0.93 K, by Maeno et al. [527] in 1994 while searching for new high $T_c$ perovskite-structure oxides. As sample quality has improved, $T_c$ has reached 1.48 K while $\Delta C/\gamma T_c$ has only reached 0.74 (NishiZaki et al. [528]) in a sample with a low residual resistivity, $\rho_0$, of 0.1 µΩ-cm. Sr$_2$RuO$_4$ is an excellent example of an UcS being sensitive to non-magnetic defects (✓9). Even samples with a respectable RRR value of 240, have depressed $T_c$ (1.2 K) and $\Delta C/\gamma T_c$ (0.43) values, as well as C/T (T→0) residual values in the superconducting state equal to half of the normal state value (vs $\gamma_{res} \sim 0$ in the NishiZaki et al. sample). Samples with a residual resistivity value above 1 µΩ-cm are not superconducting at all (Mackenzie et al. [529]), while $T_c$ rises monotonically for $\rho_0$ below this value.

   Already less than a year after the discovery by Maeno et al., Rice and Sigrist [530] had suggested that Sr$_2$CuO$_4$ might be a triplet (p-wave) superconductor in analogy to $^3$He and suggested NMR measurements of the Knight shift as a way to check their prediction. (See also G. Baskaram [531].) Ishida et al. [532] measured the Knight shift and indeed found no change in the spin susceptibitlity when cooling through $T_c$, leading to the definite conclusion that Sr$_2$RuO$_4$ is a spin triplet superconductor. More measurements supporting this point of view will be discussed below. Before these experimental evidences for UcS in Sr$_2$RuO$_4$ are discussed, a short discussion of the theory of p-wave superconductivity in Sr$_2$RuO$_4$ is in order. (A more complete discussion may be found in the reviews by Mackenzie and Maeno [533] and by Sigrist and Ueda [3]. See also Balian and Werthamer [58].).

   In a p-wave ($\ell$=1) superconductor, the spatial part of the wave function is antisymmetric and thus the spin part must be a symmetric triplet state, with the spin dependence described by three complex functions. (Balian and Werthamer) The spin dependence of the pairing is typically given in a 2 x 2 matrix for the momentum dependent gap function $\Delta(\mathbf{k})$. However, Balian and Werthamer proposed the use of a three component complex vector $\mathbf{d}(\mathbf{k})$ for triplet pairing. Discussing now specifically (Liu and Mao [534]) tetragonal Sr$_2$RuO$_4$ believed to be in

the spin triplet state, there are five possible spin triplet pairing states (five possible $\mathbf{d(k)}$) allowed. As we will see after further discussion of the experimental data (in particular the time reversal symmetry breaking, section 7.6) below, the preferred one of these pairing states for $Sr_2RuO_4$ is the $E_u$ ($\Gamma_5^-$), with $\mathbf{d(k)} = \mathbf{z}(k_x \pm ik_y)$ – the only one of the five with $\mathbf{d(k)}$ along the c-axis. The analog with the same $\mathbf{d(k)}$ in the other well known triplet pairing material, superfluid $^3$He, is the A-phase (ABM state).

The $\Gamma_5^-$ pairing state in a single band superconductor (which $Sr_2RuO_4$ – just like the IBS – is not) would be <u>a priori</u> fully gapped, with (Liu and Mao [534]) gapless chiral edge states. The $\Gamma_5^-$ pairing state has the possibility of domains ($k_x + ik_y$ or $k_x - ik_y$) and chiral surface currents. Using scanning SQUID microscopies, Kirtley et al. [535] saw no evidence of the expected spontaneously generated surface supercurrents, with their limit of detection a factor of 100 smaller than the predicted magnitude. This important inconsistency with $Sr_2RuO_4$ being in the $\Gamma_5^-$ pairing state has recently been sidelined by the theoretical work of Scaffidi and Simon [536] who, in a three band, weak coupling model, predict a large reduction in surface currents from previous predictions. (This is a demonstration of the importance of using multi-band models where appropriate, rather than simplifying down to a one band model which can lead to incorrect conclusions. This was also shown to be important in the IBS in numerous cases, as discussed above in section 6.)

$Sr_2RuO_4$ has been reviewed numerous times, the most recent review was by Liu and Mao [534] in 2015. Some other important reviews, in addition to those mentioned above, include those by Maeno et al. [537], Maeno, Rice and Sigrist [538], and Bergemann et al. [539]. We present here, as we did for the individual heavy fermion compounds, a brief synopsis of the evidence for UcS in $Sr_2RuO_4$, with discussion of where controversy/measurement difficulties still exist. Also discussed is how $Sr_2RuO_4$ is similar/different to the other UcS discussed so far.

### 7.1. *Superconductivity forms out of a non-Fermi liquid normal state, implying that quantum critical fluctuations are involved in the Cooper pairing.*

This is not the case for $Sr_2RuO_4$ – both the in-plane and c-axis resistivity behave as a Fermi liquid, $\rho = \rho_0 + AT^2$ up to about 25 K (Maeno et al. [540]). The resistivity is strongly two dimensional (similar to the layered cuprates) – in the sample in Maeno et al. [540] $A_c/A_{ab} > 500$.

### 7.2. *Neutron spin resonance peak develops in the superconducting state*

Such a peak is an indication of a sign change in the superconducting energy gap $\Delta$ on different parts of the Fermi surface, as would be consistent with d-wave or s± pairing. Since the (<u>a priori</u> nodeless) energy gap in p-wave $Sr_2RuO_4$ varies as $\Delta_0 (k_x^2 + k_y^2)^{1/2}$, where $k_x$ and $k_y$ are wave vectors on the Fermi surface (Mackenzie and Maeno [533]), there is no neutron spin resonance peak in $Sr_2RuO_4$.

### 7.3. *Pseudogap*

$Sr_2RuO_4$ is not believed to exhibit pseudogap behavior in any of its properties. In the interests of completeness, it should be mentioned that Jin, Liu and Lichtenberg [541] claim that their magnetoresistance data imply a pseudogap, but only over a small portion of the Fermi surface making it hard to observe.

### 7.4. *Power laws*

Power laws in various measured parameters like specific heat, spin-lattice relaxation time $1/T_1$, or penetration depth $\lambda$ - as discussed above in section 2.1.5. and Table 1 – can, if measured over a broad range of temperature and down to at least 0.1 $T_c$, indicate nodal behavior and assist in identifying the pairing symmetry in an UcS. Such power laws have been found in $Sr_2RuO_4$ and offer some evidence for nodal behavior. The question of whether $Sr_2RuO_4$ has nodal behavior remains controversial.

#### 7.4.1. *Specific heat*

Although C/T of $Sr_2RuO_4$ varies (NishiZaki, Maeno and Mao [542]) as T below 0.5 K (consistent with line nodes, Table 1), a fit [533] of the data up to $T_c$ over the whole temperature range to a full line node model (described in [543]) shows rather poor agreement.

#### 7.4.2. *Thermal conductivity*

Izawa et al. [544] measured the in-plane thermal conductivity in the [110] and [100] directions as a function of an applied magnetic field rotated in the ab plane. They concluded that what little oscillation as a function of angle that they observed in their $\kappa_{ab}$ ($\Theta$, H) data was not due to the effect on the density of states due to line nodes along the Fermi surface cylinder (from the 2 dimensional character of $Sr_2RuO_4$ inferred from the strong anisotropy in the resistivity). Similar work by Tanatar et al. [545] also found very little anisotropy in their $\kappa_{ab}$ ($\Theta$, H) data. Thus, in order to explain the evident presence of quasiparticles in the gap implied by the measured $\kappa \propto T^2$ (Table 1, line nodes) behavior, both sets of authors suggest line nodes *perpendicular* to the Fermi surface cylinder (horizontal nodes).

#### 7.4.3. *NMR*

The absence of the Hebel Schlichter peak in $1/T_1$ (Ishida et al. [532]) just below $T_c$ is already evidence of non s-wave superconductivity in $Sr_2RuO_4$. Using high quality samples (with low impurity pair breaking) of $Sr_2RuO_4$, Ishida et al. [546] measured $1/T_1$ down to 0.09 K and found $1/T_1 \propto T^3$ between 0.15 and 1 K – consistent with "line-node-like models." However, as discussed above for the heavy fermion superconductors, such a $T^3$ dependence (taken as implying line nodes, see Table 1) for the spin lattice relaxation rate is only indicative as carefully phrased by Ishida et al. [546].

#### 7.4.4. *Penetration depth*

Bonalde et al. [547] find $\Delta\lambda(T) \propto T^2$ (not the line node linear-in-T behavior from table 1) from 0.04 to 0.8 K in good quality ($T_c$ between 1.39 and 1.44 K) samples. They propose that this temperature dependence is indicative of line nodes with non-local electrodynamic effects (Kosztin and Leggett [548]), arguing *against* all the proposed (Sigrist and Ueda [3]) p-wave states that are suggested by the other data, and in particular the $\Gamma_5^-$ pairing state.

### 7.4.5. *Ultrasonic attenuation*

Lupien et al.'s [549] ultrasonic attenuation data follow a power law ($T^{1.4}$ to $T^{1.8}$ – depending on mode) from $T_c/30$ up to 0.7 K – clear sign of nodal (or very deep minima) behavior. However, just like the other power law measurements, these data are not consistent with these nodes being vertical.

### 7.4.6. *Summary*

Liu and Mao, in their recent review [534] of Sr$_2$RuO$_4$, state that whether or not the order parameter $\Delta(\mathbf{k})$ has nodes remains unresolved. If one - using the presence of the three bands at the Fermi energy in Sr$_2$RuO$_4$ – argues that the $\Gamma_5^-$ pairing state can have an admixture of another state which brings in $k_z$ dependence (for example the triplet f-wave pairing state $\mathbf{d}(\mathbf{k}) = \mathbf{z}(k_x \pm ik_y)\cos k_z c$ – see Hasegawa, Machida and Ozaki [550]), then one can have horizontal nodes [534]. Or, indeed, the multiple bands (just as discussed above for the IBS for the field dependence of the specific heat $\gamma$) can mimic nodal behavior without there being any nodes.

### 7.5. *Josephson tunneling/STM/Phase sensitive measurement that shows non-BCS order parameter symmetry (e. g. triplet p-wave in Sr$_2$RuO$_4$)*

Josephson tunneling work by Jin et al. [551] between good quality ($T_c$=1.45 K) single crystals of Sr$_2$RuO$_4$ and a BCS superconductor (In) were consistent with the $\Gamma_5^-$ pairing state, but did not rule out d-wave pairing. Similar work by Liu et al. [552] found their Josephson tunneling results consistent with either the $\Gamma_5^-$ pairing state or with the f-wave pairing state with admixture of $k_z$ and horizontal nodes proposed by Hasegawa et al. [550] mentioned above. Andreev reflection on lower quality single crystals ($T_c$=1.02 K) of Sr$_2$RuO$_4$ (F. Laube et al. [553]) was interpreted to imply triplet pairing.
    Firmo et al. [554], using STM measurements of, conclude that Sr$_2$RuO$_4$ is a magnetically mediated pairing, odd parity (p- or f-wave) superconductor with either nodes or deep minima.
    Finally, Nelson et al. [555] succeeded in performing a phase sensitive tunneling measurement (joining, as discussed above, the cuprates, section 5.2.3., and the IBS, section 6.2.2.) Their results proved conclusively that the pairing state in Sr$_2$RuO$_4$ is triplet (either p- or f-wave.) A nice overview of these measurements in given in Liu [556]. At the current time, the question of $\mathbf{d}(\mathbf{k}) = \mathbf{z}(k_x \pm ik_y)\cos k_z c$ vs the $\Gamma_5^-$ pairing state (horizontal node f-wave vs vertical node p-wave) appears to be still open. The lack of the ability to make superconducting thin films of Sr$_2$RuO$_4$ (due to the defect sensitivity of the superconductivity) at this time hampers the effort to distinguish between these two pairing states.

### 7.6. *Breaking of Time Reversal Symmetry: Polar Kerr Effect or μSR measured spontaneous appearance of an internal magnetic field below $T_c$*

Luke et al. [557] performed muon spin relaxation measurements on high quality ($T_c$=1.46 K) single crystal Sr$_2$RuO$_4$ and found an internal magnetic field appearing upon cooling below $T_c$. This implies that the superconducting state in Sr$_2$RuO$_4$ breaks time reversal symmetry. (✓8). Also, Kerr effect measurements (Xia et al. [558] and Kapitulnik et al. [12]) in Sr$_2$RuO$_4$ observed definite evidence of time reversal symmetry breaking.

### 7.7. $C(H,\Theta)/\kappa(\Theta,H)$

Deguchi et al. [559] measured the field and orientation dependence of a high quality ($T_c$=1.48 K) single crystal of $Sr_2RuO_4$. Their analysis of the low temperature, H<1.2 T (below the field inversion (Vorontsov and Vekhter [77]) as discussed in 2.1.11.) data was found to be in good agreement with the p-wave order parameter $d(k) = \hat{z} \Delta_0 (sink_x + isink_y)$, which indicates the absence of nodes but a very small gap.

Hassinger et al. [560] measured the thermal conductivity of a single crystal of $Sr_2RuO_4$ ($T_c$=1.2 K, residual resistivity $\rho_0$=0.24 $\mu\Omega$-cm) in fields down to $H_{c2}/100$ and temperatures down to $T_c/30$ both in-plane and along the c-axis. Their analysis indicates vertical line nodes on all three sheets of the Fermi surface in $Sr_2RuO_4$.

### 7.8. *Summary*

**Table 8: Strongly Suggestive Evidence for UcS for $Sr_2RuO_4$.** The symbol '✘' means that that property was specifically looked for but not found

|  | 1 | 2 | 3 | 4 | 5 | 6 | 7 | 8 | 9 | 11 | 14 |
|---|---|---|---|---|---|---|---|---|---|---|---|
|  | $T_F \ll \Theta_D$ | QC | Spin res | PG | $T^\alpha$ | >1 sc phase | tunneling ⇒ non-BCS | TRSB | $T_c$/non-mag imp | $C(H,\Theta)$ $\kappa(H,\Theta)$ | $\gamma_{residual}$ large |
| $Sr_2RuO_4$ | ✘ | ✘ | ✘ | ✘ | ? | ✘ | ✓ | ✓ | ✓ | ✓ | ✘ |

$Sr_2RuO_4$ displays several clear signs of UcS: strong sensitivity of $T_c$ to non-magnetic impurities, breaking of time reversal symmetry, and phase sensitive tunneling proving triplet superconductivity. Evidence for non-exponential power laws in various measurements (e. g. penetration depth, $1/T_1$) are consistent with nodal behavior, which remains controversial. The ability to prepare thin films of $Sr_2RuO_4$ (until now impossible due to the extreme sample dependence on impurities) would be helpful for furthering understanding.

Although the theory of the pairing state is well advanced, the actual mechanism that causes this pairing is at this point not well understood. In the IBS discussed above, there seems to be a preference (although no proof) for considering the exchange of antiferromagnetic spin fluctuations as the pairing mechanism. Although this is also one of the proposed pairing mechanisms in $Sr_2RuO_4$ (Kuwabara and Ogata [561]), there are other proposals:

1.) Agterberg, Rice and Sigrist [562], before higher quality samples were prepared and characterized, proposed a model in which the primary superconductivity occurred in just one of the Fermi surface sheets ($Sr_2RuO_4$ is strongly two dimensional), the $\gamma$ sheet with Ru $d_{xy}$ orbital character. This orbital dependent form of superconductivity explained various results, including the large finite $\gamma_{res}$ in the low temperature specific heat. With the improvement in the sample quality, and $\gamma_{res}$ approaching zero, this theory required modification. For a review of the Agterberg, Rice and Sigrist orbital dependent model for the microscopic mechanism in $Sr_2RuO_4$, see Maeno, Rice and Sigrist [538].

2.) Theorists soon after Maeno et al.'s discovery in 1994 of superconductivity in $Sr_2RuO_4$ have speculated that $Sr_2RuO_4$ is near a ferromagnetic instability (e. g. Rice and Sigrist

[530]), partially at least since the related compound $SrRuO_3$ is ferromagnetic. Such early speculation of parallel spin correlations helped fuel the consideration of the parallel spin, p-wave ($\ell=1$) pairing state, which as we have seen in this section is believed to be the case for $Sr_2RuO_4$. Although there is not a lot of experimental evidence for ferromagnetic spin fluctuations, Yoshida et al. [563] argued that their resistivity under pressure data support "in a qualitative sense" 2D ferromagnetic spin correlations in $Sr_2RuO_4$ and also note that the NMR Knight shift data of Ishida et al. [564] are consistent with ferromagnetic spin correlations. Mazin and Singh [565] calculated $T_c$ in $Sr_2RuO_4$ assuming the pairing mechanism is due to exchange of paramagnons. Monthoux and Lonzarich [566], using a one Fermi sheet model, also calculated $T_c$ with exchange of ferromagnetic fluctuations as the pairing 'glue.'

## 8. Non-centro-symmetric superconductors

In UcS, one or more symmetries in addition to U(1), the one dimensional global gauge symmetry, are broken at the onset of superconductivity. Breaking of time reversal symmetry (TRS) is not infrequently seen in UcS as discussed in several places already in this review starting in section 2.1.8., e. g. for the transition to the B-phase (cf. Figs. 4 and 10) in $UPt_3$, $PrOs_4Sb_{12}$, $PrPt_4Ge_{12}$, $Sr_2RuO_4$, $URu_2Si_2$, and the cuprates.

    The lack of inversion symmetry ("non-centro-symmetric") in superconductors is another important breaking of symmetry, which leads to non-s-wave parity pairing to some extent (enhanced by the electronic correlation strength) being mixed in with conventional s-wave pairing. (For additional theoretical discussion of this class of UcS, see section 2.1.10. above.) This class of UcS began with the discovery in 2004 of superconductivity ($T_c$=0.75 K) in non-centrosymmetric $CePt_3Si$ (Bauer et al. [567], which has $\gamma$=390 mJ/molK$^2$ (i. e. a strongly correlated system). (The fact that $CePt_3Si$ is also antiferromagnetic below 2.2 K is less important for the properties.) The field of non-centrosymmetric superconductivity has been recently reviewed (Bauer and Sigrist [568] and Kneidinger et al. [67]). Although, as we will discuss, there have been many non-centrosymmetric superconductors discovered stretching back to the 1960's, most of them display conventional superconductivity, with only a handful with UcS properties.

    Based on a result of Anderson in 1984 [120], it was commonly believed at the time of discovery of $CePt_3Si$ that a superconductor without a center of inversion symmetry would *not* exhibit spin triplet pairing. Theoretically, Frigeri et al. [569] showed that triplet superconductivity is not necessarily excluded in non-centrosymmetric systems, which was an important result considering how the upper critical field in the recently discovered $CePt_3Si$ so exceeded (Bauer et al. [567]) (~factor 4) the Pauli paramagnetic limit, $H_P$ (=1.8 $kT_c/\mu_B$) – usually taken as a sign of possible triplet pairing since an applied magnetic field does not break Cooper pairs with parallel spins. Kneidinger et al. [67] discuss how anomalous spin fluctuations, which tend to cause triplet pairing, occur in non-centrosymmetric materials (original theory from Takimoto and Thalmeier [570]). Such spin fluctuations are enhanced by strong electronic correlations, but vanish in the weakly correlated limit. Thus, most (the weakly correlated, low $\gamma$ examples) non-centrosymmetric superconductors display conventional, phonon-coupled s-wave pairing symmetry superconductivity, while the (relatively few) more strongly correlated examples like $CePt_3Si$, $CeRhSi_3$, and $CeIrSi_3$ (the latter two are superconducting only under pressure) display mixed pairing symmetry behavior and definite UcS. Another route to enhancing the anomalous spin fluctuations that lead to triplet pairing is asymmetric spin orbit

correlations (ASOC) found in compounds like $ZrRe_6$ with heavier elements. These ASOC lift the spin degeneracy of the bands; these spin split bands can then have either spin singlet or triplet pairing. (Kneidinger et al. [67]).

We review the properties of the prototype material, $CePt_3Si$, and summarize properties for two strongly correlated, believed to be UcS, examples – $CeRhSi_3$ and $CeIrSi_3$ - (which require pressure to superconduct) and two low correlation, ASOC induced examples. Another strongly correlated non-centrosymmetric material, $CeCoGe_3$, (not discussed here) is an antiferromagnet that superconducts ($T_c$ = 0.72 K) under pressure (5.6 GPa) with a high (~7 T, or five times $H_P$) $H_{c2}$, taken as consistent with triplet superconductivity. (Settai et al. [571] and Kneidinger et al. [67])

### 8.1. *CePt$_3$Si*

Yogi et al. [572] measured the Knight shift in $CePt_3Si$ and found no change in the spin susceptibitlity when cooling through $T_c$, implying that $CePt_3Si$ is a spin triplet superconductor like $Sr_2RuO_4$. (A more circumstantial indicator of triplet pairing is the measurement of the upper critical magnetic field, $H_{c2}$, which exceeds the paramagnetic limit. (Bauer et al. [567]) as discussed above.) Yogi et al. also surprisingly found in $CePt_3Si$ a peak (presumably the Hebel-Schlichter coherence peak consistent with simple BCS s-wave) in $1/T_1$ (the spin lattice relaxation time) just below $T_c$. Yogi et al. note that the observed peak is "much smaller" than in a conventional BCS superconductor.

It may be that this minor indication of BCS behavior is just simply consistent with the predicted mixture of pairing symmetries (a two component order parameter), and the multi-band nature of $CePt_3Si$. Mukuda et al. [573] offer another interpretration for the Yogi et al. small size coherence peak below $T_c$ in their $1/T_1$ data. In the Mukuda et al. $1/T_1$ measurements, they proposed two kinds of physical domains in their single crystal (which was crushed into coarse powder for the NMR measurments), whose NMR relaxation signals they can separate in their data analysis. (They quantify the quality of their single crystal sample by the small residual γ of 34 mJ/molK$^2$, < 10% of $\gamma_n$, in the superconducting state.) In the ordered domains (in one sample, 70%), Mukuda et al. propose that coherent f-electron bands are responsible for spin triplet behavior, with no peak in $1/T_1$ below $T_c$=0.35 K. In disordered domains, non-f-electron bands present in $LaPt_3Si$ are posited to be responsible for conventional s-wave superconductivity below $T_c$=0.6 K, and the Mukuda et al. $1/T_1$ data from these disordered domains show a small peak below $T_c$. This explanation helps clarify the rather broad superconducting transition width in the discovery paper [567], where $\Delta T_c$ in the bulk specific heat of $CePt_3Si$ is >20% of $T_c$.

In terms of power law measurement that imply UcS in $CePt_3Si$, Mukuda et al. [573] analyze their NMR $1/T_1$ data in terms of a line node gap. Bonalde et al.[574] find that the penetration depth $\lambda \propto T$ in $CePt_3Si$ in the low temperature limit from 0.12 K down to 0.049 K (admittedly a limited temperature range), also consistent with line nodes. Izawa et al. [575] analyze their low temperature (down to 0.04 K) thermal conductivity data as consistent with line nodes.

Thus, in summary, $CePt_3Si$ qualifies on several counts as an UcS, with the caveat that the power law measurements are over limited temperature ranges.

**Table 9: Strongly Suggestive Evidence for UcS for CePt₃Si.** The symbol '☒' means that that property was specifically looked for but not found. Note that time reversal symmetry breaking has been found in the low correlation non-centrosymmetric superconductors ZrRe₆ and LaNiC₂.

| | 1 | 2 | 3 | 4 | 5 | 6 | 7 | 8 | 9 | 10 | 11 | 14 |
|---|---|---|---|---|---|---|---|---|---|---|---|---|
| | $T_F \ll \Theta_D$ | QC | Spin res | PG | $T^\alpha$ | >1 sc phase | tunneling ⇒ non-BCS | TRSB | $T_c$/non-mag imp | Non-centro-sym | $C(H,\Theta)$ $\kappa(H,\Theta)$ | $\gamma_{residual}$ large |
| CePt₃Si | ✓ | | | | ~✓ | ☒ | ☒ | ☒ | ✓ | ✓ | ✓ | ☒ |

### 8.2. *Other strongly correlated non-centrosymmetric superconductors: CeRhSi₃, CeIrSi₃*

At zero pressure CeRhSi₃ orders antiferromagnetically at 1.8 K and has a specific heat γ=120 mJ/molK² (Muro et al. [576]). Applied pressure creates a superconducting dome, and the $T_N$ vs pressure line joins the dome at 2 GPa (Kimura et al. [577]). The peak $T_c$ is 1.05 K at 3 GPa. CeRhSi₃ has an extremely large upper critical field – estimated to be ~30 T in 2.9 GPa, far above the paramagnetic limit ($H_p = 1.8$ kT$_c$/μ$_B$). This is cited [568] as an indication of possible triplet behavior, and is apparently (Kneidinger et al. [67]) the only measurement which implies UcS in CeRhSi₃.

CeIrSi₃ is similar to CeRhSi₃. In zero applied pressure, $T_N$=5 K and γ=125 mJ/molK² (Muro et al. [576]). 2.5 GPa applied pressure corresponds to the peak in the pressure-induced superconducting dome, with $T_c$=1.6 K. (Sugitani et al. [578]). The upper critical field, although larger than the paramagnetic limit, is reduced from that of CeRhSi₃ (despite the higher $T_c$ of the Ir compound), with $H_{c2}(0)$=11 T at 2.5 GPa. The resistivity at zero pressure is proportional to $T^2$ (Fermi liquid like), and behaves linearly (non-Fermi liquid like) with temperature from $T_c$ up to 18 K in 2.5 GPa. (Sugitani et al. [578]) Again like in CeRhSi₃, except for the large upper critical field (and the possibility of a quantum critical point in the phase diagram where $T_N \rightarrow 0$ and indicated by the $\rho \propto T$), CeIrSi₃ has little in the way of experimental evidence for UcS.

### 8.3. *Low correlation, ASOC induced examples*

LiPt₃B is a superconductor (T. Badica et al. [579]) with $T_c$=2.7 K. Nishiyama et al. [580], using NMR data, propose that this non-centrosymmetric compound is a spin triplet superconductor with line nodes, with weak electron-electron correlations. They state that the strong spin orbit coupling causes the spin triplet state to dominate, rather than the spin singlet pairing found in the partner compound, LiPd₃B.

μSR data in the non-centrosymmetric superconductor ZrRe₆, $T_c$=6.75 K, by Singh et al. [581] showed spontaneous static magnetic fields below $T_c$, indicating (see section 2.1.8. above) breaking of time reversal symmetry.in the superconducting state, and UcS with strong singlet-triplet pairing symmetry mixing. The specific heat γ=26 mJ/molK² [581], which is definitely not a highly correlated value, especially considering that six times the elemental γ of Re plus the elemental γ of Zr would (Stewart [19]) already be 16.5 mJ/molK². Similar μSR data (Hillier et al. [582]) in the non-centrosymmetric superconductor LaNiC₂, $T_c$=2.7 K, also indicate the appearance of spontaneous magnetic fields below $T_c$, and thus time reversal symmetry breaking and UcS with dominant triplet pairing symmetry.

Thus, non-centrosymmetric superconductors add two members ($ZrRe_6$ and $LaNiC_2$) to our list (B-phase $UPt_3$, $Sr_2RuO_4$, $PrOs_4Sb_{12}$, and $PrPt_4Ge_{12}$) of UcS with time reversal symmetry breaking so far in this review, with the non-centrosymmetric $ZrRe_6$ and the filled skutterudite $PrPt_4Ge_{12}$ having the largest $T_c$'s (6.75 and 7.9 K respectively) by more than a factor of two.

Finally, there are a large number (see Kneidinger et al. [67] for an overview) of known non-centrosymmetric superconductors that, due to weak correlations and a lack of strong spin orbit coupling, are simply normal BCS superconductors. Such materials predate the discovery of UcS in the non-centrosymmetric, highly correlated $CePt_3Si$ by over three decades, and include the high $T_c$ (17 K) $Y_{0.7}Th_{0.3}C_{1.5}$ (discovered by M.C. Krupka et al. [583]), where the specific heat $\gamma$ is only (Stewart, Giorgi and Krupka [584]) 4.7 mJ/molK$^2$.

Theoretically, non-centrosymmetric superconductors have been calculated to show a rich variety of complex phenomena, including – e. g. in quasiparticle tunneling – multigap features and spin currents carried by surface states. For reviews, see M. Sigrist [585] and several articles in Bauer and Sigrist [568].

## 9. Organic superconductors

The discovery of organic superconductors was published in early 1980, shortly after the heavy fermion $CeCu_2Si_2$. Jérome et al. [7] reported superconductivity, $T_c$=0.9 K, in the quasi-one dimensional $(TMTSF)_2PF_6$ under 1.2 GPa pressure (which suppresses the metal insulator (Mott localization) transition at 12 K). Several anions other than $PF_6$ cause similar superconducting behavior around 1 GPa, with $ClO_4$ resulting (K. Bechgaard et al. [586]) in $T_c$~1.4 K at room pressure.

Two dimensional organic superconductivity, $T_c$=1.4 K, at zero pressure was discovered in 1984 (Yagubskii et al. [587]) in β-phase $\beta$-$(BEDT-TTF)_2I_3$ (also called $\beta$-$ET_2I_3$). Soon thereafter, Tokumoto et al. [588] discovered a high $T_c$ phase ($T_c$=8 K) of the same compound, $\beta_H$-$(BEDT-TTF)_2I_3$, that could be stabilized at zero pressure by cycling up to 0.13 GPa and back to ambient. Further progress in increasing $T_c$ in 2D organic superconductors was made by Jack Williams' group at Argonne National Laboratory bringing $T_c$ up to 11.6 K in the ambient pressure superconductor κ-phase $\kappa$-$(ET)_2Cu[N(CN)_2]Br$ (Kini et al. [589]) and to 12.8 K under 0.03 GPa in κ-phase $\kappa$-$(ET)_2Cu[N(CN)_2]Cl$ (Williams et al. [590]).

The restricted-dimension organic superconductors, where the superconductivity comes from p-electrons from the chalcogens, have many interesting and unique properties. These properties can be measured in a single sample (i. e. no change of chemical purity) simply by varying pressure. For example, the 12.8 K superconductor $\kappa$-$(ET)_2Cu[N(CN)_2]Cl$ can be tuned (S. Lefebvre et al. [591]) between insulating (P=0), metallic, antiferromagnetic, and superconducting (P=0.03 GPa) states within a few hundreths of a GPa. While in the cuprates superconductivity occurs near (hole-doped) or coexistent with (electron-doped) antiferromagnetism (Fig. 14), in the organic superconductors superconductivity occurs in the phase diagram near to Mott localization due to a balance between the correlation strength and the strength of the interchain coupling.

In the early years of research on the 1 and 2D organic superconductors, although there were data consistent with UcS (e. g. the work of Choi et al. [592] where a very low level of

controlled defects from proton irradiation suppressed superconductivity in (TMTSF)$_2$PF$_6$ under pressure – consistent with triplet pairing (✓9), see also **S** Bouffard et al. [593]), in general the belief (Brown et al. [594]) was that these new materials were conventional superconductors.

However, theorists began to propose UcS. Abrikosov [595], citing the Bouffard et al. T$_c$ suppression with low irradiation result, proposed triplet pairing for (TMTSF)$_2$PF$_6$. Gorkov and Jerome [596], citing H$_{c2}$ values that exceed the Pauli paramagnetic limit (see the discussion of non-centrosymmetric superconductors above), also proposed triplet pairing for the organic superconductors. Bulaevskii [597] discussed unconventional pairing for the 2D organic superconductors, and Bourbonnais and Caron [598] proposed a spin fluctuation mediated coupling between the chains in 1D organic superconductors. More recently, 2D organic superconductors like β''-(ET)$_2$SF$_5$CH$_2$CF$_2$SO$_3$ have been discovered where there are apparently no low frequency antiferromagnetic fluctuations and coupling is believed to be mediated by charge fluctuations (G. Koutroulakis et al. [599]).

By the early to mid-1990's, a number of physical measurements had been gathered that – with some exceptions – confirmed UcS in the organic superconductors. These measurements mostly concentrated on the highest T$_c$ ambient pressure (2D) superconductor, κ-(ET)$_2$Cu[N(CN)$_2$]Br and, for the 1D materials, on (TMTSF)$_2$X (X=PF$_6$ under pressure and X=ClO$_4$ at ambient pressure).

## 9.1. *1D*

For X=ClO$_4$, <u>line nodes</u> concluded from: measurements (Takigawa et al. [600]) of NMR 1/T$_1$ (down to T$_c$/2)~T$^3$; measurements (Shinagawa et al. [601]) of the Knight shift implying singlet (d-wave) pairing; measurements (Yonezawa et al. [602]) of C(H, Θ) (d-wave pairing). Bourbonnais and Sekeki put forward a theory [89] for these materials with d-wave superconductivity caused by the exchange of spin correlations. Doiron-Leyraud et al. [603] applied pressure to cause superconductivity in (TMTSF)$_2$PF$_6$. At 1.18 GPa, where T$_c$ is a maximum and T$_{SDW}$→0, they found that the resistivity was linear with temperature between 0.1 K (T$_c$ suppressed by 0.05 T) and 8 K. This implied quantum critical behavior (✓2) at the critical point where the second order magnetic transition is suppressed to T=0.

## 9.2. *2D*

For κ-(ET)$_2$Cu[N(CN)$_2$]Br (κ-Br for short): measurements of scanning tunnel spectroscopy (STS) were interpreted as strongly supporting d-wave pairing with line nodes (Ichimura et al. [604]); measurements (Mayaffre et al. [605]) of NMR 1/T$_1$ ~ T$^3$ indicate strongly anisotropic superconducting gap or maybe nodes, with no Hebel Schlichter peak in 1/T$_1$ below T$_c$, i. e. arguing against s-wave; measurements (De Soto et al. [606]) of NMR indicate unconventional pairing, possible nodes; measurements (Kanoda et al. [607]) of NMR indicate UcS with strong gap anisotropy and possibly nodes; measurements in κ-Br *and* κ-(ET)$_2$Cu(NCS)$_2$ (κ-NCS for short) (Le et al. [608]) of penetration depth λ~T consistent with line nodes; measurements (Achkir et al. [609]) in κ-NCS of penetration depth imply line nodes and UcS; measurements (Dressel et al. [610]) in κ-Br *and* κ-NCS of surface impedance indicate an anisotropic superconducting gap but without nodes in the gap; measurements (Taylor, Carrington and Schlueter [611]) in κ-Br *and* κ-NCS of the specific heat C~T$^2$ plus a fit of C from 0.1 T up to T$_c$

both indicate d-wave pairing symmetry; measurements (Milbradt et al. [612]) in κ-Br of surface impedance indicate d-wave behavior.

In summary, most of these measurements support a superconducting energy gap with nodes, i. e. UcS.

Besides these measurements, there is not much other evidence for UcS in the organic superconductors. There is apparently no pseudogap, 3✗ (Brown review [74]), and the small masses of crystals available prohibit neutron scattering investigations of a possible resonance, 4✗. Further, to date there are no phase sensitive determinations of the pairing symmetry, although there are STS measurements 7✓ as already mentioned. Specific heat investigations of (TMTSF)$_2$ClO$_4$ (Garoche et al. [613]) give γ=10.5 mJ/molK$^2$ and on κ-(ET)$_2$Cu[N(CN)$_2$]Br (Andraka et al. [614]) provided a value for γ of 22 mJ/molK$^2$. Thus, γ of neither organic superconductor is anywhere near large enough to satisfy strongly suggestive evidence #1, 1✗). Also, there was no evidence of a significant residual γ value in (TMTSF)$_2$ClO$_4$ or κ-(ET)$_2$Cu[N(CN)$_2$]Br, thus ✗14.

Finally, although there is strong evidence for the FFLO phase (where high magnetic field in a clean singlet, Pauli-paramagnetic-limited superconductor causes an inhomogeneous, finite momentum state) in several organic superconductors, e. g. in **κ-NCS** (Wright et al. [615]; Lortz et al. [616]; Singleton et al. [617]; Bergk et al. [618]; in **λ−(BETS)$_2$FeCl$_4$**, a field induced superconductor (Uji et al. [619]) above 17 T (Uji et al. [620]) and in **β''-(ET)$_2$SF$_5$CH$_2$CF$_2$SO$_3$**, T$_c$=4.3 K, (G. Koutroulakis et al. [621]), this is not necessarily evidence for UcS.

**Table 10: Strongly Suggestive Evidence for UcS for Organic Superconductors.** The symbol '✗' means that that property was specifically looked for but not found

|  | 1 | 2 | 3 | 4 | 5 | 6 | 7 | 8 | 9 | 11 | 14 |
|---|---|---|---|---|---|---|---|---|---|---|---|
|  | T$_F$ ≪ Θ$_D$ | QC | Spin res | PG | T$^α$ | >1 sc phase | tunneling ⇒ non-BCS | TRSB | T$_c$/non-mag imp | C(H,Θ) κ(H,Θ) | γ$_{residual}$ large |
| organics | ✗ | ✓ |  | ✗ | ✓ | ✗ | ✓ |  | ✓ | ✓ | ✗ |

## 10. Layered Nitrides

This superconducting class consists of electron-doped layered metal nitride halides MNX, with M= Ti, Zr, Hf and the halides X = Cl, Br, I. The undoped parent compounds are 2D layered structure band insulators, with a large (>2.5 eV) gap (Kasahara et al. [622]) and two distinct structures, see Fig. 27. Unlike many of the other UcS discussed herein, there is no magnetism in the phase diagram of either structure. The α structure has an orthogonal metal-nitride layer network with two halogen layers in between, while the β structure has a double honeycomb metal-nitride layer with two close packed halogen layers in between. (See Zhang et al. [623], Yamanaka [624] and Kasahara et al. [622] for more structural details.) To cause superconductivity, electron doping is carried out by intercalation of atoms such as Li, Ca, Na, K and Yb between the twin halogen layers, sometimes with co-intercalation of organic molecules to spread the halogen layers further apart. As well, Zhu and Yamanaka [625] discovered that de-intercalating the Cl in β-(Zr,Hf)NCl$_{1-x}$, x~0.3, causes sufficient electron doping to result in approximately the same superconducting transitions as when doping with, e. g., Li.

Fig. 27 (color online) From Zhang et al. [623]. Pictorial representation of the two layered metal nitride halides MNX structures. The nitrogen, N, is blue, the metal atoms, M, are red, and the halide atoms X are large light brown spheres. (figure is reproduced from ref. [623]. © IOP Publishing. Reproduced with permission. All rights reserved.)

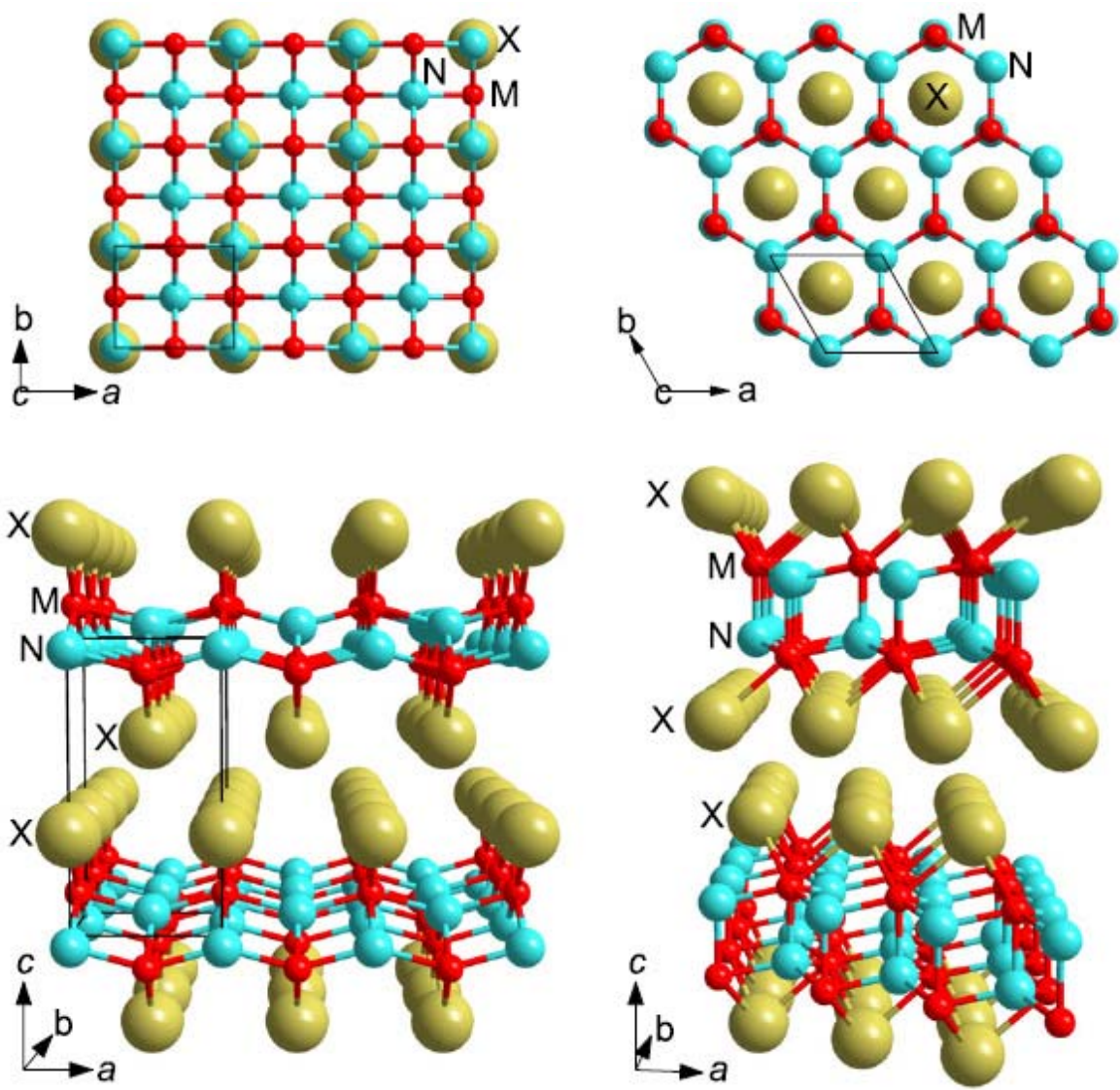

Superconductivity in this class of superconductor was first discovered (Yamanaka et al. [626]) in 1996, with $T_c^{onset}$=12.5 K in β-ZrNCl intercalated with Li: $Li_{0.16}ZrNCl$. This first layered nitride superconductor, with its $T_c$ higher than the corresponding 3D metal nitride ZrN ($T_c$=9.05 K (Matthias and Hulm [627])) was surpassed in $T_c$ two years later with β-HfNCl intercalated with $Li_{0.48}(THF)_{\sim 0.3}$ (where THF is tetrahydrofuran), $T_c$=25.5 K for $Li_{0.48}(THF)_{\sim 0.3}HfNCl$. The Hf compounds are quite air sensitive. (Kasahara et al. [628])

The α form of MNX has been intercalated between the halide layers with various alkali metals, with or without additional organic molecules (PC, propylene carbonate, and BC, butylene carbonate) besides THF, with success at causing superconductivity up to 16.3 K via electron doping into the insulating parent compound first in 2009. (Yamanaka et al. [629]). See Fig. 28 for $T_c$ of intercalated α-TiNCl, where $T_c$ for this sequence is plotted as a function of the inter-MN layer spacing d.

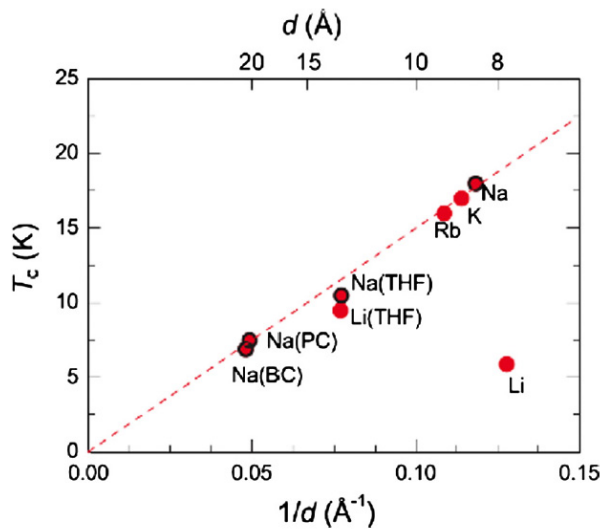

Fig. 28 (color online) $T_c$ vs log(d) and 1/d for intercalated α-TiNCl (figure from Zhang et al. [630]). The outlier, $Li_{0.13}TiNCl$, is thought to have a lower $T_c$ because the Li ions are small enough to penetrate into the Cl layer, while the process of co-intercalation with the THF molecule is thought to retain the Li ions between the Cl layers. A similar graph for intercalated α-TiNBr (Zhang et al. [623]), without the Na(BC) and the Li points, shows very similar behavior of $T_c$ vs 1/d. Both the α-TiNCl and the α-TiNBr intercalated compounds are air sensitive. For the β-form, $T_c$ increases with d at small d and then flattens out with increasing d (K. Hotehama et al. [631]), rather than increasing linearly with 1/d as shown here for the α-form. Reprinted figure with permission from Zhang et al., Phys. Rev. B 86 (2012), p. 024516 [630] Copyright (2012) by the American Physical Society.

### 10.1. *Superconducting properties that impinge on possible UcS, β-form*

The upper critical field, $H_{c2}$, of this class of superconductors is quite high and gives convincing evidence for quasi-2D behavior, as expected based on the structure (Fig. 27). For H⊥c in de-intercalated $β-ZrNCl_{0.7}$, $H_{c2}(T=0)$= 27 T, with $H_{c2}(T=0)$ for H∥c equal to 7 T. (H. Tou et al. [632]) Another way to quantify anisotropy using $H_{c2}$ measurements that avoids the need for such high fields is to compare the *slopes* $dH_{c2}/dT$ ($\equiv H_{c2}'$) near to $T_c$, which from Tou et al.'s work at H=1 T is $H_{c2}'_{\perp c}/H_{c2}'_{\parallel c} \sim 4.5$.

The isotope effect in Li-intercalated β-HfNCl decreases the 25.5 K $T_c$ upon substitution of $^{15}N$ for $^{14}N$ by only 0.1 K, implying that the exponent in $T_c \propto M^{-\alpha}$ is α=0.07±0.02. (H. Tou et al. [633]) The isotope effect in Li-intercalated β-ZrNCl decreases $T_c$=11.4 K upon substitution of $^{15}N$ for $^{14}N$ by only 0.06 K, implying that the exponent in $T_c \propto M^{-\alpha}$ is α=0.07±0.04. As will be discussed below, both these works argued that the data suggest considering the relevance of something instead of/in addition to the conventional electron-phonon pairing interaction although, as discussed above in section 2.1.13., the *absence* of an isotope effect can be misleading.

NMR measurements of $1/T_1$ and the Knight shift in $\beta$-Li$_x$ZrNCl show singlet behavior with no Hebel-Schlichter peak. (H. Kotegawa et al. [634]; Kasahara et al. [622]).

Using band structure calculations of the electronic density of states at the Fermi energy and their specific heat data (where the rather low value of $\gamma=1.1$ mJ/molK$^2$ is found, similar to the value of 2.7 mJ/molK$^2$ for conventional electron-phonon coupled superconductor MgB$_2$, T$_c$=39 K) for $\beta$-Li$_{0.12}$ZNCl, T$_c$ = 12.7 K, Taguchi et al. [635] argue that the electron phonon coupling parameter $\lambda$ is only 0.22. They then argue that this value used in the McMillan formula gives T$_c$ too low by a factor of five.

Kasahara et al. (Y. Kasahara et al. [636]) report the specific heat of $\beta$-Li$_x$ZrNCl for various x, and find for 0.10≤x≤0.40, 15 K≤T$_c$≤10 K, that $\gamma$ as a function of magnetic field H rises quickly at first, and then more slowly. Such behavior is indicative of either two s-wave gaps (Bang [83])) or strong anisotropy/possible nodal behavior in a single gap (Wang et al. [526]). The calculation of large $2\Delta/k_BT_c$ values (up to 4.5) by Kasahara et al. [636], which they claimed as evidence for an increase in coupling strength with doping above x=0.07, was done using an empirical theory (Padamsee, Neighbor, and Shiffman [637]) that assumed an "isotopically [sic] gapped state," which is questionable based on the evidence from their own $\gamma$(H) data which implied strong anisotropy.

μSR measurements (Hiraishi et al. [638]) on $\beta$-Li$_x$ZrNCl were interpreted to mean an anisotropic gap, in agreement with the specific heat $\gamma$(H) data already discussed. μSR measurements (T. Ito et al. [639]) on $\beta$-Li$_x$(Hf,Zr)NCl, both with and without co-intercalated THF/PC, were interpreted as strongly favoring 2D superconductivity with only weak interlayer coupling. (This is of course in contrast to the data for $\alpha$-TiNCl in Fig. 28, where T$_c$ is a strong function of interlayer distance.)

Due to the air sensitivity of $\beta$-ZrNCl$_{0.7}$, tunneling data were measured (Takasaki et al. [640]) by breaking a pressed-pellet sample in an inert atmosphere glove box to expose a fresh surface ("break junction" tunneling spectroscopy.) More recently, the same composition was measured (T. Ekino et al. [641]) by the same group, but using a single crystal facet on a polycrystalline sample as the probing surface. In both works there were two gap features, one similar to the BCS value ($2\Delta/k_BT_c \sim 3.52$) plus a rather large gap feature ($2\Delta/k_BT_c \sim 10$), with both gaps disappearing as the sample is warmed to T$_c \sim 13$ K, leading the authors to conclude that this compound displays "unusual" superconducting properties. This large gap apparently has not been confirmed by other measurement techniques – unlike a similar measured very large (factor of two to three larger than the BCS $2\Delta/k_BT_c$=3.52) gap in the cuprates, see Fig. 21 for the cuprate superconducting energy gap and the multiple measurement techniques.

## 10.2. *Superconducting properties that impinge on possible UcS, $\alpha$-form*

The $\alpha$-form, where superconductivity was discovered 13 years after its discovery in the $\beta$-form, is much less well characterized. Based both on the structural differences (Fig. 27) and on the different behavior of T$_c$ with interlayer spacing d (Fig. 28), the superconductivity in the $\alpha$-form may well be different than in the $\beta$-form of the layered metal nitrides. Certainly, the $\alpha$-form's T$_c$ $\propto$ 1/d dependence suggests that the Coulomb interlayer coupling in these materials is more central to the pairing mechanism than in the $\beta$-form.

Unlike in the $\beta$-form, where H$_{c2}$(T=0) was different by ~factor of four in the planar and c-axis directions (and H$_{c2}'$$_{\perp c}$/ H$_{c2}'$$_{\parallel c}$ at 1 T was 4.5), in the $\alpha$-form of intercalated TiNCl H$_{c2}'$$_{\perp c}$/

H$_{c2}'_{\|c}$ ~ 1.2 to 1.5 (Zhang et al. [630]), depending on the intercalate. This lack of anisotropy in H$_{c2}$ is also observed in intercalated TiNBr (Zhang et al. [623]), with T$_c$ values up to 17.2 K. Thus, the α-form is more isotropic than the β-form, which agrees with bandstructure calculations (Kasahara et al. [622]).

## 10.3. *Discussion of the Possible Superconducting Pairing Mechanism in the β-form (honeycomb lattice) Layered Metal Nitrides*

Unfortunately, much of the theoretical discussion is focused on the β-form with its higher (25.5 vs 16.3 K) T$_c$. The α-form is experimentally significantly different in two important aspects: it is more isotropic, and T$_c$ is a linear function (see Fig. 28) of the inverse of the interlayer spacing, 1/d.

### 10.3.1. *Phonon coupled*

Phonon-mediated pairing in β-Na$_x$HfNCl, T$_c$=25 K, was considered by Weht, Filippetti, and Pickett [642]. In their band structure calculations, the electron-phonon coupling was considered too weak to explain the relatively high T$_c$, and the authors concluded that "the origin of the high T$_c$ remains mysterious." Heid and Bohnen [643] performed calculations for β-Li$_x$ZrNCl, and came to the same conclusion: "the calculated electron-phonon coupling is too weak to explain the observed T$_c$."

### 10.3.2. *Low energy conduction electron collective modes (plasmons)*

Bill, Morawitz, and Kresin [644] calculate a strong coupling phonon-plasmon scheme to explain the T$_c$ in this class of materials, where the dynamic screening of the Coulomb interaction in the superconducting layered metal nitrides is the dominant interaction.

### 10.3.3. *Spin fluctuations in a honeycomb lattice (β-form, Fig. 27) doped band insulator*

Despite the weakness of the spin fluctuations, Kuroki [645] considers this a possible model to explain the superconductivity in the β-form layered metal nitrides due to the disconnectivity of the Fermi surface in the honeycomb lattice. If this model applies, Kuroki predicts d + id gap symmetry and breaking of time reversal symmetry.

### 10.4. *Summary*
The layered metal nitride superconductors may well be UcS, although they are rather understudied – particularly in the α-form, at least partially due to air sensitivity and lack of single crystals. Also, a T$_c$ of 25 K discovered in 1998 did not draw the attention that it would have in 1973, when A15 structure Nb$_3$Ge held the record (Stewart [2]) for the highest T$_c$ at 22.8 K. It is certainly still under discussion whether this class is an example of UcS, with no conclusive strong substantiating evidence as yet.

**Table 11: Strongly Suggestive Evidence for UcS for Layered Nitrides.** The symbol '☒' means that that property was specifically looked for but not found

|   | 1 | 2 | 3 | 4 | 5 | 6 | 7 | 8 | 9 | 11 | 13 | 14 |
|---|---|---|---|---|---|---|---|---|---|----|----|----|
|   | $T_F \ll \Theta_D$ | QC | Spin res | PG | $T^\alpha$ | >1 sc phase | tunneling $\Rightarrow$ non-BCS | TRSB | $T_c$/non-mag imp | $C(H,\Theta)$ $\kappa(H,\Theta)$ | Lack of isotope effect | $\gamma_{residual}$ large |
| $Li_{0.1}ZrNCl$ | ☒ |   |   |   |   |   |   |   |   |    | ✓  |    |

## 11. Unconventional Superconductivity caused by Coexistence with Ferromagnetism, has to break TRS

This class is similar to the non-centrosymmetric class (section 8) in that there are only a few known UcS examples in this class, along with several conventional superconductors that are coexistent (due to special circumstances as will be described) with ferromagnetism. However, the ferromagnet-superconductor UcS's have a large amount of theoretical interest despite the limited number of compounds. In order to avoid the magnetic field breaking anti-spin aligned Cooper pairs, triplet (e. g. p-wave) pairs are favored in cases where $T_{Curie} > T_c$ – unless of course (for reasons of, e. g., separation of the moment-bearing ions from the ions supplying the superconducting electrons) the local B field is less than $H_{c2}$ for the superconducting electrons.

### 11.1. *Conventional coexistent superconducting/ferromagnetic compounds*

A long-standing example of conventional superconductivity coexisting with ferromagnetism is the Chevrel phase $HoMo_6S_8$. With special care given to forming a single domain using external fields to reduce the induction B through the demagnetizing field, the local exchange field, $B_{loc}$, from the Ho ions can be made less than the upper critical field of the superconducting Mo electrons, and the sample, $T_{Curie} \sim 0.7$ K, is superconducting below 0.1 K. (Burlet et al. [646]). Without such special effort, the superconductivity in $HoMo_6S_8$ (which forms below $T_c$=1.82 K) is destroyed below $T_{Curie}$=0.61 K, i. e. the only coexistence is between superconductivity and an oscillatory magnetic state between 0.71 and 0.61 K. (Lynn et al. [647])

$RuSr_2GdCu_2O_8$ achieves (Tallon et al. [648]) coexistent superconductivity, $T_c$=37 K, and ferromagnetism, $T_{Curie}$=132 K, in a fashion similar to $HoMo_6S_8$, by spatially separating two sets of ions. In the case of $RuSr_2GdCu_2O_8$, the Ru (ferromagnetic, ~ 0.1 $\mu_B$/Ru (Yang et al. [649]) is in $RuO_2$ planes and causes a rather small (~ 10 G) magnetic dipole field in the superconducting $CuO_2$ planes. Of course, since the superconducting $CuO_2$ planes can be expected to exhibit UcS just as in a regular, non-ferromagnetic cuprate, this is a somewhat different example of a mix between superconductivity and ferromagnetism. This compound is mentioned here together with $HoMo_6S_8$ in the 'conventional' section because the Ru ferromagnetism in $RuSr_2GdCu_2O_8$ is not causing the UcS.

Huxley, in his recent review of ferromagnetic superconductors (Huxley [85]), although he does not mention $RuSr_2GdCu_2O_8$, does mention $AuIn_2$, in which Herrmannsdoerfer et al. [650] discovered ferromagnetic ordering of the [113,115]In *nuclear* moments at 37 µK while the

AuIn$_2$ compound's electrons undergo a superconducting transition at 207 mK. Any coupling/interaction between the In nuclei and the superconducting compound's electrons is through polarization of the conduction electrons by the local In nuclear moments. Thus the opening of the superconducting gap in AuIn$_2$ at 207 mK (>> 37 µK) implies that there are essentially *no* conduction electrons at the Fermi energy left to polarize by the time ferromagnetic order between the In nuclei sets in. Thus, AuIn$_2$ is the ultimate in separating the fermions (in this case, the In nuclei) that are coupled ferromagnetically from the superconducting electrons, i. e. the superconductivity in AuIn$_2$ is entirely without influence from ferromagnetism and therefore entirely conventional.

**11.2.** *Unconventional coexistent superconducting/ferromagnetic compounds*

The more interesting, UcS, examples of coexistent superconductivity and ferromagnetism (UGe$_2$ (S.S. Saxena et al. [651], add references Aoki and Flouquet, JPSJ 74, 705 2005).and its two derivatives UCoGe and URhGe) were discovered starting in 2000. UGe$_2$ is a ferromagnet, $T_{Curie}$=52 K, at ambient pressure which has a resistive/bulk specific heat superconducting transition (which is sensitive to defects) at 0.8 K/0.6 K under 1.2 GPa pressure (Pfleiderer [70]/ N. Tateiwa et al. [652]), see Fig. 29. The phase diagram (Fig. 29) also shows a second ferromagnetic (FM2) phase (dashed line) within the higher temperature, FM1 phase, where the transition FM1 → FM2 becomes first order [653] around the peak of the superconducting dome. (This transition has also been discussed as metamagnetic, see ref. [654].) Thus, UGe$_2$ is the first homogeneous coexistent ferromagnet/superconductor with its Curie temperature greater than the superconducting transition temperature, joined later by UCoGe (single crystalline values $T_{Curie}$=2 K and $T_c$=0.6 K) and URhGe ($T_{Curie}$=9.6 K, $T_c$=0.25 K) – both at ambient pressure. These three ferromagnetic superconductors all break time reversal symmetry (✓8).

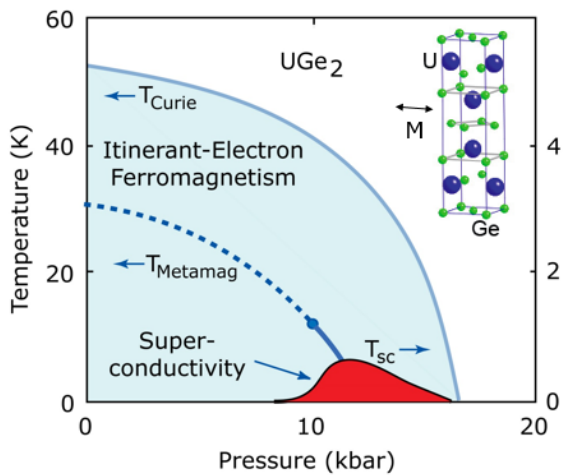

Fig. 29. From Monthoux, Pines and Lonzarich [101]. Phase diagram of ferromagnetic ordering and superconductivity as a function of applied pressure for UGe$_2$. Note that the superconducting dome's $T_c$ values are multiplied by a factor of 10 to make them more visible in the plot. Reprinted by permission from Macmillan Publishers Ltd: Nature [101], copyright (2007).

For UGe$_2$, the specific heat $\gamma_n$=32 mJ/molK$^2$ at zero pressure and ~100 mJ/molK$^2$ under pressure, with the residual $\gamma$ in the superconducting state under pressure equal to 72 mJ/molK$^2$ (Tateiwa et al. [652]). Indications for UcS (besides the coexistent magnetism/superconductivity (with breaking of time reversal symmetry ✓8), sensitivity of T$_c$ to impurities ✓9 (as indicated by residual resistivity), and the large residual $\gamma$ ✓14) are mostly lacking – partially due to the restricted temperature range below T$_c$ and partly due to the difficulty of making measurements under pressure. The fact of coexistent superconductivity and ferromagnetism in UGe$_2$ has engendered numerous theoretical models, including spin triplet superconductivity mediated by spin fluctuations (Kirkpatrick et al. [655]); coupled CDW and SDW fluctuations (Watanabe and Miyake [656]); magnon exchange (Karchev [657]); special features of the density of states (Sandeman, Lonzarich and Schofield [658]). For a thorough (4 ½ text pages) discussion of this coexistent ferromagnet-superconductor (presumed to be p-wave), see Pfleiderer [70].

The specific heat $\gamma$ of URhGe above T$_{Curie}$ (9.6 K) is 20 mJ/molK$^2$. Below T$_{Curie}$, the term in the specific heat linear with temperature is raised to 160 mJ/molK$^2$ (Pleiderer [70]) – how much of that is due to the T$^{3/2}$ magnon contribution in a ferromagnet (difficult to distinguish from C/T $\propto \gamma$) is not known. Other than the coexistence of ferromagnetism and superconductivity, indications of UcS in URhGe include sensitivity of T$_c$ to lattice defects ✓9 ($\propto$ residual resistivity, see Aoki et al. [659].) Stated another way, as argued by Huxley [85] and discussed above in section 2.2.3., T$_c$ plotted vs mean free path ($\propto$ to sample quality through RRR) for URhGe follows the Abrikosov-Gorkov model (discussed in section 2.1.16.) like observed for UPt$_3$ and qualitatively for CeCoIn$_5$. In addition, another indicator of UcS in URhGe is the large residual $\gamma$ in the superconducting state (~ ½ $\gamma_n$). (Aoki et al. [660])

The T$_c$ of UCoGe is, unlike UGe$_2$ and URhGe, insensitive to sample quality (Aoki et al. [659]) as determined by a T$_c$ that is independent of RRR down to RRR=2. This brings into question the normal assumption that a coexistent ferromagnet/superconductor is in a spin triplet state. The normal state specific heat $\gamma_n$=57 mJ/molK$^2$ (Pfleiderer [70]). As discussed in Huxley [77], it is the upper critical field H$_{c2}$ (T=0) in the direction of the moment that is the important test of whether the critical field exceeds the paramagnetic limit. In the case of UCoGe, the upper critical field in the direction of the moment is only 0.6 T, i. e. less than the Pauli limit of 1.8 kT$_c$/$\mu_B$. (Huxley [85]). Thus, both the low critical field in the direction of the moment and the independence of T$_c$ on sample quality argue against triplet pairing in UCoGe.

In summary, UGe$_2$, URhGe, and UCoGe exhibit coexistent superconductivity and magnetism, with evidence for UcS fairly conclusive for the first two.

**Table 12: Strongly Suggestive Evidence for UcS for UGe$_2$ and URhGe.** The symbol '☒' means that that property was specifically looked for but not found

|  | 1 | 2 | 3 | 4 | 5 | 6 | 7 | 8 | 9 | 11 | 14 |
|---|---|---|---|---|---|---|---|---|---|---|---|
|  | T$_F$ << $\Theta_D$ | QC | Spin res | PG | T$^\alpha$ | >1 sc phase | tunneling $\Rightarrow$ non-BCS | TRSB | T$_c$/non-mag imp | C(H,$\Theta$) $\kappa$(H,$\Theta$) | $\gamma_{residual}$ large |
| UGe$_2$/URhGe | ☒ |  |  |  |  |  |  | ✓ | ✓ |  | ✓ |

## 12. Cobalt Oxide Hydrate

The two dimensional ($H_{c2}(T=0)^{ab} \sim 4\, H_{c2}(T=0)^c$) Co oxide material $Na_x(H_3O)_zCoO_2 \cdot yH_2O$ was discovered (K. Takada et al. [661]) to be a superconductor in 2003. Further work has refined the initial estimate of the sample's composition to include the ($H_3O$) listed here, and $T_c$ values are about 4.7 K for Na concentrations x between 0.26 and 0.28 (Zheng et al. [662].) For further details on the subtleties of the phase diagram, where the Co valence s=4 – (x+z) = 3.48 gives one superconducting phase vs Na content x (see Fig. 30b) while s=3.40 gives two superconducting phases separated by a non-superconducting phase vs Na content as shown in Fig. 30a, see the recent review by Sakurai, Ihara and Takada, [663]. The proper description of the non-superconducting phase is still open, with Sakurai, Ihara and Takada preferring magnetic ordering over charge modulations.

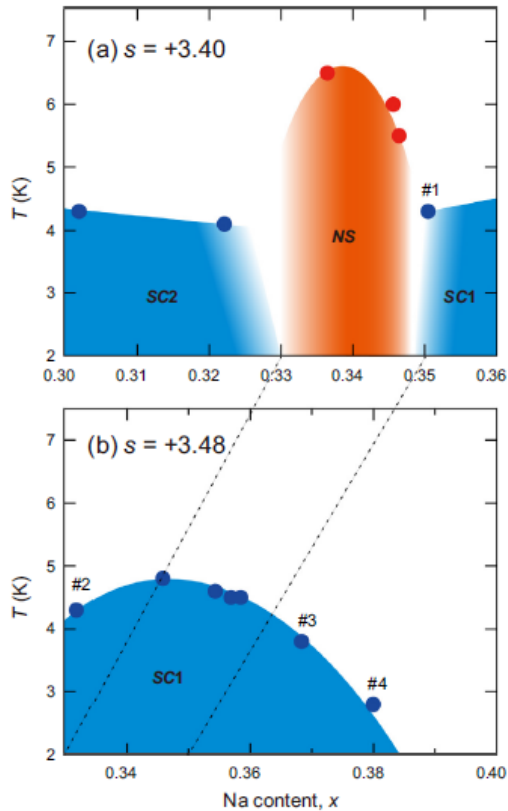

Fig. 30 (color online) (From Sakurai, Ihara and Takada [663].) Two phase diagrams, $T_c$ vs Na content, in $Na_x(H_3O)_zCoO_2 \cdot yH_2O$, with valence s=3.40 in Fig. 30a with two superconducting phases below Na concentration x=0.33 and above x=0.35. In Fig. 30b, with valence s=3.48, only one superconducting phase occurs, with no non-superconducting phase centered at x=0.34 as seen for s=3.40. Reprinted from Physica C, vol 514, H. Sakurai et al., p. 378, copyright (2015), with permission from Elsevier.

Sufficient water molecules (a double layer) intercalated between the Co oxide planes are necessary for superconductivity, and this composition is known as the bilayered hydrate, while the mono-layered hydrate structure is not superconducting. This dependence on the water content, and the tendency of the water to evaporate at ambient conditions, makes this compound the material with the most sample dependence of any in this review. It displays also the extremal aging dependence of any sample discussed herein: while $^{239}PuCoGa_5$, $T_c$=18.5 K, due to self-irradiation damage, shows [431] a diminishing $T_c$ ($\Delta T_c \sim$ -0.5 K in one month), $Na_x(H_3O)_zCoO_2 \cdot yH_2O$ shows a superconductivity that changes on the scale of (two) days (Oeschler et al. [664].) Within 40 days at ambient conditions, the sample is entirely non-superconducting.

Despite the sample difficulties, some measurements indicative of UcS have been made. Photoemission (Shimojima et al. [665]) and NMR (Ihara et al. [666]) indicate a pseudogap (✓4) at about 20 meV. NMR measurements (Zheng et al. [662]) of $1/T_1T$ at one composition, x=0.26 (but not the neighboring ones with approximately the same $T_c$) indicate a $T^2$ dependence from $T_c$ down to $T_c/6$, consistent with line nodes (✓5) in the superconducting gap (see Table 1). Zheng et al. also report the presence of antiferromagnetic spin fluctuations. Ihara et al. (Ihara et al. [667]) report that their NQR data track the magnetic ordering temperature to T=0 at the point of the maximum $T_c$ in the phase diagram and argue for quantum critical fluctuations (✓2) being the cause of the superconductivity. µSR data (Kanigel et al. [668]) are also consistent with some sort of nodal behavior. Specific heat gives (Oeschler et al. [664]) a large $\gamma_{residual}$ (6.4 or 11.0 mJ/molK² depending on aging vs 16.1 mJ/molK² for $\gamma_{normal}$), ✓14. Isotope effect experiments are unclear (Yokoi et al. [669]), since the measured changes are a.) expected to be small (substitution for $^{16}O$ in the $CoO_2$ planes) and b.) uncertain due to sample dependence. Impurity doping (Ir and Ga for Co) effects on $T_c$ result in a too small effect (Yokoi et al. [670]) (1 K/%) for non-magnetic impurities (☒9) in an $\ell \neq 0$ UcS. µSR measurements (Higemoto et al. [671] and Sugiyama [672]) rule out time reversal symmetry breaking ☒8.

Discussion of theory: Using a multiorbital Hubbard model and the random phase approximation, Mochizuki and Ogata [673] find that the non-superconducting phase at s=3.4 (Fig. 30a) is magnetic, and that the two superconducting phases on either side in the phase diagram vs Na content are s± and a triplet p-wave. For the higher Co valence (s=3.48) where there is just a single superconducting phase present in the phase diagram (Fig. 30b) vs Na content, they predict that this has p-wave symmetry. Yada and Kontani [674] argue for an anisotropic (nodal) s-wave state brought about by the interaction between the electron-phonon interaction and antiferromagnetic fluctuations.

In summary, $Na_x(H_3O)_zCoO_2 \cdot yH_2O$ – with its complex phase diagram and extreme sample dependence – seems clearly to be an UcS.

**Table 13: Strongly Suggestive Evidence for UcS in $Na_x(H_3O)_zCoO_2 \cdot yH_2O$.** The symbol '☒' means that that property was specifically looked for but not found

| | 1 | 2 | 3 | 4 | 5 | 6 | 7 | 8 | 9 | 11 | 14 |
|---|---|---|---|---|---|---|---|---|---|---|---|
| | $T_F \ll \Theta_D$ | QC | Spin res | PG | $T^\alpha$ | >1 sc phase | tunneling ⇒ non-BCS | TRSB | $T_c$/non-mag imp | $C(H,\Theta)$ $\kappa(H,\Theta)$ | $\gamma_{residual}$ large |
| $Na_x(H_3O)_zCoO_2 \cdot yH_2O$ | ☒ | ✓ | | ✓ | ✓ | | | ☒ | ☒ | | ✓ |

## 13. Other Unusual Superconductors: Toplogical, H₂S, Interfacial FeSe on SrTiO₃

Here at the end of the review of UcS, we mention some interesting new superconducting classes, one ($H_2S$ under pressure) with values of $T_c$ over 200 K. At present, these emerging classes might properly be described as "possible" UcS. If $H_2S$ is finally adjudged to be conventional, this would run counter to the 30 year old experience that the route towards higher $T_c$ is via UcS based on the example of the cuprates (which up until recently held the record for high $T_c$: 166 K in fluorinated Hg-1223 at 23 GPa pressure – a slight, 2 K, increase over the 1994 value of Gao et al.) (Monteverde et al. [675]). However, the question of conventional vs unconventional superconductivity is still the subject of intensive study in the emerging classes in this section.

## 13.1. Topological Superconductors

There are several thorough (mostly theoretical due to lack of extensive experimental results as yet) reviews on this subject, e. g. Qi and Zhang [676], Sasaki and Mizushima [677], and Alicea [678]. See also the special focus issue in Superconducting Science and Technology, ref. [679]. Topological superconductors are predicted to host Majorana fermions, which is a fermion that is its own antiparticle hypothesized by E. Majorana in 1937. Majorana fermions are theorized to exist as quasiparticle excitations in superconductors (either as gapless bound surface states or in vortices). Although some researchers talk about having found topological superconductors, others state that "experimentally realizing topological superconductivity is not trivial." (Fleet [680]) Thus, the question of possible other explanations for the few experimental systems as yet identified as topological superconductors continues to be a subject of investigation.

Current discussion of strong experimental candidates for topological surface states on bulk topological insulators include [676] BiSb alloys, and $Bi_2Te_3$ and $Bi_2Se_3$ crystals. Possible topological superconductivity may occur at the interface between a topological insulator (where the bulk is an insulator and there are conducting surface states that are symmetry protected) and a conventional superconductor (theory by Fu and Kane [681]) In addition to such interfaces, hybrid devices involving InSb nanowire and normal and superconducting contacts have also produced evidence for Majorana fermions (Mourik et al. [682].)

More recently, scanning tunneling microscopy (Nadj-Perge et al. [683]) found convincing evidence for Majorana bound states at the edge of a chain of Fe atoms on superconducting Pb. Doping Cu into the topological insulator $Bi_2Se_3$ causes superconductivity at 3.8 K (Hor et al. [684]) with the topological surface states still present, as determined by ARPES, at the Fermi level of the superconductor (Wray et al. [685].) Point contact spectroscopy on Cu-doped $Bi_2Se_3$ show a zero-bias conductance peak, indicative of Majorana fermions and topological superconductivity (Sasaki et al. [686].) Instead of manipulating the topological surfaces states in $Bi_2Se_3$ via proximity to a superconductor, this discovery of superconductivity in Cu-doped $Bi_2Se_3$ offers another route to structuring novel hybrid devices.

Although the superconducting properties of the materials which are thought to be candidate topological superconductors (theoretically predicted (Nagai [687]) to be robust against non-magnetic disorder) are certainly unusual, whether or not they are unconventional is still under discussion. On one side of this continuing discussion, β-phase $PdBi_2$ – identified as a topological superconductor by Biswas et al. [688] – as characterized by μSR exhibits preservation of time reveral symmetry in a nodeless single band s-wave isotropic gap superconductor. The topologically protected surface states do not affect the bulk properties of the superconductor. On the other hand, Matano et al. [689], using NMR, find that Cu-doped $Bi_2Se_3$ has odd pairing (triplet) symmetry and state that their results have a "potential impact on establishing topological superconductivity" in this compound.

## 13.2. $H_2S$ at high pressure

The pathway to discovery of superconductivity above 200 K in $H_2S$ under pressure was an interesting and somewhat convoluted one. Fundamentally, theorists were trying to help in establishing a pathway to higher $T_c$ values using conventional Eliashberg theory for electron-phonon coupling. Three necessary conditions were listed: high characteristic phonon frequency ($<\omega_0>$), large electronic density of states at the Fermi energy ($N(0)$), and strong electron-phonon coupling. For example, Ashcroft [690], as well as others, discussed pressurizing insulating compounds (to metallize them and create a finite $N(0)$) with large hydrogen atom content (e. g. $MH_4$, or ternary compounds, with M a heavier atom), which then satisfies the high characteristic phonon frequency since $<\omega_{log}> \propto 1/(mass)^{1/2}$. Based on infrared measurements, $H_2S$ was known (Sakashita et al. [691]) to become a metal under a pressure of 96 GPa (using diamond anvil techniques available since the mid 1970's.) Various theoretical predictions were made, including one by Li et al. (Li et al. [692]) that $H_2S$ would superconduct at $T_c$=80 K under a pressure of 160 GPa. Another prediction was by Duan et al. [693] for a partially dissociated $H_2S$ (stoichiometry $H_3S$) under 200 GPa pressure with a $T_c$ between 191 and 204 K.

Experimentally, Drozdov et al. [694] reported a lengthy set of experiments. A first (unpublished) report, Drozdov, Eremets, and Troyan [695], at the end of 2014 reported $T_c$ = 190 K on $H_2S$ under pressures up to 225 GPa. Seven months later, the same team expanded on their earlier work with the highest $T_c$ reported to be ~203 K at 155 GPa. $D_2S$ was reported to have a large isotope effect with $T_c$ decreased by about 30 K. This is large compared to BCS, depending on what mass is considered appropriate in the relationship $T_c \propto M^{-\alpha}$ (i. e. is M=34 g/mol for $H_2S$ and 36 g/mol for $D_2S$ – i. e. $\alpha$ is unphysically large, or are the S modes unimportant and the vibrational frequency has only to do with the masses of $H_2$ and $D_2$, i. e. $\alpha$~0.3 ?) Already the earlier, 2014 initial report of $T_c$=190 K in $H_2S$ under pressure caused numerous theoretical analyses, e. g. Papaconstantopoulos et al. [696] who find that $H_3S$ is "basically an atomic hydrogen high temperature superconductor" involving electron-phonon (conventional) coupling. In this analysis, $H_2S$ under high pressure is a conventional, electron-phonon coupled superconductor with an electron-phonon coupling parameter $\lambda$ between 1.5 and 2.2 ($\lambda$ is around 1.7 for $T_c$~20 K A15 superconductors [2]) and a characteristic phonon frequency $<\omega_{log}>$ of 1500 to 1770 K [696]. If one takes a phenomenological formula from 1987 from Bourne et al. [697] ($T_c=<\omega_{log}>(exp(2/\lambda)-1)^{-1/2}/(2\pi)$), sometimes called the Kresin-Barbee-Cohen formula, one obtains about $T_c$=182 K for $\lambda$=2 and $<\omega_{log}>$=1500 K.

Although this view that superconductivity with $T_c$~200 K in $H_2S$ under pressure is conventional (including the predictions of Li et al. and Duan et al.) remains the consensus, there are opposing views that hold that such a high $T_c$ (and such a large isotope effect) cannot come from electron-phonon coupling. For example, Hirsch and Marsiglio [698] propose the so-called "hole superconductivity" approach, where a non-phononic mechanism pairs highly-renormalized-mass hole carriers. It has been commonly assumed (for an early paper on the subject, see Cohen and Anderson [699]) that electron-phonon coupling could not cause superconductivity at high temperatures due to the required large $\lambda$ causing lattice instabilities. Much has changed since then, and $MgB_2$, discovered (Nagamatsu et al. [700]), in 2001 with $T_c$=39 K, is accepted as a conventional electron-phonon coupled superconductor.

Also since the early work by Cohen and Anderson, numerous works on predicting $T_c$ in metallic hydrogen under pressure (germane to the discussion of $H_3S$) have been published, e. g. by Papaconstantopoulos and Klein [701] - where pure electron-phonon coupling was assumed resulting in $T_c$=234 K at 460 GPa - and by Richardson and Ashcroft [702] where both electron-electron and electron-phonon coupling contributes to a high $T_c$. How this discussion over the pairing mechanism in $H_3S$ under pressure will be resolved awaits further work.

### 13.3. *Interfacial FeSe on SrTiO$_3$*

The unfolding reports on high T$_c$ in monolayer films of FeSe (bulk T$_c$~8 K) on SrTiO$_3$ (STO) substrates were spread over almost a year. Wang et al. [463] had the initial report, with T$_c$~53 K. Submitted 9 months later, He et al. [703] reported T$_c$=65±5K. The highest T$_c$ reported so far, using four point resistivity measurements, is 109 K [704]. Various work on FeSe on STO substrates ensued, including an ARPES work (Lee et al. [465]) that indicated strong coupling between FeSe electrons and longitudinal optical phonons in STO. For recent reviews, see refs. [705-706]. Recently TiO$_2$ has also been found to work as a substrate, with monolayer FeSe on TiO$_2$ resulting in T$_c$=63 K (Rebec et al. [707].)

Obviously, much theoretical work has been devoted to understanding this unusually high T$_c$ interfacial superconductivity. One representative work is the quantum Monte Carlo study by Li et al. [708]. (See also the comment on this work by Kivelson [709].) Li et al. take the approach that they should separately calculate various (three) interactions which they estimate might be *primarily* responsible for the superconducting 'glue', and then add on the additional influence on the superconducting order parameter of either electron-phonon interactions or nematic fluctuations. The three possible primary interactions are: exchange of one of two types of antiferromagnetic fluctuations (either nearest neighbor or next nearest neighbor exchange of antiferromagnetic fluctuations) or antiferro-orbital fluctuation mediated pairing. Li et al.'s graphs show that the secondary interactions can increase the calculated order parameter (i. e. the energy gap, presumed to be proportional to T$_c$) by around 60%. The nearest neighbor antiferromagnetic fluctuations produced nodeless d-wave pairing, while the next-nearest neighbor antiferromagnetic fluctuations and the orbital fluctuations both produce s-wave pairing. In their more speculative final discussion, Li et al. argue for the d-wave pairing (i. e. nearest neighbor AFM fluctuations), enhanced either by electron-phonon exchange or nematic fluctuations (or by both.)

Clearly, the theory of Li et al. suggests UcS for interfacial FeSe on STO (or TiO$_2$). The correct explanation remains the subject of intense ongoing discussions.

### 14. Summary and Conclusions

We stated in the beginning that a goal of this review was to assist in forming a common description of the various diverse UcS classes, an equivalent to the BCS theory that described what for four and a half decades (1911-1957) had seemed a host of unconnected properties in the electron-phonon coupled superconductors. As clear from Tables 2-3, 5, and 7-13 in this review, which describe the indications for UcS of each class, the breadth of behavior in these unconventional superconductors is indeed very large. It may be the case that the classes described herein are so disparate that a common understanding – in the manner of a BCS theory – is not possible. Can coexistent ferromagnetism and superconductivity in UGe$_2$ be understood by a theory that also describes the cuprates?

We espouse the belief that a common description of a *significant fraction* of these disparate classes, which contain pairing symmetries from s± to f-wave, which see superconductivity in p-, d-, and f-electron (both 4f and 5f) systems, *must* be possible. Preparing a theory focussing on commonalities in the various UcS classes may be *easier* than describing in depth all the special features of just one class. On the other hand, perhaps, as claim various theorists, the explanation for the superconductivity in a particular class (e. g. the cuprates or the IBS) or subclass (e. g. UPt$_3$ or Th-doped UBe$_{13}$) is already published and awaits merely a wider consensus to be recognized. Such a consensus could be the starting point for a more general theory. In the end, the various indications for UcS for the 9 (up to 12) UcS classes described herein have been analyzed for reproducibility and agreement with other measurement techniques. The hoped-for overarching theory awaits only the proper connection of the various – sometimes seemingly disparate - properties together.

With that optimistic note expressed, let us review some of the extant "puzzles" and difficulties in forming an overarching theoretical description which this review of UcS has identified.

1. Chief amongst the important questions is of course: what interaction is coupling the electrons? Certainly spin fluctuations is a popular answer for a number of UcS classes where antiferromagnetism is present in the phase diagram near (e. g. in the hole-doped cuprates and in some heavy fermions) or coexistent with (e. g. in the electron-doped cuprates, in heavy fermion UPd$_2$Al$_3$ and UNi$_2$Al$_3$, and most of the IBS) superconductivity. Fig. 5 shows a linear relationship between log($T_c$) and log($T_0$), where $T_0$ is a characteristic temperature/energy of the spin fluctuations, for various heavy fermion and cuprate superconductors. Further classes of UcS discussed herein whose pairing could be describable by exchange of antiferromagnetic spin fluctuations would include high specific heat γ non-centrosymmetric superconductors, where the lack of inversion symmetry should cause anomalous spin fluctuations, which are then enhanced by strong correlations ($\Leftrightarrow$ high γ) in some materials.

If spin fluctuations are the cause of superconductivity, whence do they come? One of the commonly discussed sources is quantum criticality near in the phase diagram to where an antiferromagnetic transition is suppressed to T=0. (There are of course other possibilities for causing a QCP.) As one example, neutron scattering data (Stockert et al. [35]) on CeCu$_2$Si$_2$ have been interpreted as showing that spin fluctuations near an antiferromagnetic QCP are the <u>primary cause</u> of superconducting pairing in that compound.

It is worth noting that this review finds evidence (primarily linear-with-temperature low temperature resistivity) for quantum criticality ✓2 in the majority of the UcS classes discussed herein. As Si and Steglich [710] point out, "antiferromagnetic quantum criticality can provide a mechanism for superconductivity" in not only heavy fermions and cuprates, but also in the organics and the IBS. Moriya and Ueda [13], in their review before the discovery of the IBS, took the same point of view for the cuprates, organics, and heavy fermion UcS. Scalapino [14], making similar arguments for an overarching theory, points out that pairing mediated by spin fluctuations is not like the BCS virtual exchange of a phonon with a sharp, well defined relation between energy and momentum. Instead, spin fluctuations have a spread out spectral weight in ω and q. Also worth noting is that, as pointed out by Millis, Sachdev, and Varma [711], lower frequency spin fluctuations are pair breaking. Thus the frequency range for the optimal spin fluctuation mediated pairing corresponds to an energy that is larger than twice the superconducting $T_c$.

Despite the oft stated proposal that spin fluctuations may provide the pairing glue in various UcS, this still remains very much an open question (although there is no clear second explanation with any sort of broad acceptance as we have discussed herein.)   Anderson [102] has argued (as discussed in section 3) against exchange of antiferromagnetic spin fluctuation as responsible for pairing in the particular case of the cuprates -  stating that the frequency of such an AFM spin fluctuation interaction would be too high for the Eliashberg formalism.  Anderson's argument for the exchange of *ferro*magnetic spin fluctuations (FMSF) offering the low frequency energy scale required by the Eliashberg theory has the problem that in general, as discussed herein, with minor exceptions (e. g. possibly UPt$_3$) the experimental data show support only for AFM fluctuations being present.

As a final word, Norman [103] states –  in the specific instance of the cuprates – that in any case magnetic correlations play "a prominent role" in the superconductivity, with the Resonating Valence Bond model of Anderson [350], the afore discussed spin fluctuations, orbital currents (Varma [712]), or some mixture being the appropriate manifestations of such correlations.

2.  Perhaps the next most important puzzle is one that indeed does not appear to lend itself to linking the various UcS classes together with one description:  the pseudogap.  This phenomenon is strongly present in the cuprates, and is called "key" by Norman for understanding the pairing in the cuprates.  In the other UcS classes, a pseudogap is *possibly* present in CeCoIn$_5$,  is present in URu$_2$Si$_2$ and Na$_x$(H$_3$O)$_z$CoO$_2$ · yH$_2$O and – in the IBS - only appears to be present with any certitude in two cases:  BaFe$_2$(As$_{0.7}$P$_{0.3}$)$_2$ and Ba$_{0.75}$K$_{0.25}$Fe$_2$As$_2$.  (In fact, even the word "pseudogap" does not appear in the recent review of the IBS by Hosono and Kuroki [27].) Thus, it is incorrect to say that the phenomenon of a pseudogap is characteristic of any class of UcS other than the cuprates and perhaps cobalt oxide hydrate (the latter based on two measurements).  As discussed herein and summarized in the tables, a pseudogap has been searched for in both Sr$_2$RuO$_4$ and the organic superconductors with no success.  Thus, a theory to describe all classes of UcS must account for the cuprates showing a pseudogap while the other classes in general do not.  Perhaps the few instances of a pseudogap occurring outside the cuprate class provide an important clue, just as the presence or absence of the isotope effect in the elemental superconductors provided an experimental clue for the importance of the retarded Coulomb interaction µ* in the BCS theory.

3.  Rather than considering the non-unifying pseudogap, is there a strong indication of UcS – other than quantum criticality in the phase diagram - that is more or less commonly present/universal in the classes discussed herein?  Rather trivially, the UcS commonly have anisotropy in their gap function Δ(k), ranging from a sign changing s± gap function in the IBS; to triplet p-wave in Sr$_2$RuO$_4$, Na$_x$(H$_3$O)$_z$CoO$_2$ · yH$_2$O, UGe$_2$, URhGe, the field induced phase in CeCoIn$_5$, and UBe$_{13}$; to d-wave in the cuprates, organics, possibly heavily overdoped Ba$_{1-x}$K$_x$Fe$_2$As$_2$, several of the heavy fermions including CeCoIn$_5$; to possible f-wave in UPt$_3$ and Sr$_2$RuO$_4$ (which would then have mixed pairing symmetries).  (As thoroughly discussed herein, specific assignments of pairing symmetry, with a few exceptions, are still under discussion.)  Thus, any theory which attempts to include a broad spectrum of UcS must deal with this strong, but diverse anisotropy in the energy gap.

4. We turn now to a short list of a few of the more unique puzzles of the individual classes mentioned herein that also appear to defy the quest for a common theory of the full set of classes.

a. The second, lower $T_c$ bulk $\Delta C$ anomaly in the specific heat of $U_{1-x}Th_xBe_{13}$ (Fig. 7) around x=0.03 seems unique. No other dopant besides Th into superconducting, $T_c$=0.95 K, $UBe_{13}$ causes such a second transition, and such a second transition – which has its own phase diagram vs composition, Fig. 8 – is seen in no other material. (The second transition in $UPt_3$, according to data to date, has a fixed $\Delta C$ and fixed separation between $T_{c1}$ and $T_{c2}$ in optimally annealed samples.) The pairing symmetry at $T_{c2}$ in $U_{1-x}Th_xBe_{13}$ is under debate. Perhaps an investigation into time reversal symmetry breaking would be revealing.

b. The linear relation between $T_c$ and the c/a ratio in the 115 structure $Ce(M_{1-x}M'_x)In_5$ and $Pu(Co_{1-x}Rh_x)Ga_5$ superconducting family is an interesting puzzle. This $T_c \propto$ c/a correlation argues for some common explanation for a sub-class of superconductors which have quite disparate material's properties (e. g. the sample dependent Curie Weiss behavior of the normal state susceptibility all the way from 300 K down to $T_c$=18.5 K in $PuCoGa_5$).

c. The many differences between $La_{x-2}(Sr,Ba)_xCuO_4$ and the other hole doped cuprates certainly need to be further understood.

In summary, the classes of UcS reviewed herein show both many similar properties, as well as many properties that may be unique to an individual material. A challenge for theory is to forge a common explanation for the salient points of at least a majority of the classes. Such a starting point would help select directions for further exploration of the remaining unique puzzles and, hopefully, show the way to even higher $T_c$.

Acknowledgements: The author gratefully acknowledges helpful discussions with Y. Bang, A. Biswas, S. Brown, M. Cohen, P. Coleman, I. Eremin, R. Greene, P. Hirschfeld, S. Johnston, A. Kapitulnik, H.-Y. Kee, P. Kumar, D. Morr, T. Shibauchi, O. Stockert, J. Thompson, I. Vekhter, I. Vishik and H.-H. Wen. Several of these discussions took place while the author was at the Aspen Center of Physics, supported by National Science Foundation grant PHY1066293. Work at Florida performed under the auspices of the United States Department of Energy, Office of Basic Energy Sciences, contract no. DE-FG02-86ER45268.